\documentclass[11pt]{article}

\usepackage{amsfonts}
\usepackage{amsmath}
\usepackage{amssymb}
\usepackage{amsthm}
\usepackage[mathscr]{eucal}
\usepackage{bbold}
\usepackage{braket}
\usepackage{color}
\usepackage{comment}
\usepackage{dsfont}
\usepackage{framed}
\usepackage{jheppub}
\usepackage{mathtools}
\usepackage{physics}
\usepackage[normalem]{ulem}
\usepackage{tensor}
\usepackage{thmtools}
\usepackage{thm-restate}
\usepackage{enumitem}
\usepackage{pifont}
\usepackage{ulem}
\usepackage{tikz}
\usetikzlibrary{math} 
\usepackage{graphicx}
\usepackage{caption}
\usepackage{subcaption}
\usepackage{hyperref}

\newtheorem*{ex}{Exercise}

\newcommand{\M}{\mathcal{M}}
\newcommand{\B}{\mathcal{B}}

\makeatletter\def\@fpheader{~}\makeatother
\DeclareMathOperator{\AdS2}{\text{AdS}_{2}}

\newcommand{\BraKet}[2]{\langle #1 | #2 \rangle}

\makeatletter
\newcommand*{\defeq}{\mathrel{\rlap{%
                     \raisebox{0.3ex}{$\m@th\cdot$}}%
                     \raisebox{-0.3ex}{$\m@th\cdot$}}%
                     =} 
\makeatother

\def\be{\begin{eqnarray}}
\def\ee{\end{eqnarray}}

\newcommand{\bea}{\begin{eqnarray}}
\newcommand{\eea}{\end{eqnarray}}
\def\ben{\begin{equation}}
\def\een{\end{equation}}

     \let\r=v

\def\be{\begin{equation}}
\def\ee{\end{equation}}
\def\ba{\begin{eqnarray}}
\def\ea{\end{eqnarray}}

\def\bal#1\eal{\begin{align}#1\end{align}}
\def\bs#1\es{\begin{split}#1\end{split}}

\allowdisplaybreaks  

\numberwithin{equation}{section}

\def\be{\begin{equation}}
\def\ee{\end{equation}}
\def\ba{\begin{eqnarray}}
\def\ea{\end{eqnarray}}
\def\bal#1\eal{\begin{align}#1\end{align}}

\def\r{\rightarrow}

\def\r{\right}

\title{Euclidean and complex geometries from real-time computations of gravitational R\'enyi entropies}

\author[a]{Jesse Held,} \emailAdd{jheld@ucsb.edu}
\author[a]{Xiaoyi Liu,} \emailAdd{xiaoyiliu@ucsb.edu}
\author[a]{Donald Marolf,} \emailAdd{marolf@ucsb.edu}
\author[b]{Zhencheng Wang} \emailAdd{zcwang1@illinois.edu}

\affiliation[a]{Department of Physics, University of California, Santa Barbara, CA 93106, USA}
\affiliation[b]{Department of Physics, University of Illinois Urbana-Champaign, Urbana, IL 61801, USA}

\abstract{Gravitational R\'enyi computations have traditionally been described in the language of Euclidean path integrals. In the semiclassical limit, such calculations are governed by Euclidean (or, more generally, complex) saddle-point geometries.  We emphasize here that, at least in simple contexts, the Euclidean approach suggests an alternative formulation in terms of the bulk quantum wavefunction.  Since this alternate formulation can be directly applied to the real-time quantum theory, it is insensitive to subtleties involved in defining the Euclidean path integral. In particular, it can be consistent with many different choices of integration contour.

Despite the fact that self-adjoint operators in the associated real-time quantum theory have real eigenvalues, we note that the bulk wavefunction encodes the Euclidean (or complex) R\'enyi geometries that would arise in any Euclidean path integral.  As a result, for any given quantum state, the appropriate real-time path integral yields both R\'enyi entropies and associated complex saddle-point geometries that agree with Euclidean methods.  After brief explanations of these general points, we use JT gravity to illustrate the associated real-time computations in detail.}

\begin{document}

\maketitle

\section{Introduction}
\label{sec:intro}

Studies of quantum gravity are often formulated using Euclidean path integrals, with calculations then being performed at the semiclassical level.  Such computations typically amount to identifying interesting stationary points of the associated action and computing their saddle-point contributions, perhaps with some discussion of whether and when such saddles are expected to dominate.  A small selection of the many interesting quantities studied in this way include the entropy of black holes \cite{Gibbons:1976ue}, tunneling from metastable vaccua \cite{Coleman:1980aw,Witten:1981gj}, the Page curve and replica wormholes \cite{Penington:2019kki,Almheiri:2019qdq}, Schwarzian-mode contributions to near-extremal black hole entropy \cite{Iliesiu:2020qvm,Heydeman:2020hhw}, and tunneling between black holes and white holes \cite{Stanford:2022fdt,Blommaert:2024ftn}.

Nevertheless, the Euclidean gravitational path integral famously suffers from the conformal factor problem \cite{Gibbons:1978ac} through which the Euclidean action can be made arbitrarily negative.  The integral over real Euclidean metrics is thus manifestly divergent and cannot be taken as the fundamental definition of the theory. Without such a fundamental definition, even in the semiclassical approximation one is left to determine the relevance of saddle-points and to compute loop effects based only on what one might call physical intuition (see e.g. \cite{Halliwell:1989dy} for a discussion in the context of cosmology).   While the results of such computations are often physically satisfying (see e.g. \cite{Allen:1984bp,Prestidge:1999uq,Kol:2006ga,Headrick:2006ti,Monteiro:2008wr,Monteiro:2009tc,Monteiro:2009ke,Anninos:2012ft,Benjamin:2020mfz,Cotler:2019nbi,Marolf:2018ldl,Cotler:2021cqa}), a less ad hoc recipe is clearly desired.  A first-principles approach would be especially useful in contexts involving more controversial results such as the axion wormholes studied in \cite{Giddings:1987cg} (see also comments in \cite{Hertog:2018kbz,Loges:2022nuw,Hertog:2024nys}).

Many works \cite{Hartle:2020glw,Schleich:1987fm,Mazur:1989by,Giddings:1989ny,Giddings:1990yj,Marolf:1996gb,Dasgupta:2001ue,Ambjorn:2002gr,Feldbrugge:2017kzv,Feldbrugge:2017fcc,Feldbrugge:2017mbc} have therefore argued that one should thus fundamentally define the gravitational path integral as an integral over real Lorentz signature metrics.  Since the Lorentzian path integral is generally oscillatory, it does not converge absolutely.  But it may nevertheless converge in an appropriate conditional or distributional sense as in the case in both quantum mechanics and quantum field theory.  Furthermore, such convergence does not require a definite sign for the Lorentzian action.   Cauchy's theorem from complex analysis then suggests that integrating over the real Lorentzian contour should be equivalent to using a  ``Euclidean'' path integral that comes equipped with a specific contour of integration (which would necessarily be a complex contour from the Euclidean perspective).

We will be interested here in the specific version of this proposal advocated in \cite{Marolf:2020rpm,Colin-Ellerin:2020mva,Marolf:2022ybi}  in which the contour is defined to sum over geometries that may fail to be smooth on codimension-2 surfaces $\Upsilon$ at which the Lorentzian structure may be ill-defined.  Specifically, in \cite{Marolf:2022ybi} it was proposed to include all spacetimes that could be constructed by cutting smooth geometries along two (perhaps intersecting) surfaces of codimension-1 and then sewing the pieces back together so as to give geometries that were continuous (but which were then generally not smooth).

\begin{figure}
    \centering
   \includegraphics[width=5in]{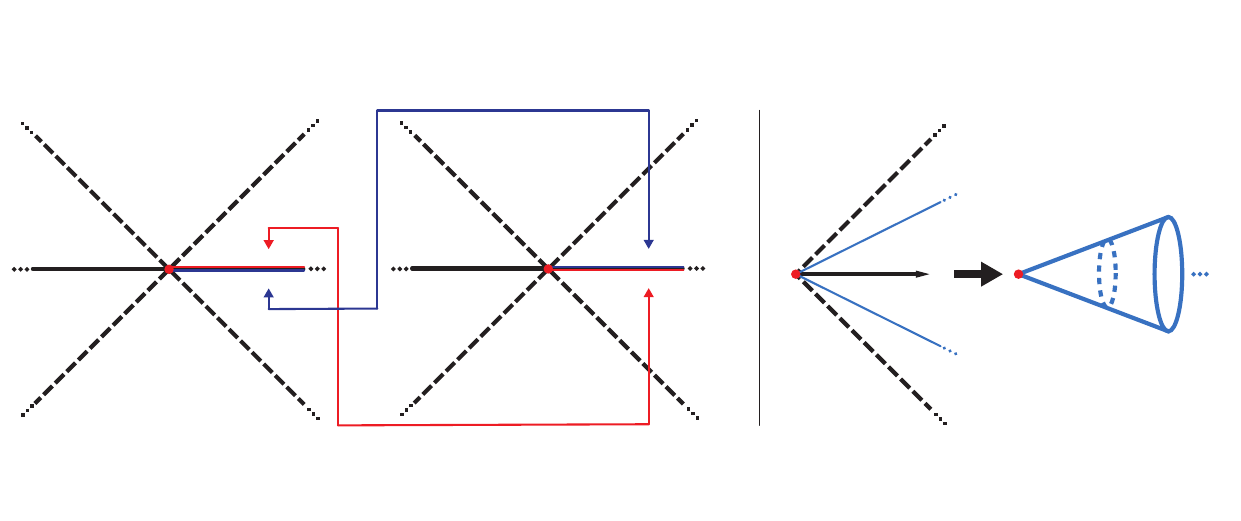}
    \caption{ {\bf Left:} Two copies of Minkowski space can be cut open along the right-half of their $t=0$ planes and glued together as indicated by the arrows.  The resulting spacetime has a special   codimension-2 surface (red dot) with 8 distinct orthogonal null congruences, 4 future-directed and 4 past-directed. {\bf Right:} Another singular spacetime obtained by taking the quotient under a boost of the right Rindler wedge. The resulting spacetime contains a singular codimension-2 achronal surface with no orthogonal null congruences.}
     \label{fig:paste}
\end{figure}
Interesting examples may be constructed by taking $n$ copies of Minkowski space and pasting them together sequentially along half-planes as shown for $n=2$ in figure \ref{fig:paste} (left).  The Lorentzian structure then becomes ill-defined at a codimension-2 surface through the origin which, instead of the usual 4  orthogonal null congruences (past/future and right/left-moving) then has $4n$ such congruences.  A different example for which the singular codimension-2 surface has no orthgonal null congruences can be obtained by taking the quotient of the right Rindler wedge under a finite-rapidity boost; see figure \ref{fig:paste} (right).  The action of such singular geometries was defined following \cite{Louko:1995jw,Colin-Ellerin:2020mva} which, as we will review, can be thought of as taking the defining contour for the path integral to be given by a certain $i\epsilon$ prescription\footnote{An interesting result of this prescription is that the Lorentz-signature action $S$ of such configurations picks up an imaginary part whose sign is given by $2-n$.  Thus for $n<2$ the integrand $e^{iS}$ is exponentially enhanced at such configurations and its magnitude can become arbitrarily large.  Nevertheless, it was suggested in \cite{Marolf:2022ybi} that the path integral should still converge when the integrations are performed in an appropriate order. A semiclassical argument that this is the case was also given in \cite{Marolf:2022ybi} for the case of gravitational partition functions.}.  This approach has been shown to reproduce familiar Euclidean results for gravitational partition functions \cite{Marolf:2022ybi}, as well as critical features of the genus expansion of the spectral form factor \cite{Blommaert:2023vbz}. A discretized version has also been shown  to reproduce the Bekenstein-Hawking entropy of de Sitter space \cite{Dittrich:2024awu}.

\begin{figure}
    \centering
   \includegraphics[width=3in]{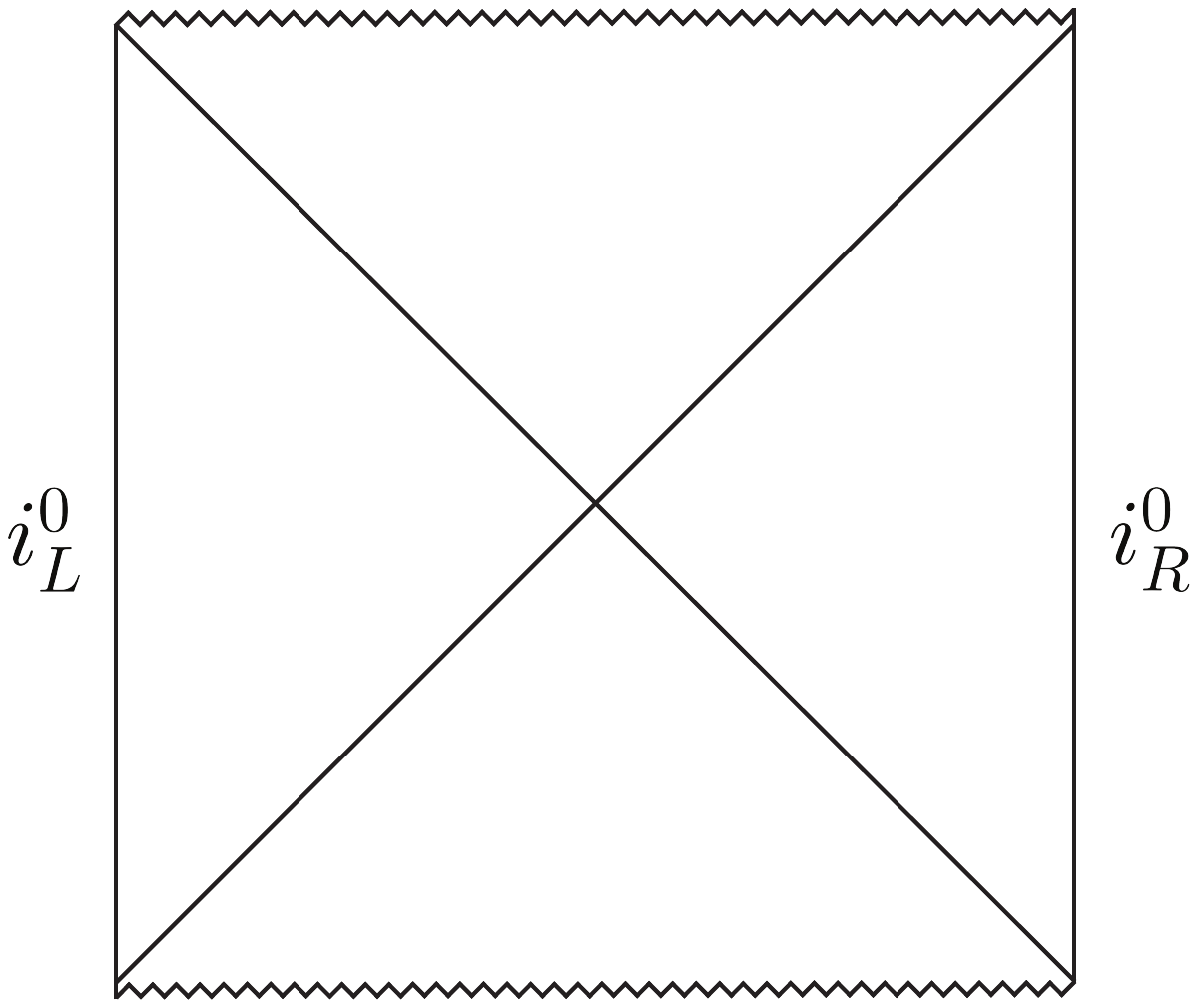}
    \caption{The conformal diagram for a 2-sided AdS Schwarzschild black hole.  In AdS/CFT, this geometry is associated with a Hilbert space $\mathcal H_L \otimes \mathcal H_R$, where the factors are associated repsectively with the left and right boundaries ($i^0_L$, $i^0_R$).}
  
    \label{fig:kruskal}
\end{figure}

Here we wish to use this scenario to compute R\'{e}nyi entropies associated with states $|\Psi\rangle$ that describe asymptotically-AdS Lorentz-signature spacetimes with two boundaries.  Such spacetimes are typified by geometries that describe two-sided black holes; see figure \ref{fig:kruskal}.  The desired R\'{e}nyi entropies are fundamentally defined by assuming the bulk to have a CFT dual or, more generally, by assuming it to have a UV completion satisfying Lorentz-signature analogs of the axioms of \cite{Colafranceschi:2023urj}.  If there is a CFT dual, it must involve a separate CFT on each boundary so that the Hilbert space takes the form ${\cal H}_{L}\otimes {\cal H}_{R}$.  We can then trace the full density matrix $\rho = |\Psi\rangle \langle \Psi|$ over either factor to define right and left density matrices $\rho_R = {\rm Tr}_L \rho,  \ \rho_L = {\rm Tr}_R \rho$; see \cite{Colafranceschi:2023urj} for the corresponding definitions in the more general case.  The R\'enyi entropies $S_n$ are given by the standard formula
\begin{equation}
\label{eq:Sn}
S_n = \frac{1}{1-n}\ln \left( \frac{{\rm Tr_L} \rho_L^n}{[{\rm Tr_L} \rho_L]^n} \right),
\end{equation}
where as usual it is convenient to allow an arbitrary normalization for $|\Psi\rangle$ (and thus for $\rho_L, \rho_R$).   Since ${\rm Tr_L} \rho_L^n = {\rm Tr_R} \rho_R^n$, \eqref{eq:Sn} may equivalently be written in terms of $\rho_R$.

The real-time path integral for such computations was first discussed in \cite{Dong:2016hjy}.
A simple class of geometries that contribute to such real-time path integrals was then discussed in more detail in \cite{Colin-Ellerin:2020mva} along with the rules for constructing  the associated saddles.   In particular, it was argued in \cite{Colin-Ellerin:2020mva} that there is always a class of saddles described by real fields at points that one may call spacelike-separated from a splitting surface $\Upsilon$, but which were complex in the future and past wedges defined by $\Upsilon$.
For the particular case where $|\Psi\rangle$ is a Hartle-Hawking state (i.e., a bulk state for which a dual description would be a thermofield-double state), such saddles are Wick-rotations of the Euclidean R\'{e}nyi saddles of \cite{Sen:2008yk,Maldacena:2016hyu} so that, when computed appropriately, their action is guaranteed to agree with familiar Euclidean results; see \cite{Colin-Ellerin:2021jev} for details. However, since such complex saddles do not lie on the real contour that defines our path integral, further analysis is required to determine if they in fact control the desired semiclassical limit.

We thus have two main goals below.  The first is to better compare Euclidean and real-time path-integral R\'enyi computations and to determine the extent to which they agree.  The second is to improve our understanding of the saddles that contribute to the real-time path integral, and in particular the sense in which such saddles are generally complex.  These goals are not independent, since agreement between Euclidean and real-time approaches would suggest that the two lead to equivalent saddles.

Comparison of the Euclidean and real-time approaches is facilitated by making use of the
bulk quantum wavefunction of the corresponding quantum state.
Indeed, it is only by showing that both compute the same bulk quantum wavefunction that one can conclude that a given set of boundary conditions for the Euclidean path integral is physically equivalent to a given set of boundary conditions for a real-time path integral.     However, the point we emphasize below is that, at least in the context of Einstein-Hilbert or JT gravity and  when the relevant spacetimes can be treated as having a unique extremal surface, both Euclidean or real-time R\'enyi admit simple reformulations that can be expressed {\it directly} in terms of this bulk quantum wavefunction.  The basic idea behind this construction was described in a number of previous works (see e.g. \cite{Dong:2016hjy,Dong:2018seb,Dong:2019piw,Takayanagi:2019tvn}), though we take care here to explicitly discuss additional gauge fixing issues and boundary terms relevant to the desired form of the bulk wavefunction.  The point we emphasize here is then that such reformulations are manifestly independent of certain subtleties involved in defining either path integral, including any details involving the choice of integration contour.  In particular, when reformulated in this way, the results of both Euclidean and real-time approaches are {\it manifestly} equivalent.  Any two such path integrals that compute the same bulk quantum wavefunction thus give identical R\'enyi entropies $S_n$.

We also observe below that, in the context of the semiclassical approximation, the reformulation in terms of bulk wavefunctions suffices to recover the saddle-point spacetimes associated with any path integral computation (in either the Euclidean or real-time formalisms).  The inclusion of appropriate boundary terms is critical for this conclusion to hold.   From this perspective it is then manifest that saddles defined by the Euclidean and real-time approaches are in some sense identical.  The construction of such saddles from the bulk wavefunction also makes clear the sense in which they are {\it necessarily} Euclidean or complex.

We provide brief  explanations of the above points in sections \ref{sec:Rfw} and \ref{sec:RTRenyi} below.  Section \ref{sec:Rfw} reformulates the standard Euclidean R\'enyi computations in terms of the bulk wavefunction and describes the association with complex saddle-point spacetimes.  Section \ref{sec:RTRenyi} then considers 
the real-time path integrals of \cite{Dong:2016hjy,Marolf:2020rpm,Colin-Ellerin:2020mva,Marolf:2022ybi} and
shows that the corresponding reformulation is identical.    The rest of the paper is then dedicated to illustrating such computations
in detail using Jackiw-Teitelboim gravity \cite{Jackiw:1984je,Teitelboim:1983ux} as a simple toy model.  In particular,
after reviewing  classical aspects of this theory,
section \ref{sec:ClassicalJT} describes the details of the corresponding complex saddles and the relationship to the saddles proposed in \cite{Colin-Ellerin:2020mva,Colin-Ellerin:2021jev}.  Section \ref{sec:JTSn} then reviews the semiclassical quantum theory and computes semiclassical R\'enyi entropies for a simple class of semiclassical bulk states specified by giving wavefunctions of dilaton-profiles at some initial time.  We then close with a summary and remarks on future directions in section \ref{sec:disc}.

\section{R\'enyi entropies from wavefunctions}\label{sec:Rfw}

Gravitational R\'enyi entropies are traditionally computed by using a Euclidean path integral; see e.g. \cite{Faulkner:2013yia,Lewkowycz:2013nqa}.  But path integrals also compute bulk wavefunctions, and the two are not independent.  Let us therefore first consider the computation of a bulk wavefunction on a slice that passes through an extremal surface.  We will suppose that the Euclidean (or, more generally, complex) path integral is well-defined, perhaps due to an appropriate choice of integration contour in the space of complex metrics.  However, our discussion will not depend on the details of that construction, so long as the contour prescription is sufficiently local in a sense to be defined below.  The ideas in this section broadly follow previous arguments of e.g. \cite{Dong:2016hjy,Dong:2018seb,Dong:2019piw,Takayanagi:2019tvn}, though we take care to explicitly discuss the relevant gauge-fixing issues and to check compatibility of the stated boundary terms with the relevant boundary conditions for path integral computations of our bulk wavefunctions.

\begin{figure}
    \centering
   \includegraphics[width=3in]{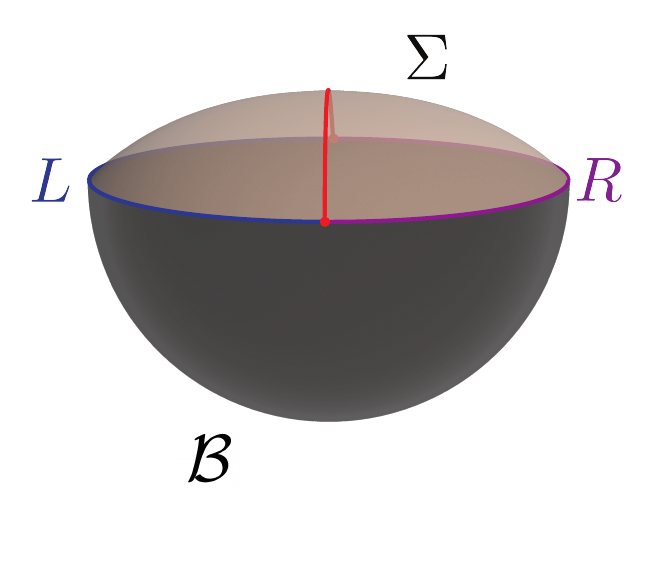}
    \caption{A state on the tan surface $\Sigma$ is defined by performing a Euclidean path integral with AlAdS boundary conditions. Here the AlAdS boundary $\mathcal{B}$ is taken to be a hemisphere (shown in grey). The surface $\partial\Sigma=\partial\mathcal{B}$ has been partitioned into two parts $L$ and $R$ which define the entropies \eqref{eq:Sn}.}
     \label{fig:EPI}
\end{figure}

Since we wish to study R\'enyi entropies, we will consider a setting that mirrors the one used for the Ryu-Takayangi (RT) entropies \cite{Ryu:2006bv,Ryu:2006ef,Headrick:2007km} or their covariant Hubeny-Rangamani-Takayanagi (HRT) generalizations \cite{Hubeny:2007xt}.  In particular, we consider a Euclidean path integral with boundary conditions defined by an asymptotically locally AdS (AlAdS) boundary $\mathcal B$ and a slice $\Sigma$ that intersect on some $\partial \Sigma = \partial \mathcal B$ as shown in figure \ref{fig:EPI}. The details of the boundary conditions at $\mathcal B$ specify the quantum state of interest, while the boundary conditions at $\Sigma$ describe the arguments of our wavefunction.   We take $\partial \Sigma$ to be partitioned into two regions $R$ and $L$ which define the entropies \eqref{eq:Sn}.  Thus $\partial R = \partial L$.

We will introduce a partial gauge fixing in terms of a preferred surface $\Upsilon$ which is codimension-1 within $\Sigma$, anchored to the boundary at $\partial R = \partial L$, and homologous to $L$ within $\Sigma$.  This $\Upsilon$ has an induced area element $\sqrt{\gamma}$.  As a partial gauge fixing of the freedom to deform $\Sigma$ in directions normal to itself, we set the corresponding Lie derivative of $\gamma$ to zero along the unit normal\footnote{This is a valid partial gauge fixing on-shell in contexts where we expect to find a non-trivial extremal surface.  It should thus be a valid partial gauge-fixing at least in the semiclassical limit studied below.}. This amounts to setting to zero the combination $K_\Upsilon = \gamma_{ij} K^{ij}$ where $i,j$ label coordinates along $\Upsilon$, $\gamma_{ij}$ is the induced metric on $\Upsilon$, and $K^{ab}$ is the extrinsic curvature of $\Sigma$.   As a partial gauge fixing of the freedom to deform $\Upsilon$ within $\Sigma$,  we then set the Lie derivative of $\gamma$ to zero along {\it any} vector field normal to $\Upsilon$ within $\Sigma$.  Here we note that, even if we extend the definition of $\gamma$ to surfaces $\Upsilon_\lambda$ that form a smooth local foliation of $\Sigma$ about $\Upsilon = \Upsilon_{\lambda=0}$, the canonical Poisson bracket $\{K_\Upsilon, \sqrt{\gamma}\}$ vanishes identically\footnote{See e.g. (2.6)-(2.13) in \cite{Kaplan:2022orm}.}, ensuring that there are no subtleties in imposing both gauge conditions simultaneously.   The corresponding construction in JT gravity replaces $\sqrt{\gamma}$ with the value $\Phi_\Upsilon$ of the dilaton at the point $\Upsilon$ as described in \cite{Harlow:2018tqv}.

We then take the remaining arguments of the wavefunction to be the full set of linear functionals of the induced metric on $\Sigma$ that have vanishing canonical Poisson Brackets with $K_\Upsilon$.  (Such functionals automatically Poisson-commute with the condition that $\sqrt{\gamma}$ be stationary within $\Sigma$ at $\Upsilon$.)  Since  $\{K_\Upsilon, \sqrt{\gamma}\}=0$, it is clear that these arguments include the codimension-2 area $A_\Upsilon$ of the surface $\Upsilon$.  While the naively-defined area diverges due to the AlAdS boundary conditions, we will choose our $A_\Upsilon$ to be the finite renormalized area defined by subtracting $c$-number counterterms appropriate to an HRT surface (see e.g. \cite{Taylor:2016aoi}).  It will be useful to denote the resulting wavefunction $\Psi[A, h]$, where $A$ is the value of the variable $A_\Upsilon$ and $h$ denotes the collection of all other arguments. In particular, we may write
\begin{equation}
\label{eq:compLPsi}
\Psi[A, h] = \int_{g\sim \mathcal{B}, g\big|_\Sigma=h,  A_\Upsilon=A} {\mathcal D} g e^{-S_E},
\end{equation}
where the notation $g\sim \mathcal{B}, g\big|_\Sigma=h,  A_\Upsilon=A$ indicates that we integrate over metrics that satisfy asymptotic boundary conditions defined by $\mathcal{B}$ and boundary conditions on $\Sigma$ defined by both $h$ and an extremal surface $\Upsilon$ with area $A$.  Here we take the action $S_E$ of a spacetime region $\mathcal{R}$ to be the usual Einstein-Hilbert action with a standard Gibbons-Hawking term, together with whatever terms $S_{\text{matter}}$ are required to describe matter fields (which we take to be minimally coupled) and the standard AlAdS counter-terms $S_{\text{counter-terms}}$:
\begin{equation}
\label{eq:SEHGH}
S_E = -\frac{1}{16\pi G} \int_{\mathcal R} \sqrt{g} R - \frac{1}{8\pi G}\int_{\partial \mathcal R} \sqrt{h_\Sigma}K + S_{\text{matter}} + S_{\text{counter-terms}}.
\end{equation}
In \eqref{eq:SEHGH}, $\sqrt{h_\Sigma}$ is the volume element on $\Sigma$.
  We emphasize that the Einstein equations are precisely the conditions for the action \eqref{eq:SEHGH} to be stationary with respect to all metric variations that preserve the stated boundary conditions on $\mathcal B$ and on $\Sigma$. This in particular uses the fact that boundary terms in $\delta S_E$ vanish at $\Upsilon$ when $K_\Upsilon=0$ and when all commuting components of the induced metric are held fixed.  This is critical for the action \eqref{eq:SEHGH} to be used in computing the wavefunction defined above for a theory with the desired classical limit.

\begin{figure}
    \centering
   \includegraphics[width=5in]{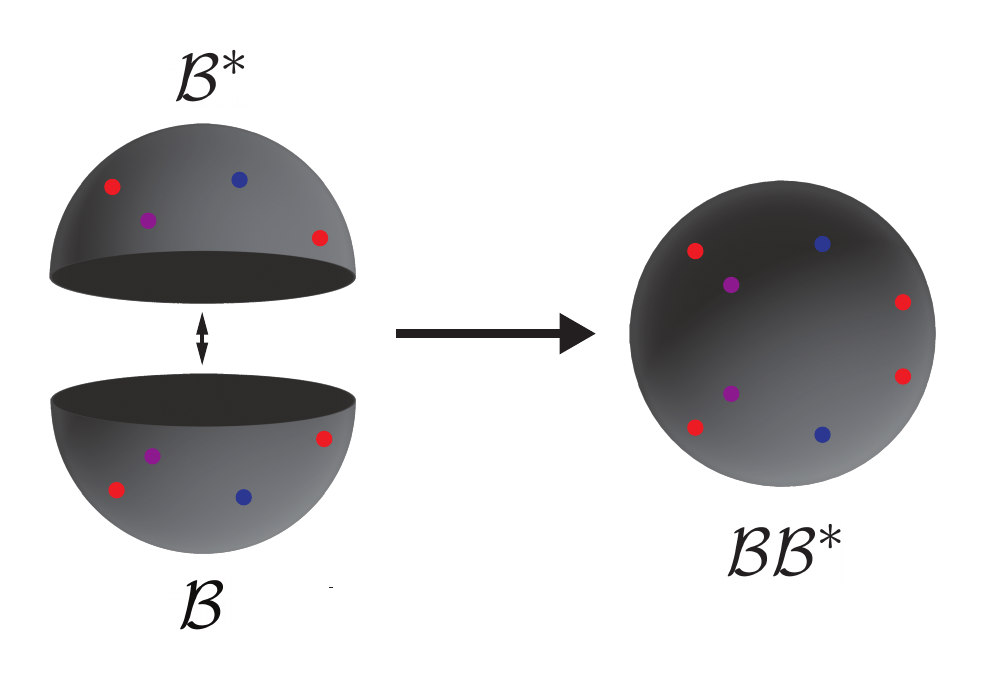}
    \caption{ Given a boundary condition $\mathcal{B}$ we can form the CPT conjguate boundary condition $\mathcal{B}^*$.  The two can then be stitched together to construct a closed boundary $\mathcal{BB}^*$,}
        \label{fig:BBstar}
\end{figure}

The probability of finding an extremal surface of area $A_\Upsilon=A$ is then given by
\begin{equation}
P(A_\Upsilon=A) = \int \dd h \ |\Psi[A, h]|^2,
\end{equation}
for an appropriate measure $\dd h$.  This measure must take into account the fact that we have not yet fully fixed a gauge and, in a Euclidean path integral formalism, it may also be computed from the path integral.  In particular, we must have
\begin{equation}
\label{eq:PAUps}
P(A_\Upsilon=A) = \int\dd h \ |\Psi[A, h]|^2 = \frac{\int_{g\sim \mathcal{BB}^*, A_\Upsilon=A} {\mathcal D} g e^{-S_E}}{\int_{g\sim \mathcal{BB}^*} {\mathcal D} g e^{-S_E}},
\end{equation}
where the closed boundary manifold $\mathcal{BB}^*$ is defined by gluing $\mathcal B$ to a `conjugate boundary' ${\mathcal B}^*$; see figure \ref{fig:BBstar}.  This conjugate boundary $\mathcal B^*$ is defined by complex-conjugating all fields on $\mathcal B$.  In addition, the notation indicates that the path integral in the numerator on the right-hand-side is performed only over metrics which contain an extremal surface\footnote{\label{foot:uniqueext}For simplicity, we assume here that we consider a context where (at least in the semiclassical expansion to be studied) the spacetimes over which we integrate can be treated as if they have only one such extremal surface.  This is of course not true in general, and the existence of multiple extremal surfaces is known to lead to interesting behavior when their areas are nearly equal (see e.g. \cite{Marolf:2020vsi,Dong:2020iod,Akers:2020pmf}).  Nevertheless, we postpone a full treatment of such cases to future work.}
 of renormalized area $A$.  The path integral in the denominator is then given by integrating the numerator over all positive $A$. The proper definition of $\dd h$ in this way will also take into account any possible choices of contour of integration for the path integral on the right-hand-side.  The final result \eqref{eq:PAUps} is of course just the fixed-area path integral of \cite{Dong:2018seb,Akers:2018fow}, though we have taken care here to construct it from path integrals for $\Psi[A,h]$ which contain appropriate boundary terms.

We will now compare the above with  the Euclidean path integral computation of $\Tr_L[\rho_L^n]$.  In doing so, we will assume that the prescription for choosing the contours of integration in the above path integrals are sufficiently local that the measure $\dd h$ can be written as a product $\dd h_L \dd h_R$ of measures on arguments of our wavefunction that are respectively localized in the regions of $\Sigma$ that lie to the left and right of the surface $\Upsilon$.

\begin{figure}
\includegraphics[width=0.45\textwidth]{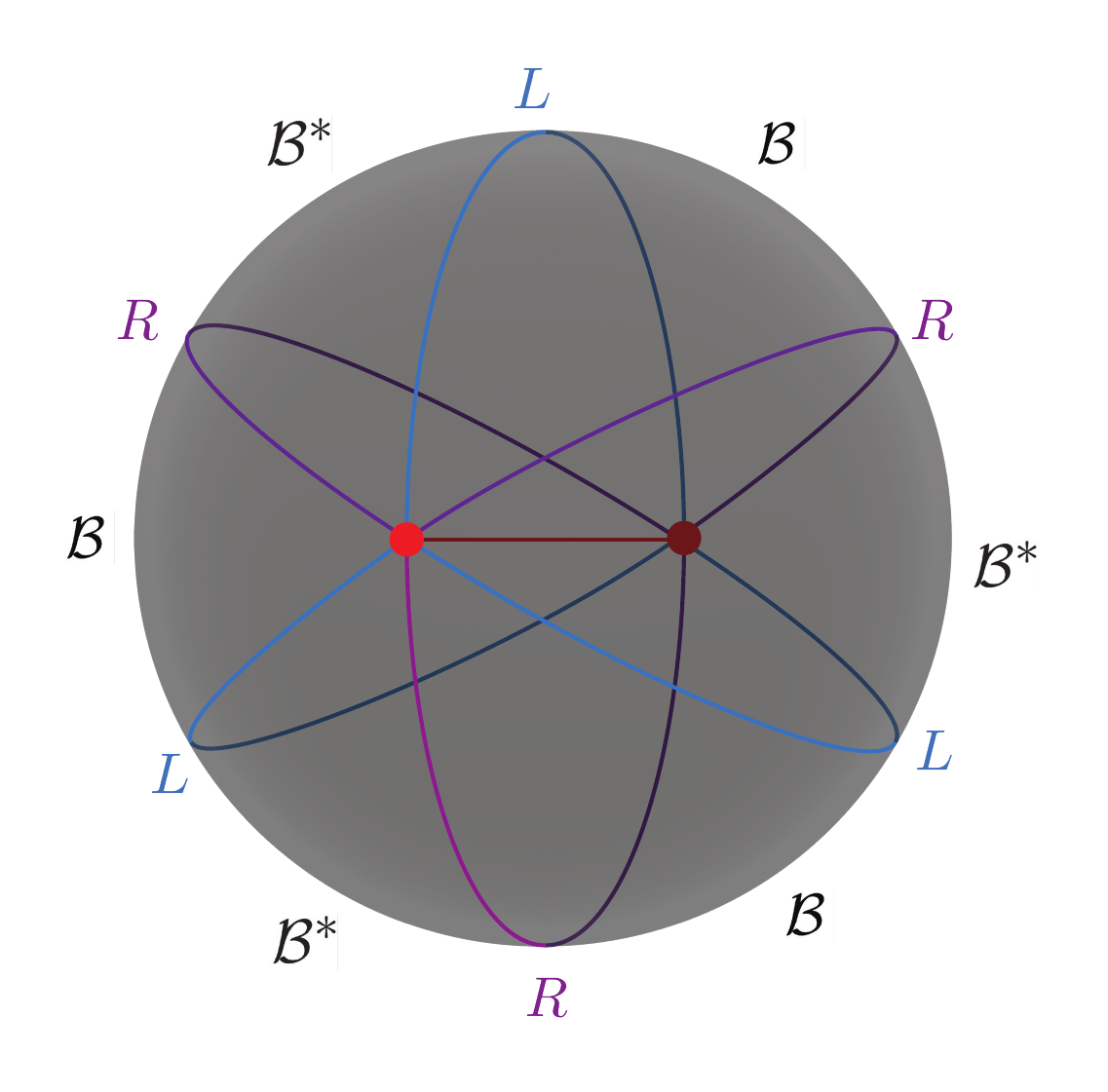}
\includegraphics[width=0.45\textwidth]{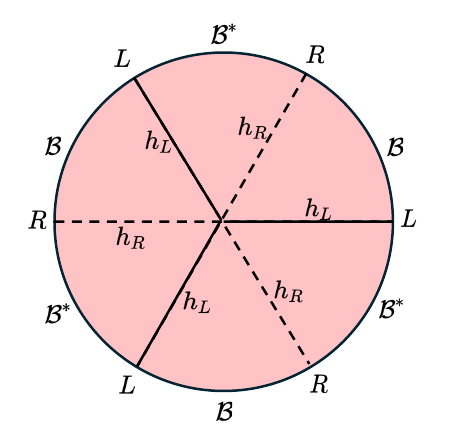} 
\caption{ {\bf Left:} A three-dimensional case with a two-dimensional boundary. The 6 wedges represent $\mathcal R_i$ for $i=1, \dots, 6$.  The figure is topologically accurate, but not metrically correct.  For each boundary region $L,R$, the boundary $\partial L = \partial R$ consists precisely of the two red dots. These dots are conical singularities, each with a conical excess of $4\pi.$  Thus each $L$ semicircle meets the neighboring $R$ semicircle at an angle $\pi$.  For the topology shown, and if the configuration that preserves replica symmetry, the geodesic connecting the two dots will be an extremal surface $\Upsilon$.
{\bf Right:} A more schematic illustration of an analogous two-dimensional case.  This panel also shows the locations of the $h_{L,R}$ degrees of freedom in the decomposition \eqref{eq:rhoLnPsi}.  For fixed $h_L, h_R$, the path integral over any given wedge with boundary $\mathcal B$ computes $\Psi[A,h_{L}, h_{R}]$, or $\Psi[A,h_{L}, h_{R}]^*$ for boundary ${\mathcal B}^*$.
 }
    \label{fig:Renyifig}
\end{figure}

Recall  that the Euclidean path integral for  $\Tr_L[\rho_L^n]$  is
\begin{equation}
\label{eq:rhoLnPI}
\Tr_L[\rho_L^n] : = \int_{g\sim [\mathcal{BB}^*]^n} {\mathcal D} g e^{-S_E},
\end{equation}
where $[\mathcal{BB}^*]^n$ is the closed boundary manifold obtained by cyclicly gluing together $n$ alternating copies of each of $\mathcal{B}$ and $\mathcal{B}^*$ around the surface $\partial L = \partial R$ as shown in figure \ref{fig:Renyifig}.    As remarked in footnote \ref{foot:uniqueext}, we will treat the spacetimes over which we integrate as if they have a unique extremal surface (at least when the surface is subject to the boundary conditions and the homology condition imposed on $\Upsilon$).  This assumption is, in particular, true for JT gravity in the semiclassical approximation. When the assumption holds, we can divide the bulk path integral into separate computations over $2n$ regions ${\mathcal R}_1, {\mathcal R}_2, \dots {\mathcal R}_{2n}$ cyclicly surrounding the extremal surface $\Upsilon$ as shown in figure  \ref{fig:Renyifig} so that the bulk manifold is $\mathcal M = \cup_{i=1}^{2n} \mathcal R_i$.

The path integral over any given region then computes either $\Psi[A, h_L, h_R]$ or $\Psi^*[A, h_L, h_R]$, where we now indicate that $h$ can be parameterized in terms of separate variables $h_{L,R}$ that are respectively localized on the parts $\Sigma_{L,R}$ of $\Sigma$ lying to the left and right of $\Upsilon$.  As also shown in figure \ref{fig:Renyifig}, the interface between a given bra region and an adjacent ket region is a copy of either $\Sigma_L$ or $\Sigma_R$, so that the associated gluing is accomplished by integrating over $h_L$ or $h_R$ with the measure $\dd h_L$ or $\dd h_R$.
As a result, we may write \eqref{eq:rhoLnPI} directly in terms of the wavefunctions $\Psi[A, h_L, h_R]$:
\begin{equation}
\label{eq:rhoLnPsi}
\Tr_L[\rho_L^n] : = \int \dd A  \prod_{i=1}^{n} \left( \dd h_{L, i}  \dd h_{R, i}\, \Psi[A,h_{L,i}, h_{R,i} ] \Psi^*[A,h_{L,i}, h_{R,i+1} ]\right) e^{-\Delta_n}\,,
\end{equation}
with $i=n+1$ being equivalent to $i=1$ and where
\begin{equation}
\Delta_n = S_E[\mathcal M]- \sum_{i=1}^{2n} S_E[\mathcal R_i]\,,
\end{equation}
with $S_E$ defined as in \eqref{eq:SEHGH}.

In writing \eqref{eq:rhoLnPsi}, we have implicitly assumed that $\Delta_n$ depends at most on $A, h_{L,i}, h_{R,i}$.   That this is the case follows by a slight generalization of the classic argument of \cite{Brill:1994mb}. At the smooth parts of the boundary where $\mathcal R_i$ meets $\mathcal R_{i+1}$, the Gibbons-Hawking terms on either side  combine to simply compute the contribution to the Einstein-Hilbert action from any delta-functions in $\sqrt{g}R$ that are localized on this interface\footnote{This computation is identical to that performed in \cite{Israel:1967zz}.}.  Thus there is no contribution to $\Delta_n$ from such regions.  However, due to the corner at $\Upsilon$, the extrinsic curvature of the boundary of each $\mathcal R_i$ contains a delta-function localized at the codimension-2 surface $\Upsilon$.  This delta-function contributes to the corresponding Gibbons-Hawking term, and thus to $S_E[\mathcal R_i]$, a term
\begin{equation}
\frac{1}{8\pi G}\int_\Upsilon \sqrt{\gamma} \,(\theta_i-\pi),
\end{equation}
where $\theta_i$ at any point of $\Upsilon$ is the angle in the plane orthogonal to $\Upsilon$ between the right and left pieces of the boundary of $\mathcal R_i$.  Thus $\sum_{i=1}^{2n} S_E[\mathcal R_i]$ includes a contribution
\begin{equation}
-\frac{nA}{4G}+ \sum_{i=1}^{2n} \frac{1}{8\pi G}\int_\Upsilon \sqrt{\gamma}\, \theta_i.
\end{equation}
This is to be compared against the contribution to $S_E[\mathcal M]$ that might come from any delta-function contribution to $\sqrt{-g}\,R$ localized on the codimension-2 surface $\Upsilon$ (as, by definition, such a term does not contribute to our $S_E[\mathcal R_i]$ for any $i$).  A standard computation shows that this contribution is $-\frac{A}{4G}+ \sum_{i=1}^{2n} \frac{1}{8\pi G}\int_\Upsilon \sqrt{\gamma}\, \theta_i$, from which we find
\begin{equation}
\label{eq:DeltaResult}
\Delta_n = (n-1)\frac{A}{4}.
\end{equation}

With this understanding, \eqref{eq:rhoLnPsi} gives a recipe for computing gravitational R\'enyi entropies directly from the bulk wavefunction.  The contribution $\Delta_n$ is essentially the same as the cosmic brane contribution of \cite{Lewkowycz:2013nqa}, though it is derived here without invoking either replica symmetry or the semiclassical approximation.  The result \eqref{eq:rhoLnPI} will come as no surprise to experts in the field, though we are not aware of a full discussion of the above points (and, in particular, explaining the important compatibility between using the Gibbons-Hawking boundary term and boundary conditions that fix to zero some components of the {\it extrinsic} curvature on the hypersurface that defines arguments of the wavfunction) in the literature to date.

\subsection{Complex saddle-point geometries from wavefunctions}

The point we wish to emphasize is that, in addition to determining the R\'enyi entropies $S_n$, the result  \eqref{eq:rhoLnPI} also contains all of the information required to construct the saddle-point geometries that dominate the  Euclidean R\'enyi path integrals for all $n$ (no matter what contour prescription is used in its definition).  If there is such a dominant saddle point, then \eqref{eq:rhoLnPsi}  will be dominated by the corresponding values $A_*, h_{Li*}, h_{Ri*}$ of $A_\Upsilon, h_{Li}, h_{Ri}$.  This, of course, is not yet enough data to fully reconstruct the given saddle.  Indeed, as we have seen, from the perspective of the classical phase space, the data $A, h$ corresponds to some maximally commuting set of gauge-invariant quantities.  The missing data is thus contained in the canonically-conjugate quantities.  But these conjugate quantities can be recovered via Hamilton-Jacobi theory from the wavefunctions $\Psi[A,h_{L,i}, h_{R,i}]$.  I.e., if the conjugates to $A_\Upsilon, h$ are $\Theta, \kappa$, then at leading order in the stationary phase approximation we have
\begin{equation}
\label{eq:HJconj}
\Theta = -i \frac{\partial}{\partial A_\Upsilon}  \ln \Psi\,,\quad \kappa = -i \frac{\partial}{\partial h}  \ln \Psi \,,
\end{equation}
where the first derivative is taken at fixed $h$, the second is taken at fixed $A_\Upsilon$, and in the $i$th region the derivatives are to be evaluated at $A_*, h_{Li*}, h_{Ri*}$.  This data should now suffice to recover all semiclassical information associated with the wavefunction $\Psi$ at $A_*,h_{Li*},h_{Ri*}$. As a result, any path integral prescription (using either a real-time formulation or an imaginary time formulation, and with any prescription for the contour of integration) that reproduces \eqref{eq:rhoLnPsi} for a given wavefunction $\Psi[A,h]$ will associate the computation of the R\'enyi entropy $S_n$
with the semiclassical spacetime described by $A_*,h_{Li*},h_{Ri*}$ and \eqref{eq:HJconj}.  In particular, any two Euclidean path integrals (with any contour prescriptions) that succeed in computing this particular state $\Psi$ semiclassically at the point $A_*,h_{Li*},h_{Ri*}$ must do so using the same (potentially-complex) semiclassical geometry.

Before turning to the real-time path integral of \cite{Colin-Ellerin:2020mva,Marolf:2020rpm,Marolf:2022ybi}, let us pause to better understand why such semiclassical R\'enyi geometries are typically Euclidean or complex.  Semiclassical states are characterized by sharp peaks and by small fluctuations about such peaks.  For simplicity, let us therefore consider a wavefunction of the factorized form
\begin{equation}
\label{eq:factorPsi}
\Psi[A, h_L, h_R] = \psi_\Upsilon (A) \psi_L(h_L) \psi_R(h_R),
\end{equation}
with $\int \dd h_L |\psi_L|^2 = \int \dd h_R |\psi_R|^2=1$.
Expression \eqref{eq:rhoLnPsi} then yields
\begin{equation}
\label{eq:rhoLnPsfactorizing}
\Tr_L[\rho_L^n] : = \int \dd A  \,|\psi_\Upsilon(A)|^{2n} e^{-(n-1)A/4G}.
\end{equation}
If \eqref{eq:rhoLnPsfactorizing} is finite, the integrand must decay at large $A$.  In the semiclassical approximation, the integral \eqref{eq:rhoLnPsfactorizing} will thus be dominated by the value $A_*$ of $A$ that maximizes the integrand.  Furthermore, if the state is indeed semiclassical then we can find $A_*=0$ only if $\Upsilon$ is the emptyset (otherwise the spacetime would have structure on scales that are not long-distance compared with the Planck scale).  The case of interest is where $\Upsilon$ is non-trivial, as it is only in that case that there is a well-defined conjugate $\Theta$ to $A_\Upsilon$.  Thus the integrand is maximized away from the endpoint $A=0$ of the integration region, so that subtle endpoint effects are not relevant.

Due to the factorizing form \eqref{eq:factorPsi} of the wavefunction,
the relations \eqref{eq:HJconj} now reduce to simply
\begin{equation}
\label{eq:HJconjfactor}
\Theta = -i \frac{\partial}{\partial A_\Upsilon}  \ln \psi_\Upsilon  \Bigg|_{A_\Upsilon=A_*}, \quad \kappa_L = -i \frac{\partial}{\partial h_L}  \ln \psi_L  \Bigg|_{h_L = h_{L*}},  \quad \kappa_R = -i \frac{\partial}{\partial h_R}  \ln \psi_R  \Bigg|_{h_R = h_{R*}},
\end{equation}
 where $h_{L*}$, $h_{R*}$ are the peaks of $|\psi_L|^2$, $|\psi_R|^2$ and are thus identical for all regions $i$.  The interesting observation is then that $A_*$ is determined by maximizing $  |\psi_\Upsilon(A)|^{2n} e^{-(n-1)A/4G}$, or equivalently by the condition
\begin{equation}
\label{eq:ImTheta}
2n \, {\rm Im} \, \Theta = \frac{n-1}{4G}.
\end{equation}
Thus for $n\neq1$ and in the presence of a nontrivial extremal surface $\Upsilon$, the saddle cannot be real.  Indeed, in general, such saddles are truly complex geometries.  However, for states that are invariant under a time-reversal symmetry (and for saddles that preserve this symmetry), the saddle point value of any quantity that is odd under time-reflection must have vanishing real part.  In such cases the saddle is a real Euclidean-signature spacetime.  It should be clear that the same conclusion holds even when $\Psi$ fails to take the form \eqref{eq:factorPsi}, as even when the values of $\Theta_k = -i \frac{\partial}{\partial A_\Upsilon}  \ln \Psi$ depend on the choice of region $\mathcal R_k$, the stationary-point condition from \eqref{eq:rhoLnPI} still ensures that 
\begin{equation}
\label{eq:ImThetasun}
\sum_{k=1}^{2n} \, {\rm Im} \, \Theta_k = \frac{n-1}{4G},
\end{equation}
so that the $\Theta_k$ cannot all be real.

Before moving on to the real-time approach, we should pause to mention a few subtleties.  First, as noted above, when the extremal surface is trivial ($\Upsilon = \emptyset$) there are no continuous deformations of $A_\Upsilon$ away from zero and thus there can be no conjugate $\Theta$ to $A_\Upsilon$.  In that context, there is no reason for the R\'enyi spacetimes to be thought of as essentially complex.  Indeed, the classic example of that case is thermal AdS, which defines a smooth spacetme in either Euclidean or Lorentzian signature.

Second, as is manifest in \eqref{eq:rhoLnPsi} our analysis above was conducted in the context of integer $n$.  It is nevertheless tempting to conclude that \eqref{eq:ImTheta},\eqref{eq:rhoLnPsfactorizing} hold at all real $n$, and in particular in the limit $n\rightarrow 1$ associated with computing von Neumann entropies to find
\begin{equation}
\label{eq:vN}
S_{\rm vN}  = \int \dd A  \,|\psi_\Upsilon(A)|^{2} \frac{A}{4G}
\end{equation}
for a normalized state $\psi_\Upsilon$; i.e., in this context the von Neumann entropy is just the expectation value of $\frac{A}{4G}$. Furthermore, the imaginary part of $\Theta$ vanishes and, for a sharply peaked semiclassical state $\Psi$,  the entropy computation is associated with the same (real) spacetime that about which the state is peaked.

That the extrapolation \eqref{eq:vN} is in fact valid follows from the fact that \eqref{eq:SEHGH} remains a valid variational principle, and that the result \eqref{eq:DeltaResult} remains also valid, in the presence of a conical singularity at $\Upsilon$.    As emphasized in \cite{Dong:2013qoa}, such conical singularities are an essential feature of the analytic continuation to non-integer $n$. In particular, in the presence of higher derivative corrections, effects associated with such singularities at non-integer $n$ give rise to new terms that remain relevant in the limit $n\rightarrow 1$ and which thus appear in the expression for gravitational von Neumann entropy.  Repeating the above arguments verbatim in the presence of higher derivative corrections and applying the conclusions for $n\neq 1$ would thus require ensuring that one fully understands the relevant variational principles in the presence of such singularities.  However, the analysis of \cite{Dong:2019piw} supports the idea that, under corresponding assumptions, there will be a direct analogue of \eqref{eq:rhoLnPsfactorizing} that replaces $A/4$ by the appropriate geometric entropy operators\footnote{Ref. \cite{Dong:2019piw} provides an action for theories with general higher derivative corrections that defines a Euclidean path integral for fixed geometric entropy $\sigma$ and which has the interpretation of computing the probability $P(\sigma)$ of $\sigma$.   This action defined a good variational principle in the presence of the desired conical singularities.   It also showed that the analogue of \eqref{eq:rhoLnPsfactorizing} is then given by $\int d\sigma P(\sigma) e^{-(n-1)\sigma}$, but it did not show that the result could be written in terms of a wavefunction computed using an action defining a correspondingly good variational principle when all arguments of the wavefunction are fixed.} (as computed e.g. in \cite{Dong:2013qoa,Camps:2013zua,Miao:2014nxa}).

\section{R\'enyi entropies from real-time path integrals}
\label{sec:RTRenyi}

Having reformulated the usual Euclidean path integral calculations of R\'enyi entropies in terms of bulk wavefunctions, we will proceed to apply similar arguments to the real-time path integral
prescription of \cite{Colin-Ellerin:2020mva}.   We again consider AlAdS spacetimes, though now with a real Lorentz-signature asymptotic boundary ${\cal B}$. This ${\cal B}$ will be taken to have a (say, future) boundary $\partial_f {\cal B}$, which is partitioned into two regions $L$ and $R$.

It will be convenient to structure our discussion in direct analogy with the Euclidean presentation of section \ref{sec:Rfw}. In particular, let us begin by noting that we can of course also use a Lorentz-signature path integral to compute wavefunctions $\Psi[A, h_L, h_R]$ of the sort described in section \ref{sec:Rfw}. We do so by using the same partial gauge fixing as above at a preferred surface $\Upsilon$ (codimension-1 in $\Sigma$) that is anchored to the boundary at $\partial R = \partial L$, and which is homologous to $L$ within $\Sigma$.  This gauge fixing again requires $\Upsilon$ to be extremal.   We again take the remaining arguments of the wavefunction to be all linear functionals of the induced metric on $\Sigma$ that have vanishing canonical Poisson Brackets with $K_{\Upsilon}$ (defined by the Lie derivative of the area element $\sqrt{\gamma}$ on $\Upsilon$ in the direction orthogonal to $\Sigma$), so these arguments again include  $A_{\Upsilon}$.  However, in Lorentz-signature we will assume that our state is defined by giving some other representation $\Psi_{in}$ of the bulk wavefunction which will generally be associated with some other Cauchy surface that we call $\Sigma_{in}$.  The corresponding $\partial \Sigma_{in}$ will then coincide with another (say, past) boundary $\partial_p {\cal B}$ of ${\cal B}$.  It is thus also natural to refer to $\Sigma, \Sigma_{in}$ as the future/past boundaries of the bulk spacetime $\mathcal M$, though this need not imply that $\mathcal M$ is time-orientable (see comments about time orientations and path integrals in \cite{Harlow:2023hjb}).  

The wavefunction on $\Sigma_{in}$ replaces some of the boundary conditions that  were imposed at the AlAdS boundary in the Euclidean context.  In particular, the desired path integral takes the form

\begin{equation}
\label{eq:LsigPsi}
\Psi[A, h] = \int_{g\sim \mathcal{B}, g\big|_\Sigma=h,  A_{\Upsilon}=A}  {\mathcal D} g \,  e^{iS}\, \Psi_{in},
\end{equation}
where the integral over $g$ will include integration over the arguments of $\Psi_{in}$, and where the above integral is over spacetimes $\mathcal M$ with boundary $\partial \mathcal M = \Sigma \cup \Sigma_{in} \cup \mathcal B$.
We take the action $S$ of a spacetime region $\mathcal{M}$ to be the usual Einstein-Hilbert action with a Gibbons-Hawking term and the with standard AlAdS counter-terms $S_{\text{counter-terms}}$:
\begin{equation}
\label{eq:SEHGHL}
S[\mathcal{M}] = \frac{1}{16\pi G} \int_{\mathcal{M}} \sqrt{-g} R + \frac{1}{8\pi G}\int_{\partial \mathcal{M}}\sqrt{h_\Sigma}K + S_{\text{matter}} + S_{\text{counter-terms}}.
\end{equation}
 In \eqref{eq:SEHGHL}, $\sqrt{h_\Sigma}$ is the volume element on $\Sigma$.  As in the Euclidean case, the Einstein equations are precisely the conditions for the action \eqref{eq:SEHGHL} to be stationary with respect to all metric variations that preserve the stated boundary conditions on $\mathcal B$ and on $\Sigma$.

Having described how to the change from an arbitrary wavefunction representation $\Psi_{in}$ to our $\Psi[A, h_L, h_R]$, we may henceforth assume that we have in fact been given the latter representation of the wavefunction.   Converting between the two representations of the state merely requires attaching the path integral \eqref{eq:LsigPsi} (or its complex conjugate) to the path integral designed for use with wavefunctions $\Psi[A, h_L, h_R]$.

\begin{figure}
    \centering
    \includegraphics[width=5in]{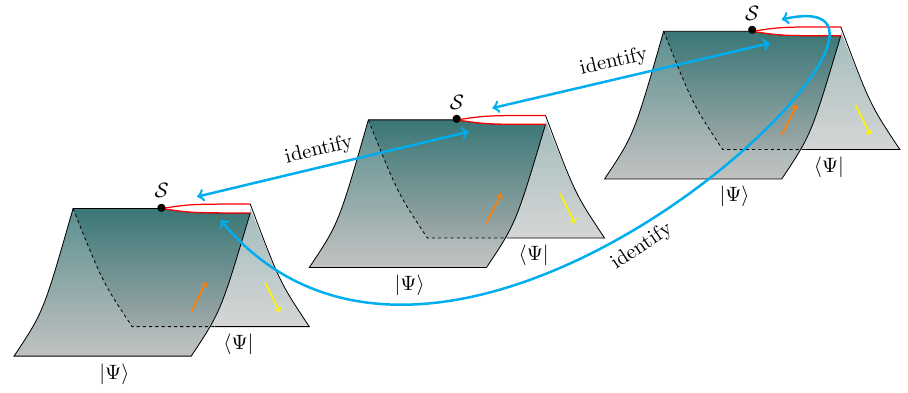}
    \caption{The bulk manifold $\mathcal{M}_n$ relevant to computing the R\'enyi entropy $S_n$ for  $n=3$ and a bulk state $|\Psi\rangle$.  The ket and bra spacetimes are identified in a manner as shown in the figure.  Following \cite{Colin-Ellerin:2020mva} and \cite{Marolf:2020rpm},  we take the interfaces between ket and bra spacetimes (black and red lines) to be the Schwinger-Keldysh timefolds, so that the timefold contains the singular splitting surface $\mathcal S$ (black dot) at which all ket and bra spacetimes are identified.  However, the timefolds can in principle be moved away from the splitting surface as described in appendix \ref{app:Rdelta}.     }
    \label{fig:replica}
\end{figure}
Let us now recall from \cite{Colin-Ellerin:2020mva} that, if $\rho_{L,R}$ is the relevant density matrix associated with $L,R$ as described above \eqref{eq:Sn}, the path integral for ${\rm Tr} \rho^n_R = {\rm Tr} \rho^n_L$ has boundary conditions defined by
$n$ copies of the ket $|\Psi\rangle$ and an additional $n$ copies of the corresponding bra $\langle \Psi|$. We integrate over geometries of the form shown in Figure \ref{fig:replica} which involve $n$ so-called `bra spacetimes' $\mathcal{M}_{\text{bra}}^i$ and $n$ so-called `ket spacetimes'  $\mathcal{M}_{\text{ket}}^i$. The union of the ket spacetimes will be denoted $\mathcal{M}_{\text{ket}}$, while  $\mathcal{M}_{\text{bra}}$ denotes the union of the bra spacetimes.
The ket spacetimes contribute to the integrand of our path integral through factors of the form $e^{iS[\mathcal{M}^i_{\text{ket}}]}$, while the ket
spacetimes contribute through factors of the form $e^{-iS[\mathcal{M}^i_{\text{bra}}]}$.
We will remind the reader of further relevant details from  \cite{Colin-Ellerin:2020mva} below.

\begin{figure}
    \centering
   \includegraphics[width=5in]{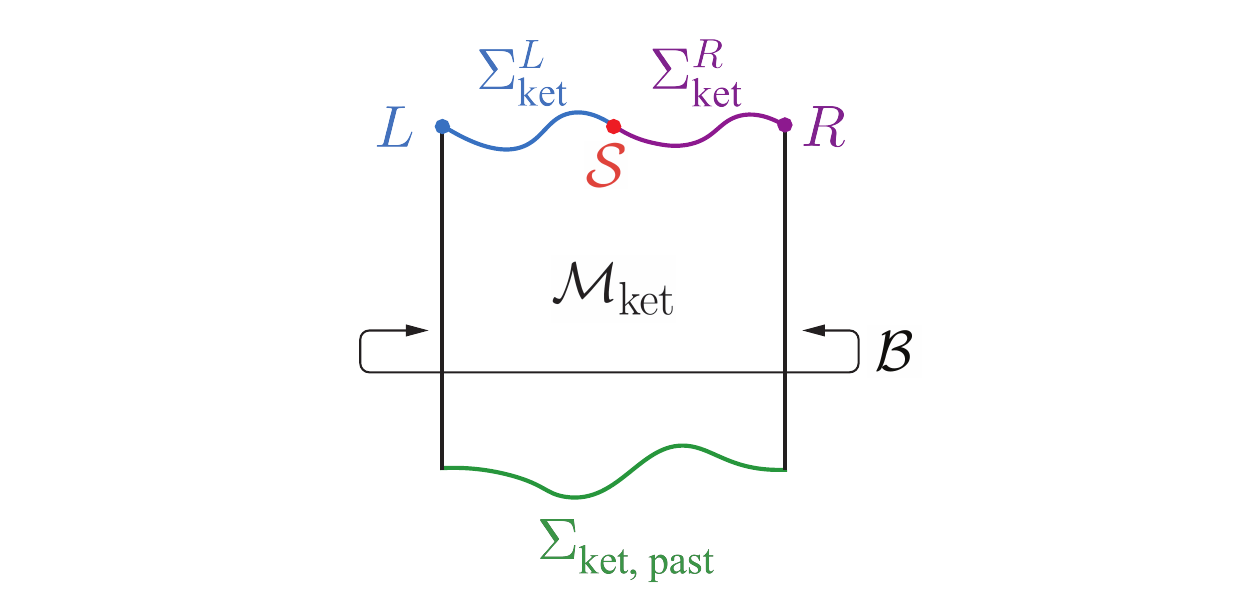}
    \caption{ A single ket spacetime $\mathcal{M}_{\text{ket}}$ is bounded by a timelike alAdS boundary $\mathcal{B}$ (which, in the figure, is the union of the left and right edges) as well as a past Cauchy surface $\Sigma_{\text{ket, past}}$ and a future Cauchy surface $\Sigma_\text{ket}=\Sigma^{L}_{\text{ket}}\cup\Sigma^{R}_{\text{ket}}$.
    The partial Cauchy surfaces $\Sigma^{L}_{\text{ket}}$ and $\Sigma^{R}_{\text{ket}}$ themselves have boundaries $\mathcal{S}\cup L$ and $\mathcal{S}\cup R$.}
       \label{fig:singleket}
\end{figure}

The boundary of each $\mathcal{M}^i_{\text{bra/ket}}$ consists of 3 components.  The first component is a past boundary\footnote{For convenience we use a language that assumes the spacetimes to be time-orientable.  But the formalism extends to the more general path integrals advocated in \cite{Harlow:2023hjb}.} $\Sigma^i_{\text{bra/ket, past}}$, on which we include a factor of $\Psi[A_i, h_{Li}, h_{Ri}]$ and integrate over all $A_i, h_{Li}, h_{Ri}$.  For ket spacetimes, the second component is a copy  ${\cal B}_i$ of ${\cal B}$, at which we impose appropriate AlAdS boundary conditions.  For bra spacetimes we instead use the conjugate boundary ${\cal B}_i^*$ defined by acting on ${\cal B}$ with a time-reversal symmetry.  Each $\partial \mathcal B_i$ or $\partial \mathcal B_i^*$ is then correspondingly partitioned into $L_i,R_i$ or $L_i^*,R_i^*$; see figure \ref{fig:singleket} .

We denote the remaining `future' boundary of $\mathcal{M}^i_{\text{bra/ket}}$ by $\Sigma^i_{\text{bra/ket}}$
and require that  $\partial \Sigma^i_{\text{ket}} = \partial \mathcal B_i$
and $\partial \Sigma^i_{\text{bra}} = \partial \mathcal B_i^*$.  Each such boundary is assumed to be partitioned into left and right parts  $\Sigma^{L,Ri}_{\text{bra/ket}}$ with $\Sigma^{i}_{\text{bra/ket}}=\Sigma^{Li}_{\text{bra/ket}}\cup \Sigma^{Ri}_{\text{bra/ket}}$ and where the intersection of $\partial \mathcal B_i$ ($\partial \mathcal B_i^*$) with
$\partial \Sigma^{Li}_{\text{bra/ket}}$ is precisely $L_i$ ($L^*_i$) and the intersection with $\Sigma^{Ri}_{\text{bra/ket}}$
is $R_i$ ($R_i^*$); see again figure \ref{fig:singleket}.   We identify $\Sigma^{Li}_{\text{ket}}$ with $\Sigma^{Li}_{\text{bra}}$, and we also identify
$\Sigma^{Ri}_{\text{ket}}$ with $\Sigma^{R(i+1)}_{\text{bra}}$.  Here we use the convention that $i=n+1$ is equivalent to $i=1$.

The above gluing conditions require that the $\Sigma^{L,Ri}_{\text{bra/ket}}$ all meet at a a special singular codimension-2 surface called a `splitting surface' $\mathcal S$ at which all of the ket and bra spacetimes intersect.  This  $\mathcal S$ is automatically  anchored at $\partial L = \partial R\subset \partial \mathcal B$ and is automatically homologous to $L$ (since any $\Sigma^{Li}_{\text{bra/ket}}$ is a valid homology surface).  The gluing construction is illustrated in figure \ref{fig:replica} where the splitting surface ${\mathcal S}$ is shown as a single black dot, though in general ${\mathcal S}$ need not be connected.  The surfaces $\Sigma^{L,Ri}_{\text{bra/ket}}$ are called timefolds due to the fact that we take the integrand of the path integral to receive a factor of $e^{iS}$ from the bra side and a factor of $e^{-iS}$ from the ket side.

The splitting surface is a singularity of both the metric and the Lorentzian structure at which the metric remains continuous and such that the splitting surface has $2n$ past-directed orthogonal null congruences.
As discussed in \cite{Colin-Ellerin:2020mva}, splitting surfaces are Lorentzian analogues of familiar Euclidean-signature conical singularities.  In particular, a splitting surface is similarly associated with delta-function contributions to the scalar curvature, though with a complex coefficient.  Any real contribution to the Lorentzian action from this delta-function cancels between the bra and ket parts of the spacetime, though the imaginary contributions add together.  Due to the factor of $i$ that accompanies the Lorentzian action, the net effect from a small disk containing the splitting surface $\mathcal S$ is to multiply the integrand by the real number

\begin{equation}
\label{eq:SUps}
e^{-S^E_{\mathcal S}}=e^{-(n-1)\frac{A_{\mathcal S}}{4}},
\end{equation}
where $A_{\mathcal S}$ is the area of $\mathcal S$.
We refer the interested reader to \cite{Colin-Ellerin:2020mva} for details of both the boundary conditions at the splitting surface and the computation of this delta-function curvature, though the curvature calculation is also reviewed in appendix \ref{app:Rdelta} where we provide further comments concerning relevant signs in the intermediate steps. 

Although the computations refer only to our real-time geometry, the notation $S^E_{\mathcal S}$ indicates that this factor has the form one would expect from a real (and, for $n>1$, positive) contribution to the Euclidean-signature action of the spacetime.
Indeed, the astute reader will note that the factor \eqref{eq:SUps} is of exactly the same form as the factor $e^{-\Delta_n}$ in the Euclidean path integral \eqref{eq:rhoLnPsi}.
This is not a coincidence. In particular,
as we review in appendix \ref{app:Rdelta},  in both the real- and imaginary-time contexts this extra factor can be derived in the same way (and in any spacetime dimension) from an appropriate version of the Gauss-Bonnet theorem \cite{Louko:1995jw}.

\subsection{Implications of stationary phase}
\label{sec:ISP}

As explained in \cite{Colin-Ellerin:2020mva}, the domain of integration for our path integral allows all metrics that are smooth away from the timefolds (which we hereafter denote $\mathbb T$); i.e., using an overline to denote closures of sets, the configurations are smooth, real, and Lorentz-signature away from the surface $\mathbb{T} : = \overline {\mathcal{M}}_{\text{bra}} \cap \overline{ \mathcal{M}}_{\text{ket}} = \cup_i \left(\Sigma^{Li}_{\text{bra}} \cup \Sigma^{Ri}_{\text{bra}} \right) = \cup_i \left(\Sigma^{Li}_{\text{ket}} \cup \Sigma^{Ri}_{\text{ket}} \right)$. Furthermore, the fields are required to be continuous at $\mathbb{T}$, and in particular at the splitting surface $\mathcal S$.  However, the extrinsic curvature need only have well-defined limits at $\mathbb{T}$ when approached from either side. In general, there need be no relation between the limits defined by opposite sides of the timefold $\mathbb{T}$. 

\begin{figure}[h!]
    \centering
   \includegraphics[width =0.6\linewidth]{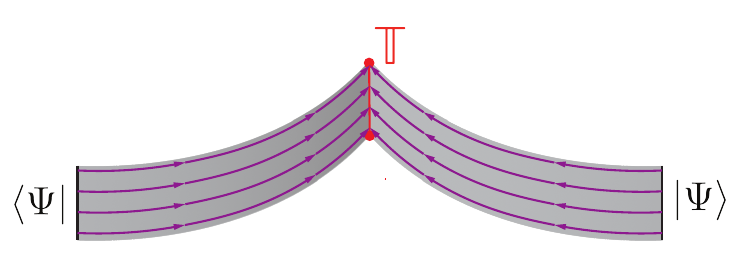}
    \caption{When a bra spacetime (left half) and a ket spacetime (right half) meet on-shell at a timefold $\mathbb T$, the equations of motion require derivatives of fields to be SK-continuous at $\mathbb T$.  By this we mean that the derivatives on the two sides of $\mathbb T$ agree at $\mathbb T$ when the derivative on each side is computed using its own outward-directed normal. I.e., if the timefold is a future boundary, then the outward-directed normal on either side is just the future-directed normal.  In the SK-continuous example shown here, the arrows represent the gradient of a scalar field that increases toward the future on either side of $\mathbb T$, so that the future-directed derivative on the left at $\mathbb T$ agrees with the corresponding future-directed derivative on the right.}
    \label{fig:normals}
\end{figure}

For higher-dimensional Einstein-Hilbert gravity, it was
advocated in \cite{Marolf:2022ybi} that one should save for last the integral over the total area $A_\mathcal S$ of the splitting surface.  (The JT analogue of this proposal is to save for last the integral over the
value $\Phi_\mathcal S$ of the dilaton at the splitting surface.)
As a result, we wish to understand stationary points of
\begin{equation}
\label{eq:I}
I := -i\sum_i S[\mathcal{M}^i_{\text{ket}}]+i \sum_i S[\mathcal{M}^i_{\text{bra}}] + S_\mathcal S^E
\end{equation}
with respect to variations that hold fixed $A_\mathcal S$ and which are subject to boundary conditions on the past boundaries defined by the quantum states.  Varying with respect to the induced metric on $\mathbb{T}$ gives the canonical momentum on $\mathbb{T}$.    And since $S[\mathcal{M}_{\text{ket}}]$, $S[\mathcal{M}_{\text{bra}}]$  appear with opposite signs in \eqref{eq:I}, setting $\delta I =0$ requires the extrinsic curvature and dilaton normal-derivative to satisfy a condition we call SK-continuity at $\mathbb T$. 
This condition states that normal derivatives at $\mathbb T$ on both $\mathcal{M}_{\text{bra}}$ and  $\mathcal{M}_{\text{ket}}$ agree when each is computed using the respective outward normal.
For example, if the timefold is a future boundary, then the future-directed derivatives on the two sides should agree;
see figure \ref{fig:normals}.

Of course, in the interior of  $\mathcal{M}_{\text{bra}}$ and  $\mathcal{M}_{\text{ket}}$, the condition $\delta I=0$ imposes the equations of motion \eqref{eq:dilEoM}. As described in \cite{Colin-Ellerin:2020mva}, combining this with the above continuity conditions shows that the quantity $I$ is invariant under deformations $\mathbb{T}\setminus \mathcal S$ so long as $\mathbb{T}\setminus \mathcal S$ remains causally separated from $\mathcal S$.  This invariance arises from the fact that, since we work in Lorentz signature, the equations of motion \eqref{eq:dilEoM} are hyperbolic.  Since we also have boundary conditions along the left and right timelike boundaries, the equations of motion determine a solution uniquely from the fields and their normal derivatives on any AdS-Cauchy surface as defined in \cite{Wall:2012uf}.  Deformations of $\mathbb{T}\setminus \mathcal S$ that remain outside the light cone of $\mathcal S$ must then lead to deformations of the ket and bra spacetimes on either side that are determined entirely by the Cauchy data on
$\mathbb{T}\setminus \mathcal S$, and which are thus identical on the two sides of $\mathbb{T}\setminus \mathcal S$
As a result, contributions to changes in $I$ from such deformations of  $\mathbb{T}\setminus \mathcal S$ simply cancel between $\mathcal{M}_{\text{bra}}$ and  $\mathcal{M}_{\text{ket}}$, and we are free to place $\mathbb{T}\setminus \mathcal S$  on any convenient spacelike surface connecting $\mathcal S$ with the specified anchor set $\partial L = \partial R \subset \mathcal{\partial B}$.

Finally, there is one further consequence of setting $\delta I=0$ while holding fixed $A_\mathcal S$. As described in \cite{Colin-Ellerin:2020mva} (using the analytic continuation of results from \cite{Dong:2019piw}), this requires the surface $\mathcal S$ to satisfy a condition that generalizes the notion of an extremal surface to spacetimes in which $\mathcal S$ is the locus of an appropriate codimension-2 singularity\footnote{\label{foot:Redelta}Ref. \cite{Colin-Ellerin:2020mva} did not directly address metrics of the most general desired form and, in particular, it did not address metrics which the rapidity parameter $\alpha$ described in appendix \ref{app:Rdelta} differs between ket and bra spacetimes.  However, this case can be included in the analysis of \cite{Colin-Ellerin:2020mva} without significant further changes.}.  We will therefore now refer to $\mathcal S$ as an extremal surface, and we will henceforth equivalently denote it by $\Upsilon$. While details of the technical analysis use the specific boundary conditions imposed by \cite{Dong:2019piw,Colin-Ellerin:2020mva} at a splitting surface, the essential idea is that we have fixed $A_\mathcal S$ but that we still freely vary the rest of the metric near $\mathcal S$.  This imposes continuity conditions on the normal derivative of the induced metric at $\Upsilon$ and, at a conical singularity, those continuity conditions require $\mathcal S = \Upsilon$ to be extremal.  However, since $A_{\mathcal S}= A_\Upsilon$ has been held fixed, conical singularities in the metric are allowed at $\Upsilon$.

Let us now consider a context where, at least in the semiclassical approximation, we can treat the spacetimes over which we integrate as having only a single extremal surface.  (We again leave the interesting more general case for future work.)  The above condition then forces $\Upsilon_i = \mathcal S$ for each $i$, where $\Upsilon_i$ is the surface associated with the argument $A_i$ of the $i$th wavefunction. Thus the path integral in a given ket or bra spacetime computes the wavefunction
\eqref{eq:compLPsi} from a given wavefunction $\Psi_{in}$.

As a result, to all orders in the semiclassical approximation the R\'enyi path integral of \cite{Colin-Ellerin:2020mva} reduces to the same recipe \eqref{eq:rhoLnPsi} for computing R\'enyi entropies from wavefunctions that was found in the Euclidean case.  Note that this agreement continues to hold even if we include contributions from multiple saddles for $(A_i, h_{Li}, h_{Ri})$.  In particular, given Euclidean and Lorentzian path integrals that define the same quantum state (as defined by having the same wavefunction $\Psi[A, h_L, h_R]$), the prescription of \cite{Colin-Ellerin:2020mva} to include \eqref{eq:SUps} in R\'enyi computations guarantees that Lorentzian computations of such R\'enyis will agree with Euclidean computations, at least to all orders in the semiclassical approximation.   Furthermore, since the complex saddles of either computation are determined by \eqref{eq:rhoLnPI} as described at the end of section \ref{sec:Rfw}, the saddle-point spacetimes associated with the Euclidean and Lorentzian path integrals are identical.  The rest of this work is devoted to using the simple case of JT gravity to illustrate such constructions in detail.  In particular, we will see how the complex saddles of \cite{Colin-Ellerin:2021jev} relate to the description given in section \ref{sec:Rfw} in terms of complex values of the conjugate $\Theta$ to $A_\Upsilon$.

\section{Classical tools for semiclassical JT R\'enyis}
\label{sec:JTgravity}

We now wish to use JT gravity as a simple toy model to illustrate the constructions described in sections \ref{sec:Rfw} and \ref{sec:RTRenyi}.  In particular, we will make a concrete link between the discussion of complex saddle points at the end of section \ref{sec:Rfw} and the saddles identified in \cite{Colin-Ellerin:2021jev}.  Since we will study the real-time path integral using semiclassial methods, it is useful to begin with a brief review of classical solutions to JT gravity in Lorentz signature in section \ref{sec:ClassicalJT}. It will also be useful to understand thee phase space structure of the theory.  We review this in section \ref{sec:PhaseSpace}, drawing on the analysis of \cite{Harlow:2018tqv}.   An important role will be played by a time-shift function $\delta$ which we choose to define in a slightly different manner than that of \cite{Harlow:2018tqv}.  This discussion will also provide an opportunity to discuss certain coordinate choices (which may be interpreted as gauge fixing schemes) that we will later use to formulate general wavefunctions $\Psi_{in}$.

\subsection{Classical JT gravity}
\label{sec:ClassicalJT}
The Lorentz signature action for JT Gravity can be written in the form
\begin{equation}
\label{eq:SJT}
    S_{\rm JT}=S_{\text{top}}+S_{\text{dyn}},
\end{equation}
where
\begin{equation}
\label{eq:Stop}
    S_{\text{top}}=\Phi_0 \Bigg[ \int_{\M}\sqrt{-g}R + 2 \int_{\partial\M}\sqrt{|\gamma|}K\Bigg]
\end{equation}
is the topological part of the action and $\Phi_0$ is a positive coupling constant.  The remaining (dynamical) part of the action is then
\begin{equation}
\label{eq:Sphi}
    S_{\text{dyn}}= \int_{\M}\sqrt{-g}\Phi (R+2) + 2 \int_{\B}\sqrt{-h}\Phi(K-1)+ 2 \int_{\Gamma}\sqrt{|q|}\Phi K+ S_{\text{corner}}.
\end{equation}
In the above equations, $\partial \M$ is the entire boundary of the spacetime $\M$, the symbol $\B$ denotes the timelike asymptotically AdS boundary, and $\Gamma$ is any finite boundary.  For example, in figure \ref{fig:singleket} this $\Gamma$ would be the union of $\Sigma_{\text{ket, past}}$, $\Sigma^L_{\text{ket}}$, and $\Sigma^R_{\text{ket}}$.  We use  $h$ and $q$ to denote the determinants of the induced metric on $\B$ and $\Gamma$ respectively\footnote{It is natural to distinguish the asymptotic and finite parts of the boundary because the asymptotic timelike boundaries  require counterterms $-\frac{1}{2}\int_{\B}\sqrt{-h}\Phi$ as shown in \eqref{eq:Sphi}.  In contrast, while a similar counterterm on $\Gamma$ would do little harm when $\Phi$ and $h$ are fixed, the addition of such a term seems unnatural and, in particular, would introduce a new divergence since $\Gamma$ extends to the asymptotic boundary.}.   The corner terms $S_{\text{corner}}$ describe further contributions from the points where $\Gamma$ meets ${\cal B}$ (e.g. the left and right endpoints of the top and bottom edges in figure \ref{fig:singleket}).  These contributions are associated with delta-function extrinsic curvatures that arise when the unit normal of the boundary changes direction discontinuously and are given by integrating $\sqrt{|q|}\Phi K$ (the same integrand as the terms on $\Gamma$) over an infinitesimal region of the boundary containing the corner.   We take ${\cal B}$ itself to be smooth.  In contrast, we require only that $\Gamma$ be continuous, though delta-function contributions from points where its tangent is discontinuous are included in the $\Gamma$ term in \eqref{eq:Sphi}.

Since we require the form of the  junction between $\Gamma$ and ${\cal B}$ to be fixed as a boundary condition, the term
$S_{\text{corner}}$ is just a constant whose value will not affect our calculations below.  However, for completeness, discussion of this constant is included in the computations of appendix \ref{app:PIprop}.   We refer the interested reader to the classic discussion of such terms in \cite{Hayward:1993my}; see also \cite{Neiman:2013ap,Colin-Ellerin:2020mva}.

We wish to fix $q$ on $\Gamma$.  The boundary conditions at $\B$ are instead defined by first imposing
\begin{equation}\label{eq:BC}
\begin{split}
    h_{t t}|_{\B} = 1/\epsilon^2,\quad
    \Phi|_{\B} = \phi_{b}/\epsilon,
    \end{split}
\end{equation}
and then taking the limit $\epsilon \rightarrow 0$.
Varying the action \eqref{eq:Sphi} yields the equations of motion
\begin{equation}
    R=-2,\quad \big( \nabla_\alpha \nabla_{\beta} - g_{\alpha\beta}\big) \Phi=0.
  \label{eq:dilEoM}
\end{equation}

The first equation in \eqref{eq:dilEoM} requires the spacetime to be $\AdS2$ or a quotient of $\AdS2$.
When $\M$ has the topology of a disk,
it is convenient to describe solutions of JT gravity  by realizing AdS$_2$ as the
 hyperboloid \ben
\mathcal{T}_{1}^{2}+\mathcal{T}_{2}^{2}-\mathcal{X}^{2}=1
\een
in the (1+2)-dimensional Minkowski space with line element
\begin{equation}
\label{eq:embmetric}
d s^{2}=-\dd \mathcal{T}_{1}^{2}-\dd \mathcal{T}_{2}^{2}+\dd \mathcal{X}^{2}.
\end{equation}
The general solution for the dilaton then takes the form
\ben
\Phi=A \mathcal{T}_1 +B \mathcal{T}_2 +C \mathcal{X},
\een
where $A,B,C$ are any real numbers. However, our boundary condition is satisfied only when the vector $n^{\mu}=(-A,-B, C)$ is timelike 
with respect to the metric \eqref{eq:embmetric}. We can perform an SO$(1,2)$ isometry and restrict the dilaton to take the form
\ben
\label{eq:Phiprofile}
\Phi=\Phi_h \mathcal{T}_1.
\een
We may also use an SO(2,1) isometry to take  $\Phi_h \ge 0$, since there is an isometry of this form that changes the sign of $\mathcal{T}_1$.  
We will consider only positive values of $\phi_b$, for which the boundary condition requires $\mathcal{T}_1 > 0$ at $\mathcal B$.  We thus restrict the solution to the parts of $\mathcal B$ where this is the case.

\begin{figure}[h!]
	\centering
	\includegraphics[width=0.4\linewidth]{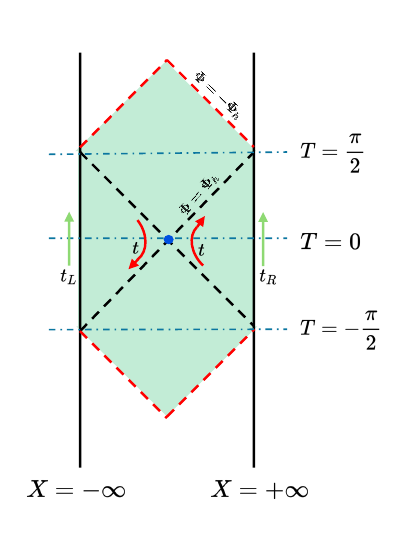}
	\caption{A two-sided black hole in JT gravity is described in global coordinates by the shaded region. The boundary conditions are satisfied on the part of the asymptotic timelike boundary with $T\in [-\pi/2, \pi/2]$, but they will fail to be satisfied on any further extension of the solution along this boundary.  It is thus natural to truncate the boundary at $T=\pm \pi/2$, and to truncate the bulk spacetime along the red dashed lines on which $\Phi=-\Phi_h$.  The blue dot marks the  bifurcation surface of the associated event  horizon.   The global time coordinate $T$ is constant on horizontal lines (such as the dotted blue lines at $T=0, \pm \pi/2$) and increases as one moves up the diagram.  We also use coordinates $t_L,t_R$ on the left and right boundaries that increase toward the future (green arrows).  In contrast, as indicated by the red arrows, the Schwarzschild coordinate $t$ increases toward the past in the left exterior wedge (though it increases toward the future in the right exterior wedge).  }
	\label{fig:TwoSidedBH}
\end{figure}

It will also be useful to introduce global coordinates $T,X$ on AdS$_2$ which are defined by the relations
\ben
\begin{aligned}
\label{eq:embedtoglobal}
\mathcal{T}_{1} =\cos T\sqrt{1+X^{2}} , \quad
\mathcal{T}_{2} =\sin T \sqrt{1+X^{2}} , \quad
\mathcal{X} =X.
\end{aligned}
\een
As a result, up to diffeomorphisms, the metric and dilaton may be written in the forms
\begin{equation}\label{eq:global}
    ds^2=-(1+X^2)dT^2+ \frac{dX^2}{1+X^2},\quad \Phi=\Phi_h \sqrt{1+X^2} \cos T.
\end{equation}
The restriction of $\mathcal B$ to regions with $\mathcal{T}_1 > 0$ implies that $T$ on $\mathcal B$ ranges only over $(-\pi/2, \pi/2)$.
As shown in figure \ref{fig:TwoSidedBH}, it is then natural to truncate the solution at the
surface where $\Phi=-\Phi_h$ and, after doing so, the remaining part of the above solutions may be interpreted as a 1+1 dimensional version of a two-sided black hole.

In the exterior regions shown in figure \ref{fig:TwoSidedBH}, one may instead choose to define Schwarzschild-like coordinates  through the relations
\ben
\begin{aligned}
\label{eq:embedtoSS}
\mathcal{T}_{1}=r / r_{s}, \quad \mathcal{T}_{2}=\pm \sinh \left(r_{s} t\right) \sqrt{\left(r / r_{s}\right)^{2}-1} , \quad
\mathcal{X}= \pm \cosh \left(r_{s} t\right) \sqrt{\left(r / r_{s}\right)^{2}-1} .
\end{aligned}
\een
In both the left and right exterior regions we find the metric and dilaton profile
\ben\label{eq:Sch}
 ds^2=-(r^2-r_s^2)dt^2+\frac{1}{r^2-r_s^2}dr^2,\quad \Phi=\phi_b r,\quad r>r_s.
\een
where $r_s$ is the Schwarzschild radius and $\Phi_h=\phi_b r_s$ is the horizon value of the dilaton.

Taking the left boundary to be at negative $\mathcal{X}$, the above signs then mean that $t$ increases toward the past (the direction of decreasing $T$) along the left boundary while $t$ instead increases toward the future (the direction of increasing $T$) along the right boundary; see again figure \ref{fig:TwoSidedBH}. On either boundary, $t$ ranges precisely once over the real line.  Comparing \eqref{eq:embedtoglobal} with \eqref{eq:embedtoSS} and taking the limit $r/r_s\rightarrow \infty$, we find that on either asymptotic boundary the global and Schwarzschild times satisfy the relation
\begin{equation}
\label{eq:gsbndyrel}
\cos T = \frac{1}{\cosh (r_st)}.
\end{equation}
We will make much use of \eqref{eq:gsbndyrel} below.

\subsection{Phase space, time translations, and the time-shift}
\label{sec:PhaseSpace}

The solutions \eqref{eq:global} described above are labelled by the single parameter $\Phi_h$.    This is much like the situation for spherically symmetric vacuum solutions in higher dimensions where, up to diffeomorphisms, the solutions are parameterized only by the mass $M$.  Of course, a one-dimensional phase space can admit only degenerate symplectic structures (as there is no room for a conjugate to $\Phi_h$ or $M$).  In the higher-dimensional context, it has long been known that a non-degenerate phase space can nevertheless be constructed by taking into account diffeomorphisms that act non-trivially at asymptotic boundaries and which the usual variational principle does not treat as gauge \cite{Kuchar:1994zk}.

The same mechanism of course works in JT gravity as well \cite{Harlow:2018tqv}. In the JT case, the relevant non-gauge diffeomorphisms are those that act as translations of  the Schwarzschild time coordinate $t$ along either boundary.  To be concrete, we may choose some time-shift $\delta$ and define time coordinates $t_L = \delta-t$ and $t_R = t +\delta$ along the left and right boundaries respectively, where the signs are chosen so that $t_L,t_R$ both increase toward the future despite the fact that $t$ increases toward the past along the left boundary; see again figure \ref{fig:TwoSidedBH}.  We then take the labels $t_L, t_R$ of points along each boundary to be part of the gauge-invariant data that distinguishes one solution from another.

The time-shift $\delta$ is then a function on the space of solutions that can be reconstructed as follows.  Consider any point $p_L$ on the left boundary.  This point is labeled by some left-boundary Schwarzschild time $t_L(p_L)$.  Starting at $t_L(p_L)$  and
moving toward the right boundary along the spacelike geodesic orthogonal to the surfaces of constant dilaton, one arrives at a point on the right boundary that has some Schwarzschild time $t_R(p_L)$.  In the solutions \eqref{eq:global}, one always finds $t_R(p_L) = -t_L(p_L)$.  But since the above-mentioned geodesics are just the surfaces $t= \mathrm{constant}$, after our redefinitions above one instead finds
\begin{equation}
t_R(p_L)+t_L(p_L) = 2\delta.
\end{equation}
Since Schwarzschild time translations are symmetries of the solution, our definition of $\delta$ does not depend on the choice of the point $p_L$.   In this sense our time shift is time-independent.

\begin{figure}[h!]
	\centering
	\includegraphics[width=0.8\linewidth]{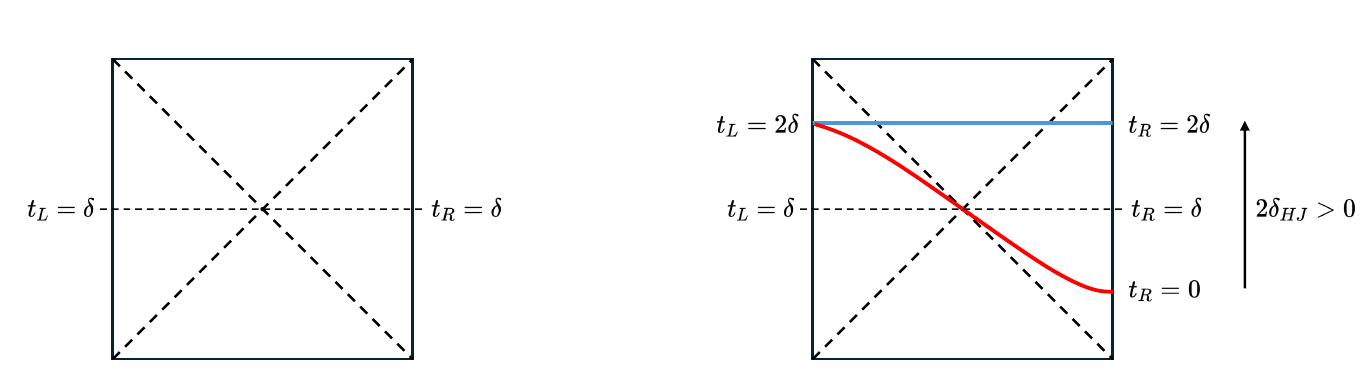}
	\caption{\textbf{Left}: A solution with non-trivial $\delta$ is shown.   The extremal-dilaton point lies on a surface of constant $\tilde T$ (dashed line) that hits the boundary at $t_L=\delta = t_R$. \textbf{Right}: An illustration of the fact  that the time shift $\delta_{\rm HJ}$ defined in \cite{Harlow:2018tqv} agrees with our $\delta$ when $\delta_{\rm HJ}$ is computed for a the Cauchy surface $\tilde T=2\delta$ (blue).    By definition, $2\delta_{\rm HJ}$ is the difference in times $t_R$ defined by the given Cauchy surface (blue) and a geodesic (red) fired from the intersection with the left boundary in the direction orthogonal to surfaces of constant dilaton.
}
	\label{fig:HJdelta}
\end{figure}

In contrast, the related notion of time-shift defined in \cite{Harlow:2018tqv} is defined to depend on a choice of Cauchy surface.  By taking the Cauchy surface to be of the form $\tilde T=\mathrm{constant}$, we may think of it as a function of time, and in particular as a function of $t_L.$  To avoid confusion, we will refer to the time-shift of \cite{Harlow:2018tqv} for such surfaces as $\delta_{\text{HJ}}(t_L)$.  In our language, their $\delta_{\text{HJ}}(t_L)$ is given in terms of the above $t_L(p_L),t_R(p_L)$ by  $\delta_{\rm HJ} = \frac{t_L(p_L) - t_R(p_L)}{2}$  so that $\delta_{\text{HJ}} =  -\delta + t_L$;
see the right panel of figure \ref{fig:HJdelta} for the case $t_L=2\delta$ .

The space of solutions described above is manifestly invariant under arbitrary separate translations of the left and right boundary times $t_L, t_R$.  There should thus be associated left and right Hamiltonians $H_L,H_R$ that generate such translations.
However, because every solution has a symmetry that exchanges the right and left boundaries, we must in fact have $H_L=H_R.$
The observation that $H_L-H_R$ vanishes identically
is consistent with the fact that every element of the phase space is invariant under Schwarzschild time translations which shift $t_L,t_R$ by equal amounts in opposite directions.  Indeed,
\cite{Harlow:2018tqv} finds
\begin{equation}
\label{eq:Hclass}
    H_L = H_R= \frac{\Phi_h^2}{\phi_b}.
\end{equation}
We will thus find it useful below to discuss only the notion of time evolution generated by the `total Hamiltonian'
\begin{equation}
    H= H_L+H_R=\frac{2\Phi_h^2}{\phi_b},
\end{equation}
which we take to translate both $t_L$ and $t_R$ by the same amount.
This $H$ is conjugate to $\delta$ in the sense that the Poisson bracket satisfies \cite{Harlow:2018tqv}
\begin{equation}
\label{eq:HdeltaCR}
    \{H,\delta\}=\{\frac{2\Phi_{h}^{2}}{\phi_b},\delta\}=\frac{4\Phi_{h}}{\phi_b}\{\Phi_{h},\delta\}=1.
\end{equation}
For later convenience, we also define a new variable
\begin{equation}\label{eq:defQ}
    Q_{0} \equiv \frac{4\Phi_h}{\phi_b}\delta,
\end{equation}
which is canonically conjugate to the extremal value of the dilaton $\Phi_h$:
\begin{equation}
    \{\Phi_h,Q_{0}\}=4\frac{\Phi_h}{\phi_b}\{\Phi_h,\delta\}= 1.
\end{equation}

It will also be convenient below to choose a particular bulk diffeomorphism that may be said to be generated by $H$.  This amounts to a form of gauge-fixing, though one may also think of the construction as introducing a globally-defined gauge-invariant time coordinate $\tilde T$ for which the slices of constant $\tilde T$ asymptote to the left and right boundaries at times $t_L(\tilde T)=t_R(\tilde T) = \tilde T$.  It turns out that this can be done by choosing $\tilde T$ to be a function of the global time $T$, so that the surfaces of constant $\tilde T$ coincide with the surfaces of constant $T$.  To see this, let us first note that the relation \eqref{eq:gsbndyrel} of course  continues to hold in solutions with non-zero $\delta$.  On the two boundaries it is natural to write this relation in the forms
\begin{equation}
\label{eq:gsbndyrelLR}
\cos T = \frac{1}{\cosh (r_s [t_L-\delta])}, \ \ \
\cos T = \frac{1}{\cosh (r_s [t_R - \delta])}.
\end{equation}
In particular, we see that we have defined the constant $T$ surfaces of any solution so that the labels $t_L,t_R$ assigned to the left and right boundaries of such surfaces satisfy $t_L - \delta = t_R -\delta$, or equivalently $t_L=t_R$.   As a result, we may define the desired coordinate $\tilde T$  through the relation
\begin{equation}
\cos T = \frac{1}{\cosh \left[r_s (\tilde T - \delta)\right]},
\end{equation}
or,  equivalently,  to set 
\begin{equation}
\label{eq:gsbndyrelR}
\tilde T: = \delta + \frac{\phi_b}{\Phi_h} \cosh^{-1} \left(\cos T \right).
\end{equation}

We will continue to use the coordinate $X$ along each surface of constant $\tilde T$. Consulting \eqref{eq:global} then shows that the profile of our dilaton is proportional to $\sqrt{1+X^2}$ along every constant $\tilde T$ slice.  However, the amplitude of that profile is a function of $\tilde T$ given by
\begin{equation}
\Phi_{\tilde{T}}=\Phi_h\sech\left[ \frac{\Phi_h}{\phi_b}\left(\tilde{T}-\delta\right)\right].
\label{eq:phitDef}
\end{equation}
Thus on any constant $\tilde T$ slice we have
\begin{equation}
\label{eq:fieldstildeT}
ds^2 = \frac{\dd X^2}{1+X^2}, \ \ \ {\rm and} \ \ \  \Phi=\Phi_{\tilde{T}}\sqrt{1+X^2}.
\end{equation}

\subsection{Complex classical solutions}
\label{subsec:JTCCS}

As described in section \ref{sec:RTRenyi}, our R\'enyi entropy computations will in fact be dominated by complex saddle-point geometries.  In particular, while we expect the relevant geometries to have real $\Phi_h$, the conjugate variable $Q_0$ will be complex.  Since $\Phi_h,Q_0$ specify classical solutions uniquely, after a complete gauge fixing there are no remaining configuration variables $h$.  Any bulk wavefunction thus trivially takes the form \eqref{eq:factorPsi}.  We may thus use the JT gravity analogue of \eqref{eq:ImTheta} which imposes
\begin{equation}
\label{eq:Imdelta}
{\rm Im} \, Q_0 = 4\pi\frac{n-1}{2n} \ \ \ {\rm or, \ equivalently,} \ \ \ {\rm Im} \ \delta = \frac{\pi \phi_b}{\Phi_h}\frac{n-1}{2n}.
\end{equation}

We therefore take this opportunity to discuss the relevant complex solutions and, in particular, to describe their relationship to the saddles identified in \cite{Colin-Ellerin:2021jev}.  Since the extremal-dilaton point occurs at $\tilde T=\delta$, a primary effect of making $\delta$ complex is to move the extremal point off of the surface defined by real values of the coordinates $\tilde T, X$.  While this is important for our analysis, we emphasize that it is a gauge choice.  Given another system of coordinates whose real surface agrees for real solutions with the surface defined by real values of $\tilde T, X$, the corresponding real surfaces generally differ for complex values of $\delta$.  There will thus be other systems of coordinates (i.e., other choices of gauge) in which the extremal-dilaton point remains at real coordinate values.

Indeed, the discussion of \cite{Colin-Ellerin:2020mva,Colin-Ellerin:2021jev} was phrased entirely in terms of coordinates $\tilde x^\pm$ taking real values at the extremal point.  
In particular, it was assumed there that a timefold $\mathbb{T}$ running through the extremal surface could be placed at real values of the coordinates.  This assumption then led to the conclusion that the fields would remain real at points spacelike separated from the extremal point (and also that the notion of `spacelike separated' was well-defined).  In contrast,  in our gauge there will be no corresponding real Lorentzian region.  In particular, as is clear from \eqref{eq:phitDef}, for $n\neq 1$ the result \eqref{eq:Imdelta} requires $\Phi_{\tilde T}$ to be complex at all real values of $\tilde T$.

Nevertheless, we noted in section \ref{sec:RTRenyi} that there is much freedom in the choice of where to place the timefold.  In the ket saddles, we will find it convenient to proceed as follows.  First, we extend the coordinates $t_L,t_R$ from the boundary into the bulk by taking them to be constant along slices of constant Schwarzschild time $t$ and making appropriate analytic continuations.  Doing so means that at all points in the bulk we may choose to set $t_L = -t_R +i \pi \phi_b/\Phi_h$.  While both of these coordinates are singular at the extremal-dilaton point, there are surfaces $t_L = 0$ and $t_R = 0$ running from the extremal point to the AlAdS boundary at points with $\tilde T=0$.  It will be convenient to take the timefold $\mathbb{T}$ to be the union of these surfaces in both the ket and bra spacetimes.  This choice may be justified by noting that it lies at real values of the coordinates in all constrained ket-saddles defined by first fixing real values of $\Phi_h$, and that we now merely consider the analytic continuation to complex $\Phi_h$.  (We will also see below that it lies at real values of the coordinates $\tilde x^\pm$ defined in \cite{Colin-Ellerin:2020mva,Colin-Ellerin:2021jev}.)

The required continuity conditions at $\mathbb{T}$ are satisfied by virtue of the fact that the bra spacetimes are the complex-conjugates of the ket spacetimes\footnote{This follows from the fact that the action for a single ket/bra region is a real function.  As a result, saddles of an integral of $e^{iS}$ will occur in complex-conjugate pairs.  Complex conjugation of the space of paths exchanges these saddles, but also maps descent curves from one saddle to ascent curves from the other. Thus at most one saddle from each pair contributes to the integral.   The integral over $e^{-iS}$ thus has the same saddles, but will receive contributions only from the complex conjugates of the saddles that contributed to integrating $e^{iS}$.}.  We may then choose any contour in the $\tilde T, X$ complex plane that connects the above $\mathbb{T}$ with the AlAdS boundaries at real $\tilde T$ and with the past boundaries at real $X$ and the specified real value of $\tilde T$.

In order to get further insight into these solutions, let us now focus on the case where the real part of $\delta$ vanishes, so that we have $\delta =i \frac{\pi \phi_b}{\Phi_h}\frac{n-1}{2n}$.  Since the timefold lines $t_L = 0$ and $t_R = 0$ correspond to setting the Schwarzschild time $t$ to $\pm \delta$ (in the right/left exterior wedges), for $\delta$ imaginary our timefold lies at real values of $\tau = -it$.  In particular, it lies in a Euclidean-signature section of our complex solution.  Furthermore, the surfaces $t_L=0, t_R=0$ meet at the extremal point at a Euclidean-signature angle defined by the difference in $-i\frac{\Phi_h}{\phi_b}t_R$ on the two surfaces.  This difference is
\begin{equation}
\label{eq:fracofEucl}
-\frac{\Phi_h}{\phi_b}\left( 2i\delta +\pi\right) =  \frac{n-1}{n}\pi -\pi = -\frac{\pi}{n}.
\end{equation}
Since the full complex solutions is composed of $n$ ket-regions and $n$ corresponding bra-regions, we see that for smooth saddles (where the sum of these angles must add to $\pm 2\pi$), we may consistently place the timefolds on the above locus in every ket-region and also in every bra-region.

Let us now compare such regions with the complex saddles for JT gravity described in \cite{Colin-Ellerin:2021jev}.  Those saddles were constructed by noting that all Euclidean solutions are diffeomorphic to Euclidean JT black holes, so that they correspond to a dilaton profile on the hyperbolic plane.  The hyperbolic plane was then described as the Poincar\'e disk using a complex coordinate $z$, in which the metric and dilaton take the form
\begin{equation}
\label{eq:Esol}
\dd s^2 = 4\frac{\dd z \dd\bar z}{(1-z\bar z)^2}, \ \ \ \phi = \Phi_h \frac{1+z\bar z}{1-z\bar z},
\end{equation}
so that the extremal-dilaton point has been placed at $z=0$.
Writing $v=z^n$ defines a complex coordinate $v$ for which a single copy of the complex $v$-plane describes a $\frac{1}{n}$ fraction of the Poincar\'e disk.  Replacing $v, \bar v$ by real coordinates  $\tilde x^+, \tilde x^-$ then yields the analytic continuation
\begin{equation}
\label{eq:prevsol}
\begin{aligned}
\dd s^{2} &=\frac{4\left(\tilde{x}^{+} \tilde{x}^{-}\right)^{\frac{1-n}{n}}}{n^{2}\left(1-\left(\tilde{x}^{+} \tilde{x}^{-}\right)^{\frac{1}{n}}\right)^{2}} \dd \tilde{x}^{+} \dd \tilde{x}^{-}, \\
\phi &=\Phi_h \frac{1+\left(\tilde{x}^{+} \tilde{x}^{-}\right)^{\frac{1}{n}}}{1-\left(\tilde{x}^{+} \tilde{x}^{-}\right)^{\frac{1}{n}}},
\end{aligned}
\end{equation}
on each of the $n$ ket-spacetimes, where the functions $\left(\tilde{x}^{+}\right)^{1/n}$ and $\left(\tilde{x}^{-}\right)^{1/n}$ are defined such that their product is real and positive for real, positive $\tilde{x}^{+}, \tilde{x}^{-}$, though where for $\tilde{x}^{+}=1$ the function $\left(\tilde{x}^{+}\right)^{1/n}$  yields the $i$th root of unity on the $i$th ket spacetime.
The functions at negative $\tilde{x}^\pm$ are then defined by analytic continuation using the $i\epsilon$ prescription $\tilde{x}^{\pm}\rightarrow \tilde x^\pm \mp i \epsilon$ for $\epsilon > 0$.
Here $\tilde{x}^\pm$ both increase in the spacelike direction to the right; i.e., it is natural to introduce time and space coordinates $\tilde t, \tilde x$ such that $\tilde{x}^\pm = \tilde x \pm \tilde t$.  The ket spacetimes are in fact taken to be the regions $\tilde t< 0$ (the lower half planes).   The upper half of the complex $v$-planes are instead analytically continued in the opposite direction to yield $n$ bra-spacetimes on which the fields are the complex conjugate of \eqref{eq:prevsol}, for which the most salient difference is that the opposite $i\epsilon$ prescription is used in defining analytic continuations.

It is then a straightforward-but-tedious exercise to compose the above maps to find
\bal
\tan T&=\frac{(\tilde x^+)^{\frac{1}{n}}-(\tilde x^-)^{\frac{1}{n}}}{1+(\tilde x^+ \tilde x^-)^{\frac{1}{n}}},\\
X&=\frac{(\tilde x^+)^{\frac{1}{n}}+(\tilde x^-)^{\frac{1}{n}}}{1-(\tilde x^+ \tilde x^+)^{\frac{1}{n}}},
\eal
where as usual
\ben
\label{eq:deltashiftagain}
\cos T=\operatorname{sech} \left[ \frac{\Phi_h}{\phi_b} (\tilde T-\delta) \right].
\een
An even more tedious computation verifies that this transformation maps our saddles to \eqref{eq:prevsol}.  Noting that the surface $\tilde x^+ = \tilde x^-$ (i.e., $\tilde t=0$) lies in the above Euclidean section and that, since $z=v^{1/n}$ is in fact a smooth coordinate, it there defines a pair of radial lines that intersect with an angle of $\frac{\pi}{n}$, we see that this surface coincides with our choice of timefolds at $t_L=0, t_R=0$.

Careful study of the above rules defining the functions $\left(\tilde{x}^{+}\right)^{1/n}$ also shows that $X,T$ are real at positive $\tilde x^+ \tilde x^-$; i.e., at what in \eqref{eq:prevsol} are spacelike separations from the extremal point.  However, due to eq
\eqref{eq:deltashiftagain} and the fact that $\delta =i \frac{\pi \phi_b}{\Phi_h}\frac{n-1}{2n}$, the coordinate $\tilde T$ is generally {\it not} real in those regions.  Thus the ket spacetimes described by real $\tilde T, X$ do not precisely correspond to those described by real $\tilde x^\pm$.  On the other hand, the two are deformable to each other within the full complex solution, even with the boundaries at $\mathbb{T}$ being held fixed.   Nevertheless we note that, in section \ref{sec:JTSn}, we will choose to impose boundary conditions associated with the choice of quantum state on surfaces $\tilde T = \tilde T_0$ with $\tilde T_0$ real, so that a description in terms of $\tilde x^\pm$ would necessarily require either using complex values of  $\tilde x^\pm$, or else evolving our quantum state to a slice on which $\tilde x^\pm$ are real\footnote{\label{foot:evolve}The use of the term `evolve' may be confusing in the context of complex spacetimes with complex coordinates.  Here this simply means to transform a given wavefunction $\Psi[\Phi_{{\tilde T}_0}]$ with real $\tilde T_0$ to a wavefunction representing our quantum state in terms of observables defined at real $\tilde x^\pm$.}.   Our current example thus shows that, while the gauge choices used in \cite{Colin-Ellerin:2020mva,Colin-Ellerin:2021jev} are certainly allowed, they may not always be convenient.  As a result, the form of the saddles presented in \cite{Colin-Ellerin:2020mva,Colin-Ellerin:2021jev} may not always be of direct use.

Let us now return briefly to the more general case where $\delta$ has a non-vanishing real part.   This case was not addressed in \cite{Colin-Ellerin:2021jev}, which considered only time-symmetric contexts. However, the generalization of their analysis is straightforward.  It merely requires replacing $1/n$ in the above formulae by
$1+i\frac{2\Phi_h \delta }{\pi \phi_b}$.  This will again lead the surfaces $\tilde x^+ = \tilde x^-$ to agree with union of  the surfaces $t_L=0$ and $t_R=0$, though a non-zero real part of $\delta$ means that neither will lie in the Euclidean section \eqref{eq:Esol}.  Again, the two real surfaces will generally not coincide, though they can be deformed to each other if one ignores the `past' boundary of our ket spacetimes at $\tilde T = \tilde T_0$.

\section{R\'enyi entropies in JT gravity}
\label{sec:JTSn}

We now wish to use the ingredients described in previous sections to compute  R\'enyi entropies for JT gravity in a quantum state specified as an arbitrary wavefunction of $\Phi_{\tilde T}$.  In particular, we wish to use JT gravity to illustrate the way in which this calculation associates complex spacetimes with semiclassical wavefunctions. Following the reasoning of section \ref{sec:Rfw}, our analysis will make use of the JT analogue of \eqref{eq:rhoLnPsi}, replacing the area $A_\Upsilon$ of $\Upsilon$ by the extremal value $\Phi_h$ of the dilaton.  Doing so requires that we understand how to transform wavefunctions of $\Phi_{\tilde T}$ to wavefunctions of $\Phi_h$.  We thus begin with a brief (and semiclassical) discussion of quantum JT gravity in section \ref{sec:scJT}.  In particular, this provides an opportunity to fix all conventions and phase factors.  It will also immediately tell us the transformation rules between wavefunctions of $\Phi_{\tilde T}$ and wavefunctions of $\Phi_h$.  With this in hand, we then compute R\'enyi entropies and R\'enyi spacetimes for a 4-parameter class of semiclassical states in section \ref{sec:scSns}.  As a further illustration, we also use the above transformations to study the $\Phi_{\tilde T}$ representation of the Hartle-Hawking state in appendix \ref{sec:setupHHW}.

\subsection{Semiclassical JT gravity}
\label{sec:scJT}

Any quantum theory of JT gravity will contain self-adjoint operators $\hat H, \hat \Phi_h, \hat \delta, \hat Q_0$, and $\hat \Phi_{\tilde{T}}$ that, in the classical limit, yield the Hamiltonian $H$, bifurcation-surface dilaton $\Phi_h$, time shift $\delta$, and quantities $Q_0,$ $\Phi_{\tilde{T}}$.  Indeed, the theory will necessarily contain many such operators that we could call e.g. $\hat \Phi_h$ and which coincide in the classical limit.  It suffices for our purposes to simply pick any one of these.

Having done so, we may introduce the (delta-function normalized) eigenstates $|\Phi;\tilde T\rangle$ of $\hat \Phi_{\tilde T}$, as well as the (delta-function normalized) eigenstates $|\Phi_h; \Upsilon\rangle$ of $\hat \Phi_{h}$.  Given a quantum state $|\Psi\rangle$, the associated wavefunctions $\Psi[\Phi;\tilde T] := \langle \Phi; \tilde T| \Psi\rangle$ and $\Psi[\Phi_h;\Upsilon] := \langle \Phi_h;\Upsilon|\Psi\rangle$ are then related by writing
\begin{equation}
\Psi[\Phi_h;\Upsilon] = \int d\Phi \langle \Phi_h;\Upsilon|\Phi;\tilde T\rangle  \Psi[\Phi;\tilde T].
\end{equation}

We are interested in studying the theory in
the semiclassical limit
\begin{equation}
\label{eq:sclim}
\phi_b \rightarrow \infty \ \ \
 {\rm with} \ \ \  \frac{\Phi_0}{\phi_b}, \frac{\Phi_h}{\phi_b}, \frac{\Phi_{\tilde{T}}}{\phi_b} \ \ \  {\rm finite}.
  \end{equation}
  In this limit it is straightforward to compute the kernel $\langle \Phi_h;\Upsilon|\Phi;\tilde T\rangle $ using either canonical or path-integral methods.  We will use the canonical method here as it turns out to be computationally simpler.  The path integral method of course gives the same result and is described in appendix \ref{app:PIprop}.

Due to our emphasis on the semiclassical limit \eqref{eq:sclim},
we need only identify the above operators to leading order in $\phi_b$, which in particular means that we study only the part of the theory in which the eigenvalues of $\hat \Phi_h$ and $H$ are large.  Under the assumption that the spectrum of energies is non-negative, we are free to impose the relation \eqref{eq:Hclass} at the quantum level by writing
\begin{equation}
\label{eq:Hhat}
    \hat{H}=\frac{2\hat{\Phi}_h^2}{\phi_b}.
\end{equation}
However, if we were to also impose
\begin{equation}
\label{eq:Hdeltacr}
[\hat H, \hat \delta] = i,
\end{equation}
then $\delta$ would be forbidden from being self-adjoint  by the familiar argument that the unitary operator $e^{i\lambda \delta }$ would then shift any eigenstate of $\hat H$ with eigenvalue $E$ to one with eigenvalue $E-\lambda$ (thus contradicting the above non-negativity of the energy spectrum). Since the classical theory requires $\hat \Phi_h \ge 0$, if we impose this constraint at the quantum level we find the same subtlety for the operators $\hat \Phi_h$ and $\hat Q_0$.  However, there is no obstacle to taking $\hat \Phi_h, \hat Q_0$ self-adjoint with $\hat \Phi_h \ge 0$ and imposing
\begin{equation}
\label{eq:PhiQcr}
[\hat \Phi_h, \hat Q_0] = i \left( 1 + {\cal O}(\phi_b^{-1})\right).
\end{equation}
One might then attempt to define
\begin{equation}
\label{eq:hatdeltasym}
    \hat \delta = \frac{1}{8} \left(\frac{\phi_b}{\hat{\Phi}_h} \hat{Q}_0  +  \hat{Q}_0 \frac{\phi_b}{\hat{\Phi}_h}\right)
\end{equation}
to realize \eqref{eq:Hdeltacr} exactly.  The operator \eqref{eq:hatdeltasym} is then symmetric but, as foreshadowed above, it fails to be self-adjoint due to issues associated with zeros of $\hat \Phi_h$.  Alternatively, however, one can add subleading terms in $\phi_b$ to \eqref{eq:hatdeltasym} to make $\delta$ self-adjoint at the expense of then also modifying \eqref{eq:Hdeltacr}. See e.g. \cite{Grot:1996xu} for the corresponding (and mathematically equivalent)  discussion of the `time-of-flight operator' in non-relativistic free-particle quantum mechanics; see also \cite{Wigner:1955zz,Smith:1960zza,Olkhovsky:1974sf,Piron:1979,Kumar1985,Gurjoy:1989,Aharonov:1997md} for a variety of viewpoints on such operators.  The upshot is that we can ignore all such subtleties in our semiclassical limit and instead treat $\delta$ as if it has a complete set of orthogonal eigenstates while still using \eqref{eq:Hdeltacr} at leading order in $\phi_b$.

Combining \eqref{eq:Hhat} with \eqref{eq:PhiQcr} yields $[H, \hat Q_0] = 4i\frac{\hat{\Phi}_h}{\phi_b}(1+ {\cal O}(\phi_b^{-1}))$, whence it is straightforward to compute the Heisenberg picture evolution  $\hat{Q}_{\tilde{T}}$ of $\hat Q_0$.   Denoting the time-evolution operator by $\mathcal{U}=e^{-i\hat{H}\tilde{T}}$, we simply write
\begin{equation}
\label{eq:QT}
    \hat{Q}_{\tilde{T}}=\mathcal{U}^\dagger\hat{Q}_0\mathcal{U}=e^{i\hat{H}\tilde{T}}\hat{Q}_0 e^{-i\hat{H}\tilde{T}}=\hat{Q}_0 - \frac{4\hat{\Phi}_h}{\phi_b}\tilde{T}.
\end{equation}
Here we have chosen to denote the time argument by $\tilde T$ since the time evolution operator
$\mathcal{U}=e^{-i\hat{H}\tilde{T}}$ may be thought of as simply shifting the time labels along both the left and right timelike boundaries by $\tilde T$.  If we think of
$\hat Q_0$ as being computed from Cauchy data on the slice $\tilde T=0$, the operator $\hat{Q}_{\tilde{T}}$ is then given by the same computation in terms of the Cauchy data on the $\tilde T$ slice. Since $\hat{\Phi}_h$ is conserved, we also find
\begin{equation}\label{eq:CommutationRelation}
    [\hat{\Phi}_h,\hat{Q}_{\tilde{T}}]=[\hat{\Phi}_h,\hat{Q}_0]=i(1 + {\cal O}(\phi^{-1}_b)).
\end{equation}

Below, we will use the notation $|Q;\tilde{T}\rangle$ to denote the eigenstate of the operator $\hat{Q}_{\tilde{T}}$ with eigenvalue $Q$, while $| Q;0\rangle$ is the eigenstate of the operator $\hat{Q}_0$ with the same eigenvalue $Q$, i.e.,
\begin{equation}\label{eq:eigenvalue}
    \hat{Q}_0| Q;0\rangle=Q| Q;0\rangle,\quad \hat{Q}_{\tilde{T}}| Q;\tilde{T}\rangle=Q| Q;\tilde{T}\rangle.
\end{equation}
In particular, we will choose the phases of these states to satisfy the natural relations
\begin{equation}\label{eq:relationship}
    |Q;\tilde{T}\rangle=\mathcal{U}^\dagger|Q;0\rangle \ \ \ {\rm and} \ \ \     |Q+\lambda;\tilde{T}\rangle=e^{i \lambda \Phi_h}|Q;\tilde{T}\rangle (1+{\cal O}(\phi_b^{-1})).
\end{equation}

Finally,
we note that the classical version of the relation \eqref{eq:QT} allows us to rewrite \eqref{eq:phitDef} in the form $\Phi_{\tilde{T}}=\Phi_h\sech\left(Q_{\tilde T}/4\right).$  At the semiclassical level, we are thus free to make the definition
\begin{equation}
\label{eq:Phithatdef}
\hat \Phi_{\tilde{T}}=\frac{1}{2} \left( \hat \Phi_h\sech\left(\hat Q_{\tilde T}/4\right) + \sech\left(\hat Q_{\tilde T}/4\right)\hat \Phi_h \right).
\end{equation}
By this we mean that we can use \eqref{eq:Phithatdef} to define the symbol $\hat \Phi_{\tilde{T}}$ in the full quantum theory, and that we can rest assured that it has the desired physical interpretation at least in the semiclassical limit \eqref{eq:sclim}.  Combining \eqref{eq:Phithatdef} with \eqref{eq:CommutationRelation} then yields
\begin{equation}\label{eq:CommutationRelation2}
    \left[\hat{\Phi}_{\tilde T},4\sinh \left(\frac{\hat Q_{\tilde T}}{4} \right)\right]=i (1+{\cal O}(\phi_b^{-1})).
\end{equation}
As a result, we will use the notation $|\Phi;\tilde{T}\rangle$ to denote the basis of $\hat \Phi_{\tilde T}$-eigenstates that satisfy
\begin{equation}
\label{eq:Phitstate}
    |\Phi;\tilde{T}\rangle=\mathcal{U}^\dagger|\Phi;0\rangle \ \ \ {\rm and} \ \ \  |\Phi-\lambda;\tilde{T}\rangle=e^{i 4 \lambda \sinh (Q_{\tilde T}/4)}|\Phi;\tilde{T}\rangle \, (1+{\cal O}(\phi_b^{-1})).
\end{equation}
The states $|\Phi_h;\Upsilon \rangle$ are similary defined to satisfy
\begin{equation}
|\Phi_h+\lambda; \Upsilon \rangle = e^{-i\lambda Q_0}|\Phi_h;\Upsilon \rangle (1+{\cal O}(\phi_b^{-1})).
\end{equation}
For later use we note that, since $Q_0$ is odd under time-reversal, this relation will be satisfied if we take each $|\Phi_h;\Upsilon \rangle$ to be even under time-reversal.  As a result, the above definition is consistent with the Euclidean path integral computations of wavefunctions in \cite{Harlow:2018tqv}.

The change of basis from $\Phi_h$ to $Q_{\tilde T}$ is of course accomplished by computing
\begin{equation}
     \tilde \Psi[Q;{\tilde{T}}]:= \BraKet{Q;{\tilde{T}}}{\Psi} =\int d \Phi_h\, \BraKet{Q;{\tilde{T}}}{\Phi_h} \BraKet{\Phi_h;\Upsilon}{\Psi}.
     \label{eq:HHpt}
\end{equation}
From \eqref{eq:CommutationRelation} and \eqref{eq:QT} we immediately find
\begin{equation}
\label{eq:BracketQPhih}
	\begin{split}
		\BraKet{Q;\tilde{T}}{\Phi_h;\Upsilon}&=\langle Q;\tilde{T}=0|\exp\left(-i\frac{2\hat{\Phi}_h^2}{\phi_b}\tilde{T}\right)|\Phi_h\rangle=\exp\left(-iQ\Phi_h-i\frac{2\Phi_h^2}{\phi_b}\tilde{T}\right) \, (1+{\cal O}(\phi_b^{-1})).
	\end{split}
\end{equation}
Similarly, \eqref{eq:CommutationRelation2} requires
\begin{equation}
     \BraKet{\Phi;{\tilde{T}}}{Q;{\tilde{T}}}=  \sqrt{\cosh Q/4} \exp\left(i 4 \Phi \sinh Q/4 \right) \, (1+{\cal O}(\phi_b^{-1})),
    \label{eq:BraKetPhiQ}
\end{equation}
where the factor $\sqrt{\cosh Q/4}$ is required for the states $|Q;\tilde T\rangle$ to be delta-function normalized.  While one would generally expect this normalization factor to be subleading in the semiclassical limit \eqref{eq:sclim}, this will not be the case when $Q/4$ is an integer multiple of $2\pi i$.  We will therefore carry this factor along in our compuations below. Path integral interpretations of the results \eqref{eq:BracketQPhih}, \eqref{eq:BraKetPhiQ} are provided in appendix \ref{app:PIprop}.

The above results then give the following expression for transforming a representation of the state in the basis $|\Phi_h\rangle$ into a representation in terms of the basis $|\Phi; {\tilde T}\rangle:$
\begin{equation}
\begin{split}
    \BraKet{\Phi_h;\Upsilon}{\Psi} &=\int d \Phi d Q
    \langle\Phi_h;\Upsilon|Q;\tilde{T}\rangle
    \langle Q;\tilde{T}|\Phi;\tilde{T}\rangle\langle\Phi;\tilde T|\Psi\rangle\\
    &=\int d \Phi d Q
     \sqrt{\cosh\frac{Q}{4}}
    \exp \left(-4i\Phi\sinh\frac{Q}{4}\right)\exp\left(iQ\Phi_h+i\frac{2\Phi_h^2}{\phi_b}\tilde{T}\right)\Psi[\Phi;\tilde T] \,  (1+{\cal O}(\phi_b^{-1}))\\
    &=\int d \Phi d Q
    \sqrt{\cosh\frac{Q}{4}}
    \exp\left[-i\left(4 \Phi \sinh{\frac{Q}{4}} -Q\Phi_h-\frac{2\Phi_h^2}{\phi_b}\tilde{T}\right)\right] \Psi[\Phi;\tilde T] \,  (1+{\cal O}(\phi_b^{-1})).
    \label{eq:PtPh}
    \end{split}
\end{equation}
 We will study this transformation in the next section for simple classes of semiclassical states.

\subsection{Semiclassical R\'enyis}
\label{sec:scSns}

With the above results in hand, we are ready to compute semiclassical R\'enyi entropies and the associated complex R\'enyi spacetimes.  To do so, we need only choose an appropriate class of initial states.   Given a wavefunction $\Psi[\Phi_h;\Upsilon]$, we may then apply the JT analogues of \eqref{eq:rhoLnPsi}  and \eqref{eq:HJconjfactor}, which at leading order in large $\phi_b$ yield
\begin{equation}
\label{eq:TrnfromPsi3}
{\rm Tr}_L \, \rho_L^n = \int_{\Phi_h>0} d\Phi_h  e^{-4\pi(n-1)(\Phi_0+\Phi_h)}\Big|\Psi[\Phi_h;\Upsilon] \Big|^{2n} =
e^{-4\pi(n-1)\left(\Phi_0 +\Phi_{h*}\right)}\Big|\Psi[\Phi_{h*};\Upsilon] \Big|^{2n},
\end{equation}
and
\begin{equation}
\label{eq:finaldelta}
\delta = -i \frac{\phi_b}{4\Phi_h}\frac{\partial}{\partial \Phi_h}  \ln \Psi[\Phi_h;\Upsilon]  \Bigg|_{\Phi_h= \Phi_{h*}}.
\end{equation}
Here, as usual, $\Phi_{h*}$ is the stationary point of $e^{-4(n-1)\pi\Phi_h}\Big|\Psi[\Phi_h;\Upsilon]\Big|^{2n}$, or equivalently of
 $e^{-4(1-1/n)\pi\Phi_h}\Big|\Psi[\Phi_h;\Upsilon]\Big|^{2}$ and $e^{-2(1-1/n)\pi\Phi_h}\Big|\Psi[\Phi_h;\Upsilon]\Big|$.

As a first simple illustration we may consider general Gaussian wavefunctions of $\Phi_h$ with real width $\sigma_h$ so that for some real $\sigma_h, \Phi_{h1}, \delta_1$ we have
\begin{equation}
\label{eq:PhihGaussian}
 \Psi[\Phi_h;\Upsilon] \propto e^{-\frac{(\Phi_h-\Phi_{h1})^2}{2\sigma_h^2}}e^{i 4 \Phi_h \Phi_{h1} \delta_1/\phi_b}.
\end{equation}
 Note that we have not attempted to normalize our wavefunction since the correct notion of norm should be defined by the path integral\footnote{If we had taken sufficient care in the normalization of the path integral, we could have chosen to make that norm agree with the standard inner product $L^2({\mathbb {R}}, d\Phi)$.}, and that normalization constants will in any case cancel in the computation of R\'enyi entropies $S_n$ through \eqref{eq:Sn} due to the factor of $\left({\rm Tr}_L \, \rho_L\right)^n$ that divides ${\rm Tr}_L \, \rho_L^n$. The above state yields
\begin{equation}
\label{eq:PhihGresults1}
\Phi_{h*} = \Phi_{h1} - 2\pi \sigma^2_h \left(1 - \frac{1}{n}\right),  \ \ \  \delta = \frac{\Phi_{h1}}{\Phi_{h*}}\delta_1 - \pi i \left(1-\frac{1}{n}\right) \frac{\phi_b}{2\Phi_h}, \ \ \ {\rm and} \ \ \
{\rm Tr}_L \, \rho_L^n = e^{-4\pi(n-1)(\Phi_0+\Phi_{h1})}e^{4\pi^2 \sigma^2_h (n-1)^2/n},
\end{equation}
 where one can now see that the notation $\Phi_{h1}, \delta_1$ was chosen to reflect the semiclassical values of $\Phi_h, \delta$ in the $n=1$ spacetimes.    The associated R\'enyi entropies are thus
\begin{equation}
\label{eq:PhihGresults2}
S_n = \frac{1}{1-n} \left(-4\pi(n-1)(\Phi_0+\Phi_{h1}) +4\pi^2 \sigma^2_h \frac{(n-1)^2}{n}\right)
= 4\pi\left(\Phi_0+\Phi_{h1}\right) - 4\pi^2 \sigma^2_h \frac{(n-1)}{n},
\end{equation}
with the von Neumann entropy being given by the case $n=1$. 
In particular, it should be no surprise that, for all $\delta_1$, the limit $\sigma_h \rightarrow 0$ gives the leading semiclassical approximation to the R\'enyi entropies of the microcanonical ensemble of JT black holes associated with the extremal-dilaton value $\Phi_h$.  In addition, we note that \cite{Harlow:2018tqv} found the Hartle-Hawking state for a two-sided black hole of inverse temperature $\beta$ to be such a Gaussian in $\Phi_h$ with $\delta_1=0$ (due to the time-reflection symmetry of the Hartle-Hawking states), $\Phi_{h1}= \frac{2\pi\phi_b}{\beta}$, and width $\sigma^2_h ={\frac{\phi_b}{\beta}}$. Using these values in
\eqref{eq:PhihGresults1} and \eqref{eq:PhihGresults2}, one obtains
\begin{equation}
\label{eq:PhihGresults}
{\rm Tr}_L \, \rho_L^n = e^{-4\pi(n-1)\Phi_0} e^{-8\pi^2 (n-1)\frac{\phi_b}{\beta}}e^{4\pi^2 \frac{(n-1)^2}{n}\frac{\phi_b}{\beta}},
\ \ \ S_n = 4\pi\Phi_0+4\pi^2 \frac{\phi_b}{\beta}\left(1{+}\frac{1}{n} \right),
\end{equation}
which of course match the standard Euclidean results \cite{Maldacena:2016hyu}. For completeness, we include a semiclassical calculation of the wavefunctions $\Psi[\Phi;\tilde T]$ for these Hartle-Hawking states in appendix \ref{sec:StatePrepartion}.

The attentive reader will note that the wavefunction \eqref{eq:PhihGaussian} has support at negative $\Phi_h$ as well as at $\Phi_h >0$.  Since our treatment is semiclassical, we remain agnostic as to whether the regime $\Phi_h <0$ is meaningful in some complete nonperturbative theory.  However, it is clear that a semiclassical treatment is unreliable if it requires this regime to play an important role\footnote{\label{foot:smallPhi}As usual, without detailed knowledge of the physics at small $\Phi_h$ it is not possible to rigorously prove the semiclassical approximation to be reliable in any regime.  Indeed, a sufficiently large and sufficiently contrived modification at small $\Phi_h$ could certainly change the result.  What can be done is instead to analyze the self-consistency of the semiclassical approximation and its sensitivity to modifications at small $\Phi_h$ associated with natural extrapolations of the given wavefunction.}.    Let us thus suppose for the moment that we instead considered a cutoff wavefunction that vanishes for $\Phi_h< \Phi_{h\flat}$ for some $\Phi_{h\flat} >0$.  Such a wavefunction is not analytic, so the resulting integral in \eqref{eq:TrnfromPsi3} is not ammenable to direct application of saddle-point methods.  However, for many values of $\Phi_h$, the difference between the integral \eqref{eq:PtPh} defined by this cutoff wavefunction and that defined by the full Gaussian \eqref{eq:PhihGaussian} would be negligible.  In such cases we may then simply replace the truncated wavefunction with \eqref{eq:PhihGaussian} and proceed with the saddle-point calculation as above.   In particular, since the integrand in \eqref{eq:TrnfromPsi3} is a Gaussian of width $\sigma_h/\sqrt{2n}$,  the effect of the above replacement is negligible for $\Phi_{h*} \gg \frac{\sigma_h}{\sqrt{2n}}$.  Furthermore, since $n>0$ we see from \eqref{eq:PhihGresults} that this condition holds in the semiclassical limit of large $\phi_b$ with $\Phi_{h1} \propto \phi_b$ and $\sigma_h \propto \sqrt{\phi_b}$ so long as $\frac{\Phi_{h1}}{2\pi \sigma_h^2} > 1$.  On the other hand, when this condition fails, the semiclassical approximation to $S_n$ is valid only for a finite range of $n$ even in the limit $\phi_b \rightarrow \infty.$

As a second illustration, let us consider the states $|\Phi; \tilde T\rangle$.  If we simply use one of these eigenstates then, to leading order in $\phi_b$, from \eqref{eq:PtPh} we obtain
\begin{equation}
|\langle \Phi_h;\Upsilon|\Phi;\tilde T\rangle|^2 = \cosh \left(\frac{Q_*}{4}\right),
\end{equation}
where
\begin{equation}
\label{eq:coshQ*}
\cosh \left(\frac{Q_*}{4}\right) = \frac{\Phi_h}{\Phi}
\end{equation}
is the value of $\cosh(Q/4)$ at which the exponent in the integrand of  \eqref{eq:PtPh} becomes stationary.  Since we see from \eqref{eq:coshQ*} that \eqref{eq:PtPh} is of order one in the limit $\phi_b\rightarrow \infty$, it should be approximated by $1$ when evaluating \eqref{eq:TrnfromPsi3}.  As a result, instead of being dominated by a stationary point, for $n>1$ \eqref{eq:TrnfromPsi3} is instead dominated by the cutoff at $\Phi_h=0$.  We thus set $\Phi_{h*}=0$ in
\eqref{eq:PtPh}  to find the leading-order results
\begin{equation}
\delta =0, \ \ \  {\rm Tr}_L \, \rho_L^n = -4\pi(n-1)\Phi_0  \ \ \ {\rm for} \ n>1.
\end{equation}
Here we have ignored contributions to $\delta$ that scale with $1/\phi_b$, and our failure to find an imaginary part of $\delta$ proportional to $(1-1/n)$ arises from the fact that the state is not semiclassical and, in particular, from the fact that the relevant integral is dominated by the $\Phi_h=0$ endpoint instead of by stationary point.

While ${\rm Tr}_L \, \rho_L^n$ is finite for $n>1$, the integral in \eqref{eq:TrnfromPsi3}  diverges for $n=1$.  Thus ${\rm Tr}_L \, \rho_L = \infty$ and $S_n = \infty$ for all $n>1$.  This rather trivial result is physically correct since $\Phi_{\tilde T}$ fails to commute with $\Phi_h$.  An eigenstate of
$\hat \Phi_{\tilde T}$ will thus have infinite fluctuations in $\Phi_h$, and so will be associated with arbitrarily large black holes.

Nontrivial results thus require us to study superpositions of the states $|\Phi;\tilde T\rangle$.  Let us in particular suppose that our state has a $\Phi_{\tilde T}$ wavefunction of the form
\begin{equation}
\label{eq:Phitsc}
\langle \Phi; \tilde T| \Psi\rangle =  e^{i\Phi \Pi_1} e^{-(\Phi-\Phi_1)^2/2\sigma^2}.
\end{equation}
where we will write $\Pi_1$ in the form $\Pi_1:= 4\sinh (Q_1/4)$. The wavefunction \eqref{eq:Phitsc} is thus determined by the four parameters $\Phi_1, Q_1, \sigma, \tilde T$.  For $\Phi_1$ of order $\phi_b$, $Q_1$ of order $1$, and widths $\sigma$  of order $\phi_b^{1/2+\epsilon}$ with $\epsilon<1/2$, 
 the state \eqref{eq:Phitsc} should describe a semiclassical state peaked at $\Phi_{\tilde T} = \Phi_1$ and $Q_{\tilde T} = Q_1$.  In particular, in that limit there should be only a negligible tail of the wavefunction in the highly-quantum region near $\Phi_h=0$.

With this assumption, evaluating \eqref{eq:PtPh} at leading order in large $\phi_b$ shows that the stationary point of the
$\Phi$ integral lies at
\begin{equation}
\label{eq:Phi*}
\Phi_* = \Phi_1 -4\sigma^2 i \left(\sinh \frac{Q}{4} - \sinh \frac{Q_1}{4}\right).
\end{equation}
Furthermore, since the $\Phi$ integral is a Gaussian with real variance, it is manifest that the integral is controlled by this stationary point and, in fact, that the semiclassical approximation is exact\footnote{The exact answer includes the first (i.e., one-loop) correction given by the van Vleck-Morette determinant.  But we can ignore this here since it gives just an overall normalization.}.
We can then study the saddle-point approximation to the $Q$ integral in \eqref{eq:PtPh}.  

Since the
action is by construction insensitive to first-order fluctuations in $\Phi_*$, the stationary point $Q_*$ of the $Q$ integral
satisfies
\begin{equation}
\label{eq:coshQ*2}
\Phi_*\cosh \left(\frac{Q_*}{4}\right) = \Phi_h.
\end{equation}
As a result, we obtain
\begin{equation}
\label{eq:almostthere}
\Psi[\Phi_h; \Upsilon] = \sqrt{\cosh\left(\frac{Q_*}{4} \right)}e^{-8\sigma^2\left(\sinh \frac{Q_*}{4} - \sinh\frac{Q_1}{4}\right)^2}
e^{iQ_* \Phi_h} e^{i\frac{2\Phi_h^2}{\phi_b}\tilde T}e^{-4i\phi_1 \left(\sinh \frac{Q_1}{4}-\sinh\frac{Q_*}{4}\right)}.
\end{equation}
However, the form of \eqref{eq:almostthere} hides the fact that solutions of \eqref{eq:Phi*}, \eqref{eq:coshQ*2} are not unique.  We should therefore determine which of the solutions should be used in \eqref{eq:almostthere}.  As is reviewed in appendix \ref{sec:SPA for GPI}, the answer is that one should include saddles for which the corresponding steepest ascent curves intersect the original contour of integration.

Our task is now to find the value of $\Phi_h$ at which the leading-order-in-$\phi_b$ part of $e^{-4\pi (n-1)\Phi_h}\Big| \Psi[\Phi_h;\Upsilon]   \Big|^{2n}$ is stationary.   This condition leads to a rather complicated transcendental equation.  However, for $n=1$ we are just finding the peak of 
$\Big| \Psi[\Phi_h;\Upsilon]   \Big|^{2}$.  Furthermore, in a semiclassical limit this peak should be given by the real saddle with $Q_*=Q_1$, $\Phi_*=\Phi_1$, and it should thus be located at the associated classical extremal-dilaton value
\begin{equation}
\label{eq:evolve to stat}
\Phi_{h1} = \Phi_1 \cosh \left( \frac{Q_1}{4} \right).
\end{equation}
In particular, near $\Phi= \Phi_{h1}$ the dominant relevant saddle should have $Q_*\approx Q_1$ and  $\Phi_* \approx \Phi_1$.  From \eqref{eq:evolve to stat}, we then also see in this limit that fluctuations in $\Phi$ of size $\sigma$ and uncorrelated fluctuations in $q:=\sinh(Q/4)$ of size $1/(4\sigma)$ will lead to fluctuations in $\Phi_h$ of size $\sigma_h$ with 
\begin{equation}
\label{eq:evolve to statfluct}
\sigma^2_{h} =  \sigma^2 \cosh^2 \left( \frac{Q_1}{4} \right) + \frac{\Phi_1^2}{16\sigma^2}\tanh^2\left(\frac{Q_1}{4}\right).
\end{equation}
For small enough $n-1$, we may thus approximate the $\Phi_h$ wavefunction by a Gaussian of the form \eqref{eq:PhihGaussian} with the above parameters and
\begin{equation}
\label{eq:evolve to statdelta}
\delta_1 = \frac{\phi_b}{4\Phi_{h1}}Q_1 + \tilde T,
\end{equation}
whence we can read off the entropies $S_n$ and the parameters of the associated complex saddles from \eqref{eq:PhihGresults1}.  These results will be valid only at first order in $n-1$, though perturbative corrections at higher orders in $n-1$ (which take into account the non-Gaussian corrections to the $\Phi_h$ wavefunction) can also be computed semiclassically.

\subsection{Semiclassical R\'enyis at finite $n-1$}

Here we again consider R\'enyi computations for the wavefunctions \eqref{eq:Phitsc}.  The case where $n-1$ is perturbatively small was discussed above.  However, if one wishes, one can also use the semiclassical approximation to compute $S_n$ numerically at finite $n-1$.   In doing so, one should realize that the results will be sensitive to the tails  of the original wavefunction \eqref{eq:Phitsc}.  As a result, this computation is only meaningful if \eqref{eq:Phitsc} is a precise characterization of the physics of interest.  

Nevertheless, as an illustration of both the techniques involved and the sort of results one can obtain, we display corresponding plots below for two sets of our parameters $Q_1, \Phi_1, \sigma, \tilde T$. At the technical level, the subtle part of the analysis is to determine which saddles $Q_*$ contribute at each $n$ (or at each $\Phi_h$) by finding the associated ascent curves and determining which ones have non-zero intersection number $n$ with the real-$Q$ axis; see again the discussion in appendix \ref{sec:SPA for GPI}.

For $\sigma^2$ of order $\phi_b$, and in the limit $\phi_b\rightarrow \infty$, our use of the semiclassical limit will give a good approximation as long 
as the Hessian of the relevant action at the saddle is non-degenerate, and as long as 
it is not sensitive to parts of the wavefunction near $\Phi=0$ or $\Phi_h=0$.
However, one can show analytically that the Hessian degenerates only when $Q_1=0$ and, moreover, this is just a case where two saddles coincide.  Thus, even at such values, the semiclassical approximation still holds at the leading-order level studied here.  We thus need not concern ourselves with further study of the Hessian.

On the other hand, for Gaussian wavefunctons of $\Phi_h$ \eqref{eq:PhihGaussian} we already saw that R\'enyi computations at finite $n-1$ can, for some values of parameters, be senstive to the wavefunction at small $\Phi_h$.  The same will clearly be true in the current context, and the appropriate diagnostic is just the saddle-point value $\Phi_{h*}$ of $\Phi_h$.  As noted above, $\Phi_{h*}$ is always real so, in our semiclassical limit $\phi_b \rightarrow \infty$ with $\Phi_h\propto \phi_b$, we avoid sensitivity to the wavefunction at small $\Phi_h$ precisely when $\Phi_{h*} >0$.

This leaves us only to understand the possible sensitivity to physics at small values of $\Phi$.  In analogy with the discussion below \eqref{eq:PhihGresults}, we need to investigate the effect of modifying the wavefunction $\Psi[\Phi;\tilde T]$ in the regime $\Phi \lesssim 0$, say by adding a term $\Delta \Psi$ supported in that region.  We will model $\Delta \Psi$ by simply using the tail of $\Psi$; i.e., by setting $\Delta \Psi = -\Psi \theta(-\Phi)$ for the usual Heaviside step-function $\theta$ so that $\Psi+\Delta \Psi$ vanishes for $\Phi<0$; see again footnote \ref{foot:smallPhi}.

Let us proceed by examining the $\Phi$ integral in \eqref{eq:PtPh} at arbitrary fixed $Q$.  This integral essentially gives the Fourier transform of the wavefunction $\Psi[\Phi ;\tilde T]$, though evaluated at $\Pi: = 4 \sinh(Q/4)$.    We should thus compare the Fourier transforms   $\tilde \Psi$ and $\widetilde {\Delta \Psi}$ of  $\Psi$ and ${\Delta \Psi}$.  Let us begin with the case of real $\Pi$.  Since $\Psi$ is a Gaussian and $\Delta \Psi$ is a cutoff Gaussian, a short computation shows that the expectation value of $\Pi$ vanishes in either state.  We also find that the expectation values of $\Pi^2$ are $\sigma^{-2}$ in the state $\Psi$ and $\sim \frac{\Phi_1^2}{\sigma^4}$ in the state $\Delta \Psi$.  Taking into account that $\tilde \Psi$ is Gaussian and that the norm of $\Delta \Psi$ is roughly $e^{-\Phi_1^2/\sigma^2}$, we see that contributions from $\Delta \Psi$ are important only for $|\Pi| \gtrsim \Phi_1/\sigma^2.$ 

Let us now consider the case where $\Pi - \Pi_1$  is imaginary (again with $\Pi_1:= 4\sinh(Q_1/4)$).  In this case, the phase of the integrand of the $\Phi$ integral in \eqref{eq:PtPh} is  independent of $\Phi$.  Thus the $\Phi$ integral is effectively real and the sensitivity to adding $\Delta \Phi$  can be analyzed by understanding the location of the peak $\Phi_*$, which is of course again given by \eqref{eq:Phi*}.  We see that (at large $\phi_b$) the peak remains far from $\Phi \lesssim 0$ for $\text{Im} \, (\Pi - \Pi_1) > -\frac{\Phi_1}{\sigma^2}$.  When combined with our analysis of real $\Pi$ above, this suggests that our results should be robust against small-$\Phi$ corrections so long as we have
\begin{equation}
\label{eq:Phiconditions}
| \text{Re} \, \Pi| < \frac{\Phi_1}{\sigma^2},  \ \ \ 
\text{Im} \, (\Pi-\Pi_1) < \frac{\Phi_1}{\sigma^2}.
\end{equation}
We will thus plot these quantities for our examples below and compare the above bounds against our numerical results.

\begin{figure}[h!]
    \centering
\includegraphics[width=0.55\linewidth]{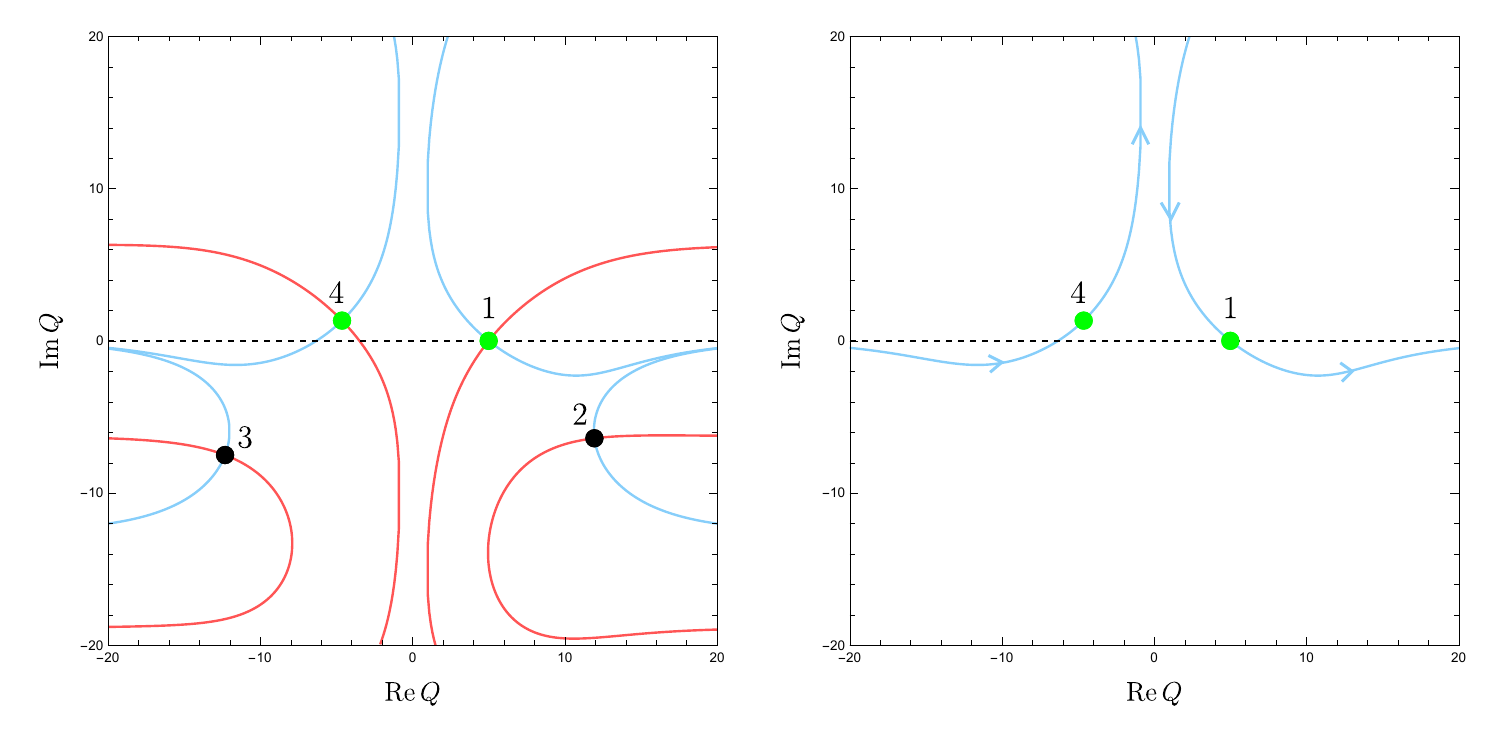}
\includegraphics[width=0.55\linewidth]{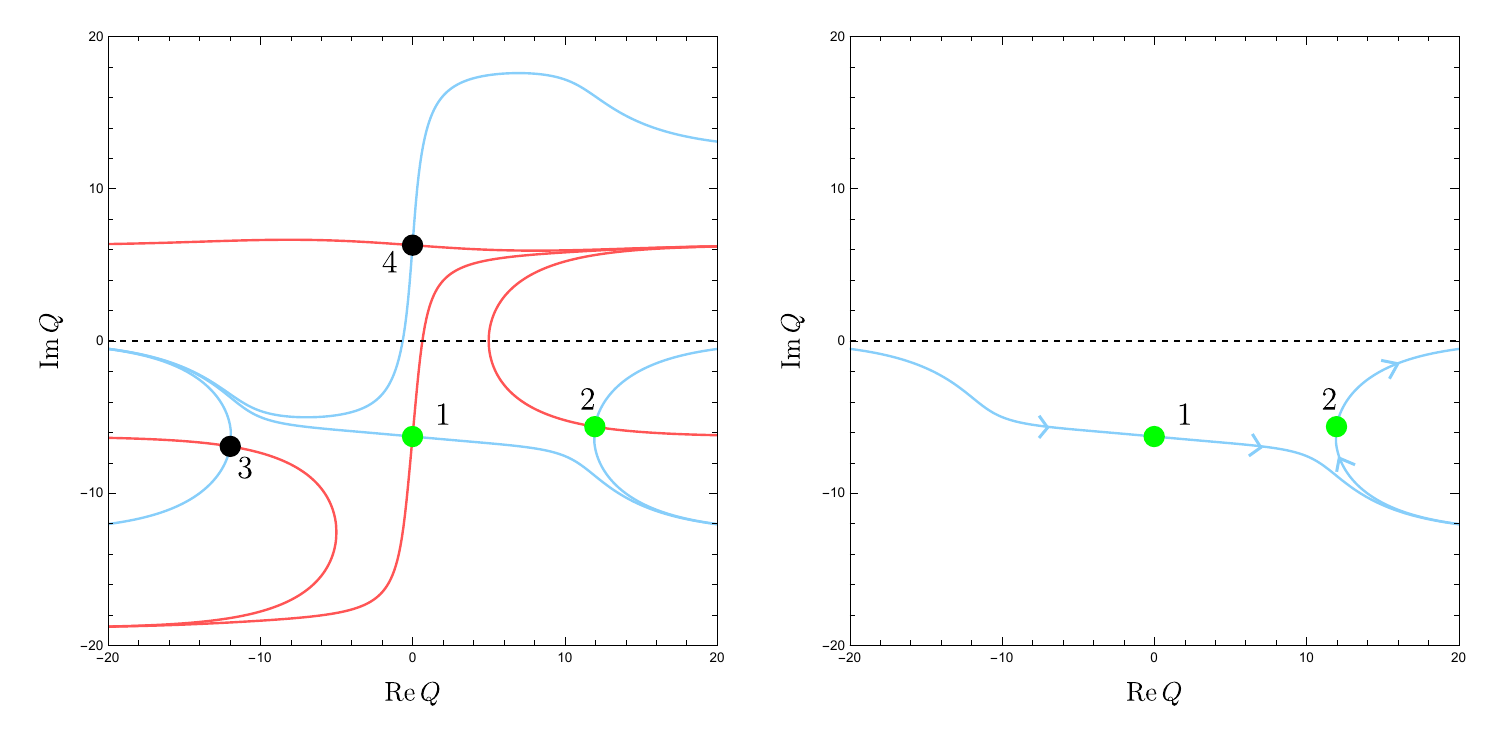}    \caption{Saddle points (dots) and their steepest descent (light blue) and ascent (red) contours for the $Q$ integral in equation \eqref{eq:PtPh} with parameters $\Phi_1=\phi_b=10^4,\tilde T=1,\sigma=\sqrt{\phi_b}/(2\pi)\sim 15.92,Q_1=5$. 
    The original (real) contour of integration is shown as a dashed black line.
    Saddles whose steepest ascent contour intersects that contour are colored  green.  Other saddles are shown in black.   As $\phi_h$ decreases from $\Phi_{h*,n=1}=\Phi_{h1}$ to zero, each of the saddles shown in the top row continuously deforms to become the correspondingly-numbered saddle in the bottom row.   In particular, saddles 1 and 4 rotate about each other by roughly $\pi/2$, while saddles 2 and 3 deform only slightly.   \textbf{Top row}: In these panels we take  $\Phi_h=\Phi_{h1}=\Phi_1\cosh(Q_1/4)$.  The full set of saddles is shown at left.  At right, we show only saddles with non-zero intersection number $n$. The original contour of integration (dashed horizontal line) can be deformed to the indicated combination of descent contours for these two saddles with the orientations shown (arrows).   The saddle labeled $1$ dominates the integral and  corresponds to the real value $Q_*=Q_1$.  \textbf{Bottom row}: Corresponding plots for $\Phi_h=0$.  The saddle labeled $1$ (the deformation of the $Q_*=Q_1$ saddle at $\Phi_h=\Phi_{h1}$) lies at $Q_*=-2\pi i$ and again dominates the integral at leading order; see the main text for comments concerning one-loop effects. }
    \label{fig:contoursvn}
\end{figure}

The first case we study corresponds to the parameters  $\Phi_1=\phi_b=10^4,\tilde T=1,\sigma=\sqrt{\phi_b}/(2\pi)\sim 15.92,Q_1=5$.  For any $Q$, the $\Phi$ integral is Gaussian and so can be performed using the saddle-point approximation and the saddle \eqref{eq:Phi*}. Saddles in the complex $Q$-plane for the remaining $Q$-integral are shown with their ascent and descent curves in figure \ref{fig:contoursvn} for the classical value $\Phi_{h1}$ of $\Phi_h$ given in \eqref{eq:evolve to stat} and for the extreme case $\Phi_h=0$.  As expected, the real saddle dominates at $\Phi_{h1}$ (figure \ref{fig:contoursvn}).   This value of $\Phi_h$ describes the R\'enyi saddle for $n=1$.  Furthermore, for the above parameters, the dominant saddle at all $\Phi_h$ turns out to be a continuous deformation of that saddle. As we will shortly explain, for this choice of parameters we find $\Phi_{h*}\rightarrow 0$ as $n\rightarrow \infty$.  The bottom row of figure \ref{fig:contoursvn} thus shows the saddles for the limiting case $n\rightarrow \infty$.

\begin{figure}[h!]
    \centering
    \includegraphics[width=\linewidth]{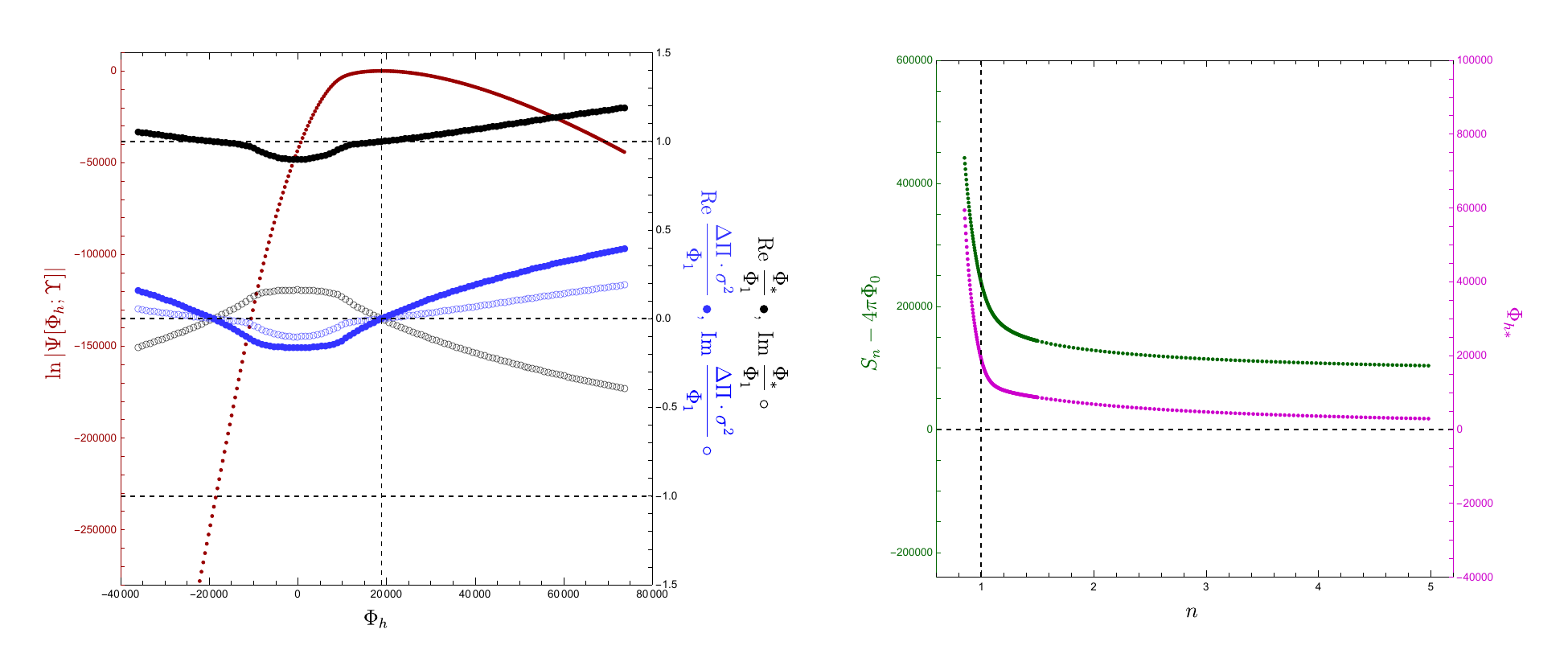}
    \caption{Results of leading-order semiclassical calculations for  $\Phi_1=\phi_b=10^4,\tilde T=1,\sigma=\sqrt{\phi_b}/(2\pi)\sim 15.92,Q_1=5$.
    \textbf{Left panel}: The logarithmic magnitude of the wavefunction $\ln|\Psi(\Phi_h;\Upsilon)|$ (shown in red) and the real/imaginary part (black dots/circles) of $\Phi_*$ and (blue dots/circles) of $\Delta \Pi:= \Pi-4\sinh(Q_1/4)$ as functions of $\Phi_h$. The vertical dashed blue line marks the classical value $\Phi_{h1} = \Phi_{h*,n=1}$, while the  horizontal dashed black lines indicate $\mathrm{Re}\,\Phi_*=0$, $\mathrm{Im}\,\Phi_*=0$, 
    $\mathrm{Re}\,\Delta \Pi=0$, $\mathrm{Im}\,\Delta \Pi=0$ and $\mathrm{Re}\,\Phi_*=\pm \Phi_1$, $\mathrm{Im}\,\Phi_*=\pm \Phi_1$, 
       $\mathrm{Re}\,\Delta \Pi=\pm \Phi_1/\sigma^2$, $\mathrm{Im}\,\Delta \Pi=\pm \Phi_1/\sigma^2$.
 \textbf{Right panel}: The R\'enyi entropy  $S_n-4\pi\Phi_0$ (shown in green) and  $\Phi_{h*}$ (shown in purple) as functions of $n$. The vertical dashed line marks  $n=1$. Note from the left panel that $\Phi_{h*}$ remains positive for all $n$,  Furthermore, $\rm{Re} \, \frac{\Phi_*}{\Phi_1}$ stays well away from $0$ and $\rm{Im} \, \frac{\Phi_*}{\Phi_1}$, 
 $\mathrm{Re}\,\Delta \Pi\frac{\sigma^2}{\Phi_1}$,  $\mathrm{Im}\,\Delta \Pi\frac{\sigma^2}{\Phi_1}$
 are  fairly small for $0< \Phi_h < \Phi_{h1}$.  However, these latter quantities become large at large $\Phi_h$ which, from the right panel, yield stationary points $\Phi_{h*}$ for $n$ close to zero.  This suggests that 
    the semiclassical approximation should be valid for all $n>1$, though  for $n<1$ it will be valid only for small $1-n$.}
    \label{fig:renyis}
\end{figure}

\begin{figure}[h!]
    \centering
    \includegraphics[width=0.5\linewidth]{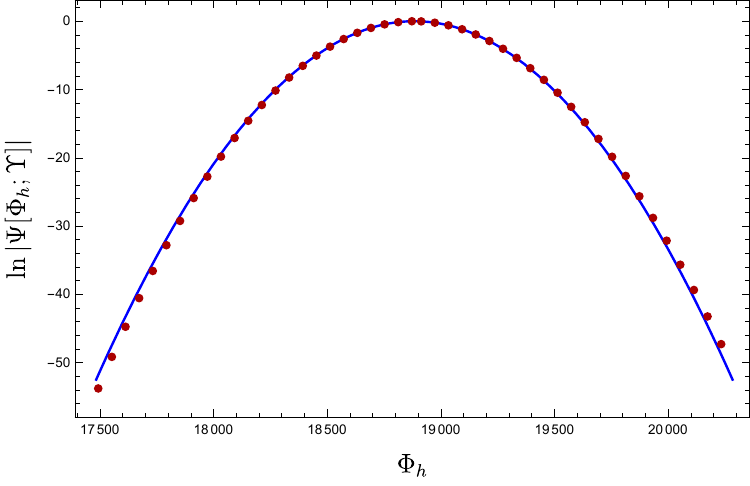}
    \includegraphics[width=0.5\linewidth]{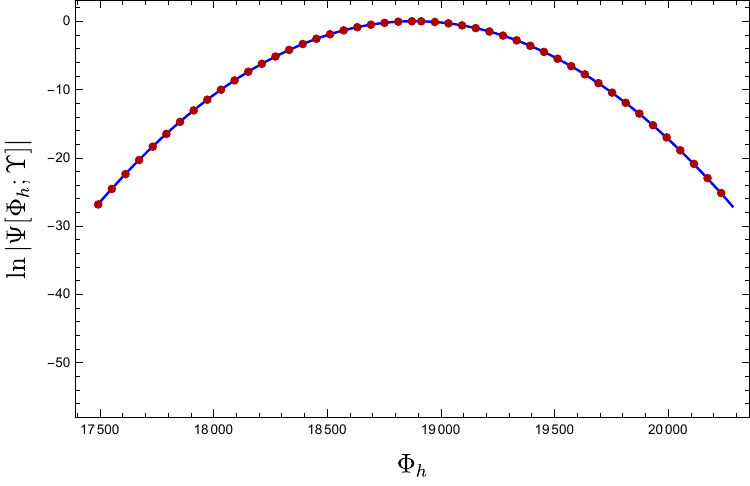}
    \caption{Data (red dots) from near the peak $\Phi_{h1} = \Phi_{h*, n=1}$ of the semiclassical wavefunction is compared with Gaussians (blue curves) of width $\sigma_h$ as given by \eqref{eq:evolve to statfluct}. On the top (bottom) the parameters match those in figure \ref{fig:renyis} (figure \ref{fig:renyis2}) so that $\sigma_h= 136.60$ ($\sigma_h=190.03$).  }
    \label{fig:approx}
\end{figure}

The results of using this saddle to compute R\'enyi entropies $S_n$ and the wavefunction $\Psi[\Phi_h;\Upsilon]$ are shown in figure \ref{fig:renyis}.  We also include figure \ref{fig:approx} as a check that the peak is well-modeled by a Gaussian of width given by \eqref{eq:evolve to statfluct}. The results also show that,  in the semiclassical limit, the semiclassical approximation to the R\'enyi entropy $S_n$ remains valid for arbitrarily large $n$. In particular, the relevant saddle-point values $\Phi_{h*}$ of $\Phi_h$ are always positive and, in the limit of large $\phi_b$, at any fixed $n$ they will satisfy $\Phi_{h*} \gg 1$. Furthermore, we see that $\rm{Re} \, \Phi_*$ never becomes small, though it appears that  $\rm{Im} \, \Phi_*$ will become large and negative at sufficiently large $\Phi_h$ (which will yield stationary points $\Phi_{h*}$ for R\'enyi indices $n<1$ that are sufficiently close to zero).  This is consistent with the behavior of $\Delta \Pi:=\Pi-\Pi_1$ which also appears to be come large at large $\Phi_h$.  Thus in this case we expect the semiclasical limit to break down at sufficiently large $\Phi_h$ or, equivalently, for $n$ sufficiently close to zero.

It is, however, interesting that we find $\Phi_{h*}\rightarrow 0$ as $n\rightarrow \infty$ for the above parameters.   From the analysis below equation \eqref{eq:finaldelta}, one sees that for $n\rightarrow \infty$ the saddle point $\Phi_{h*}$ is the value of $\Phi_h$ at which the red curve in figure \ref{fig:renyis} has slope $2\pi$. One may show analytically that this occurs at $\Phi_h=0$ so long as the integral is dominated by the obvious saddle at $Q_*=-2\pi i$ (see \eqref{eq:coshQ*2}).  It is worth remarking that computing saddles numerically for $\Phi_h$ between zero and $\Phi_{h1}$ indicates that the $Q_*=-2\pi i$  saddle for $\Phi_h=0$ continuously deforms to the real classical saddle $Q_*=Q_1$ for  $\Phi_h=\Phi_{h*,n=1}=\Phi_1\cosh(Q_1/4)$.  In any case, however, a perturbative analysis near $\Phi_h=0$ allows one to write the saddle-point value $Q_*$ in the form 
\begin{equation}
    \label{eq:pertQs}
    Q_*(\Phi_h)=-2\pi i+\frac{4\Phi_h}{(4\sigma^2-\Phi_1)i+4\sigma^2\sinh(Q_1/4)}+\mathcal{O}(\Phi_h^2).
\end{equation}
Inserting \eqref{eq:pertQs} into equation \eqref{eq:almostthere}, taking the derivative of $\ln|\Psi(\Phi_h;\Upsilon)|$, and evaluating the result at $\Phi_h=0$ yields the advertised value $2\pi$. While this does not necessarily exclude the possibility that there is another point where the derivative of $\ln|\Psi(\Phi_h);\Upsilon|$ is $2\pi$, our numerics suggests that this does not occur; see again the red curve in figure \ref{fig:renyis}. 
 The saddle $Q*=-2\pi i$ with $\Phi_{h*}=0$ gives a limiting leading-order R\'enyi entropy  $S_\infty=4\pi\Phi_0-16\sigma^2+16\sigma^2\sinh^2(Q_1/4)+8\Phi_1$.

\begin{figure}[h!]
    \centering
    \includegraphics[width=\linewidth]{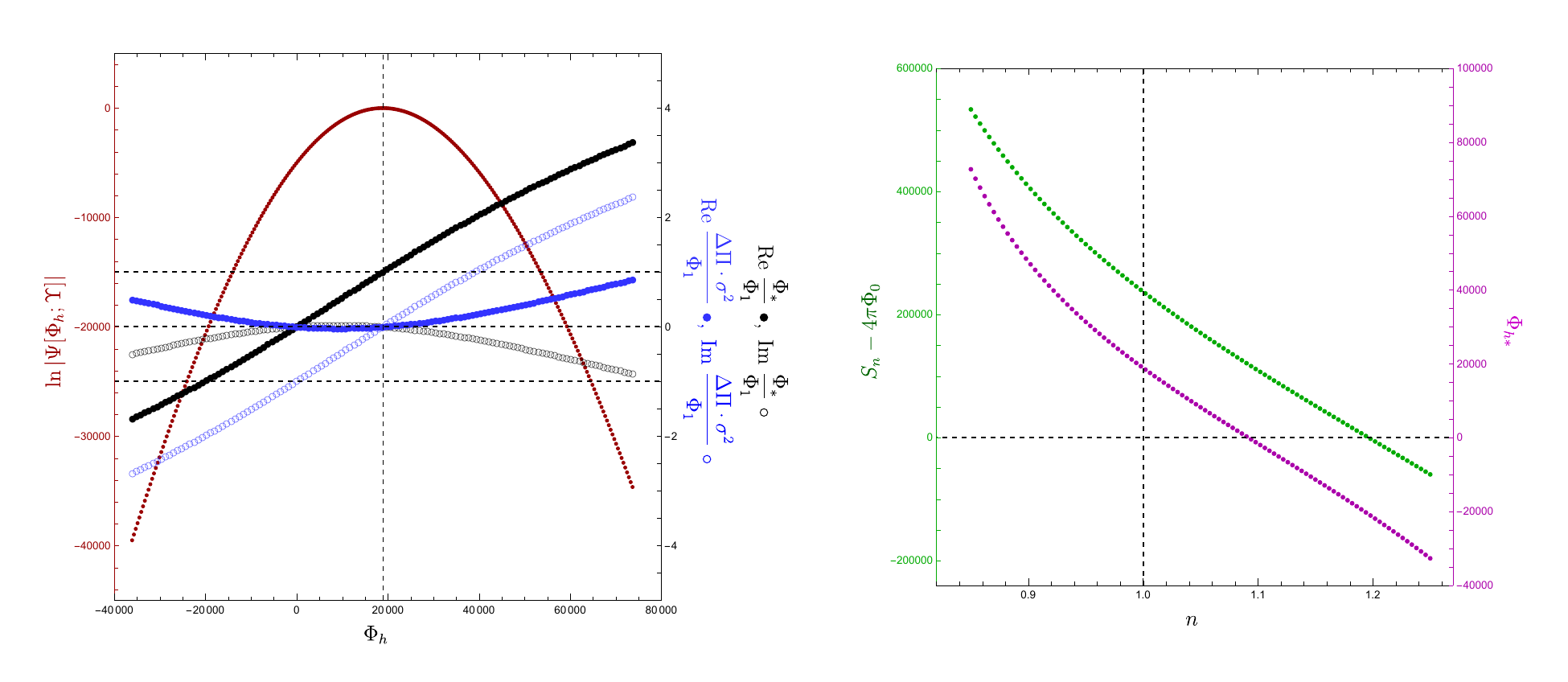}
    \caption{Results of leading-order semiclassical calculations for $\Phi_1=\phi_b=10^4,\tilde T=1,\sigma=\sqrt{\phi_b}=100,Q_1=5$.  Though the only difference from figure \ref{fig:renyis} is an increase of $\sigma$ by a factor of $2\pi$, the results are markedly different.
    \textbf{Left panel}: The logarithmic magnitude of the wavefunction $\ln|\Psi(\Phi_h;\Upsilon)|$ (shown in red) and the real/imaginary part (black dots/circles) of $\Phi_*$ and (blue dots/circles) of $\Delta \Pi:= \Pi-4\sinh(Q_1/4)$  as functions of $\Phi_h$. The vertical dashed blue line marks the classical value $\Phi_{h1} = \Phi_{*,n=1}$, while the  horizontal dashed black lines indicate $\mathrm{Re}\,\Phi_*=0$, $\mathrm{Im}\,\Phi_*=0$, 
    $\mathrm{Re}\,\Delta \Pi=0$, $\mathrm{Im}\,\Delta \Pi=0$ and $\mathrm{Re}\,\Phi_*=\pm \Phi_1$, $\mathrm{Im}\,\Phi_*=\pm \Phi_1$, 
       $\mathrm{Re}\,\Delta \Pi=\pm \Phi_1/\sigma^2$, $\mathrm{Im}\,\Delta \Pi=\pm \Phi_1/\sigma^2$.
 \textbf{Right panel}: The R\'enyi entropy  $S_n-4\pi\Phi_0$ (shown in green) and  $\Phi_{h*}$ (shown in purple) as functions of $n$. The vertical dashed line marks  $n=1$. We now find that $\Phi_{h*}$ becomes negative for $n>n_{crit}=1.091(6)$, indicating that the semiclassical approximation breaks down.  This breakdown is also further suggested by the fact that $\mathrm{Im}\,\Delta \Pi=\pm \Phi_1/\sigma^2$ is less than $-1$ in this regime.  }
    \label{fig:renyis2}
\end{figure}

However, it is important to note that an order-of-limits issue arises when discussing saddles near $\Phi_h=0$ (or, equivalently, at large $n$). Since the integrand in \eqref{eq:PtPh} has a factor of $\sqrt{\cosh(Q/4)}$ in front of the exponential, the one-loop corrected contribution from any saddle at $Q=-2\pi i$ vanishes exactly.  Thus, while a saddle near this point dominates the integral in the limit $\phi_b\rightarrow \infty$ for any fixed large-but-finite $n$, it is the other green saddle shown in the bottom row of figure \ref{fig:contoursvn} that dominates if one instead takes the limit $n\rightarrow \infty$ before taking $\phi_b \rightarrow \infty.$

\begin{figure}[h!]
    \centering
    \includegraphics[width=0.6\linewidth]{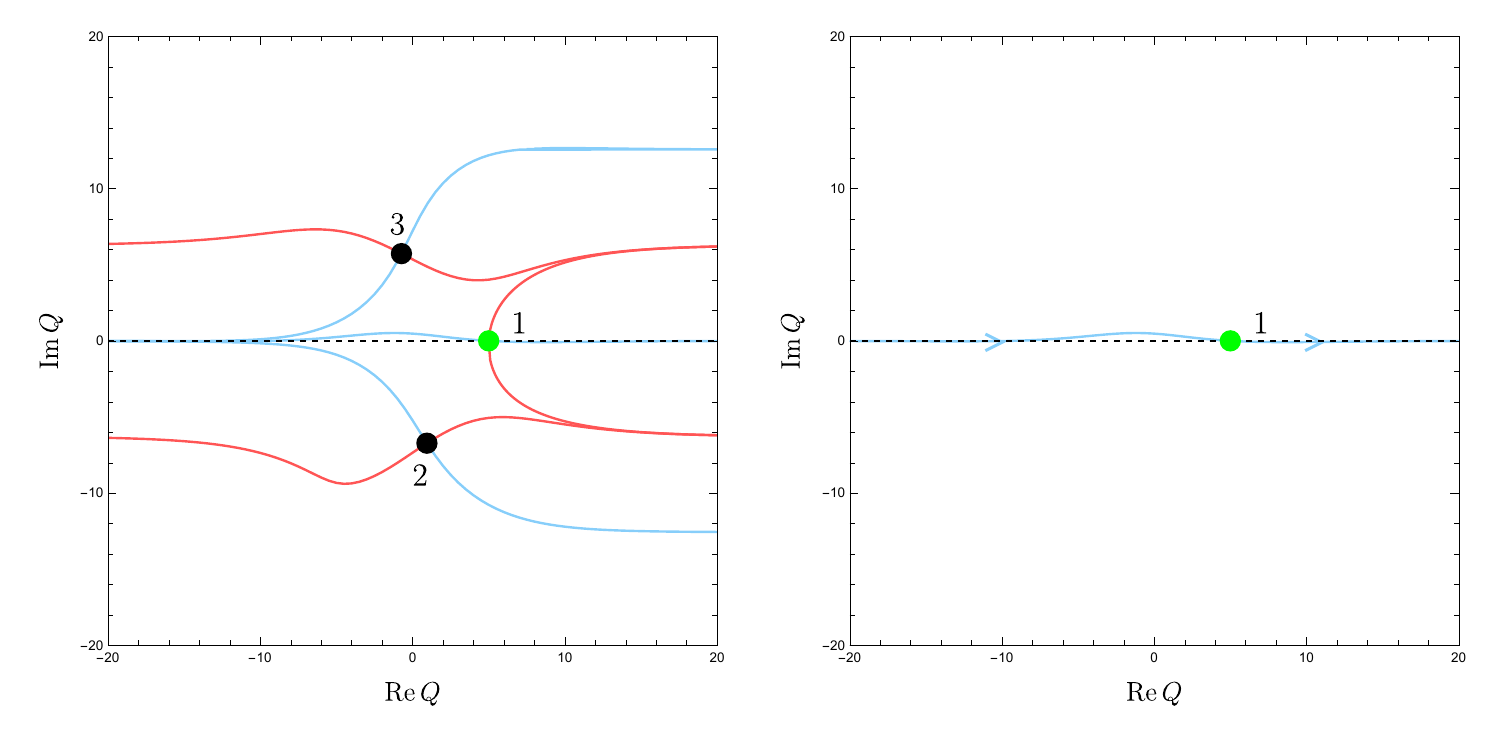}
    \includegraphics[width=0.6\linewidth]{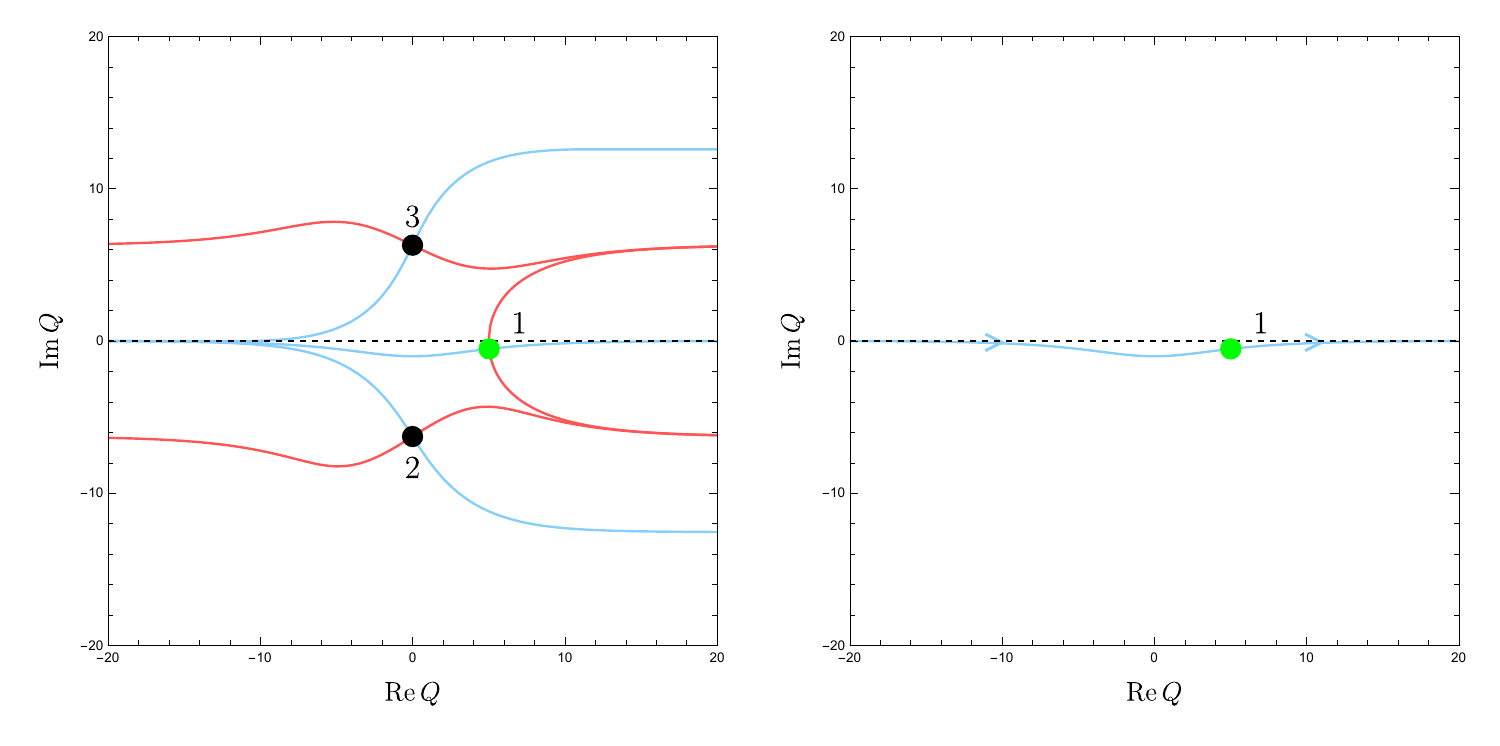}
    \caption{The analogue of figure \ref{fig:contoursvn} for parameters as in figure \ref{fig:renyis2}.  Though the only change in parameters relative to figure \ref{fig:contoursvn} is an increase of $\sigma$ by a factor of $2\pi$, there are now only $3$ saddles near the real axis and they barely move as $\Phi_h$ ranges from $\Phi_h = \Phi_{h1}= \Phi_{h*,n=1}$ to $\Phi_h=0$.  Again,  saddle $1$ in the top row is the classical saddle $Q_*=Q_1$ for  $\Phi_h = \Phi_{h1}= \Phi_{h*,n=1}$, and this saddle smoothly deforms to saddle $1$ in the bottom row as $\Phi_h$ decreases to zero.  However, for $\Phi_h=0$ it is no longer the obvious saddle at $Q=-2\pi i$.  The $Q=-2\pi i$ saddle is instead saddle $2$, which never contributes (since its ascent curve never crosses the real axis).}
    \label{fig:counter}
\end{figure}

Our second example increases $\sigma$ to $\sigma = \sqrt{\phi_b}$ while keeping all other parameters as above.
Results for this case are shown in figure \ref{fig:renyis2}. For these parameters, we may again evaluate the $\Phi$ integral using the saddle \eqref{eq:Phi*} and study the resulting integral over $Q$. 
 However,  the saddle $Q=-2\pi i$ now fails to dominate at $\Phi_h=0$ (even at leading order), though the dominant saddle (at $Q_*\neq -2\pi i$) {\it is} again a smooth deformation of the classical saddle $Q_*=Q_1$ for $\Phi_h = \Phi_{h1} = \Phi_{h*,n=1}$.  The saddles and their ascent/descent curves are figure \ref{fig:counter} below.
For these parameters We find numerically that $\Phi_{h*} \rightarrow 0$ at $n_{\mathrm{crit}} \approx 1.091(6)$, so that the semiclassical approximation becomes unreliable for R\'enyi entropies $S_n$ with $n> n_{\mathrm{crit}}$. This illustrates the above claim that for some choices of parameters that semiclassical approximation can hold only over a finite (and small!) range of $n$ near $n=1$, while for other choices (as in figure \ref{fig:renyis}) the approximation can be valid for all $n>n_{crit} > 1$.

\section{Discussion}
\label{sec:disc}

Our work above emphasized that, at least in contexts where we can treat spacetimes as containing a unique extremal surface, the gravitational R\'enyi entropies $S_n$ defined by any path integral can be computed directly from the corresponding bulk wavefunction.  Furthermore, in the semiclassical approximation, the saddle-point spacetimes of such path integrals can also be recovered directly from the bulk wavefunction.  As a result, for a given wavefunction, the saddle-points of both the  Euclidean and real-time path integrals must define equivalent complex spacetimes.  Indeed, one can also see directly from the representation in terms of the bulk wavefunction that for $n\neq 1$ these saddles must be intrinsically complex (or Euclidean).  While our arguments considered only Einstein-Hilbert and Jackiw-Teitelboim gravity, as described at the end of section \ref{sec:Rfw} it is natural to expect similar results to hold for theories with arbitrary perturabtive higher-derivative corrections.

We illustrated the above constructions in detail for a variety of semiclassical bulk wavefunctions in JT gravity. The fact that the
saddle-point geometries are complex led to an interesting subtlety.  Constructing a useful representation of the bulk wavefunction typically requires fixing a gauge or, equivalently, specifying a set of coordinates, in order to associate the bulk wavefunction with a particular hypersurface $\Sigma$ in the spacetime.  The subtlety is then that if $\Sigma$ lies at real values of a given set of coordinates, then it may not lie at real values of some other set of coordinates on the complex saddle.  In particular, saddle-point spacetimes were described in \cite{Colin-Ellerin:2020mva,Colin-Ellerin:2021jev} using coordinates $\tilde x^\pm$ that were real at the extremal-dilaton point and with respect to which the metric and dilaton satisfied certain reality properties.    Since this was not the case for the coordinates $\tilde T, X$ introduced in section \ref{sec:ClassicalJT}, our saddles were related to those of \cite{Colin-Ellerin:2020mva,Colin-Ellerin:2021jev} by a complex coordinate transformation.  Equivalently, one may say that the two saddles (each being defined by the surface on which the associated coordinates are real) were different, but that they were (almost) deformable to each other through the larger complex spacetime in which they were both embedded.  The subtlety in this deformation is that
the boundary conditions for our path integral were fixed by giving a wavefunction at finite real $\tilde T$ which, as suggested above, lies at complex values of $\tilde x^\pm$.    As a result, a better statement is that our saddle can be deformed to a surface in the larger complex spacetime that first evolves our wavefunction to a surface with real $\tilde x^\pm$, and which thereafter agrees with the explicit real-time saddle of \cite{Colin-Ellerin:2021jev}; see again footnote \ref{foot:evolve}.

For simplicity, we have focused here on bulk wavefunctions satisfying the factorization condition \eqref{eq:factorPsi}.  This condition is trivially satisfied in our pure JT examples since there are no other gauge-invariant variables $h$ that commute with $\Phi_h$.  More generally, however, effects associated with other bulk degrees of freedom can be significant. Indeed, the integrals over $h$ in \eqref{eq:rhoLnPsi} generally contribute factors of $e^{-S_n^{\text {quantum}}(A)}$, where $S_n^{\text {quantum}}(A)$ is the $n$th R\'enyi entropy of whatever fields cannot be treated semiclassically.  In order to perform whatever integrals remain in the semiclassical approximation, one must then think of $S_n^{\text {quantum}}(A)$ as a correction to the classical action.  One then expects this to modify the relevant notion of geometric entropy $\sigma = \frac{A}{4} + \dots$ and that a consistent semiclassical approximation will be simplest when the surface $\Sigma$ associated with the wavefunction is chosen to contain the extremum of $\sigma$ rather than containing the surface of extremal area; see e.g. \cite{Engelhardt:2014gca,Dong:2017xht}.

Finally, we have also  focused on contexts  where we could treat spacetimes as having only one extremal surface.  However, in analogy with the constructions of \cite{Marolf:2020vsi,Dong:2020iod,Akers:2020pmf}, in the more general case Euclidean path integral computations of R\'enyi entropies can also be described in terms of gluing wavefunctions together.  The essential ingredient in this construction is the fact that any two extremal surfaces will be spacelike related in real Lorentz-signature classical spacetimes satisfying the null energy condition \cite{Wall:2012uf}.  As a result, with sufficient care, in the semiclassical limit we can consider a wavefunction $\Psi$ associated with a hypersurface $\Sigma$ that  passes through {\it all} of the relevant extremal surfaces.  At fixed integer $n>1$, one can then write an analogue of \eqref{eq:rhoLnPsi} that expresses $\Tr_L \, \rho_L^n$ as a sum of integrals over the areas of various surfaces weighted by factors of $e^{-A/4}$ and by the modulus $|\Psi|$ of the wavefunction with various arguments.  In such contexts, it should again be possible to describe the associated complex saddles in terms of conjugate variables to the extremal areas using formulae analogous to \eqref{eq:HJconj}, though they will be correspondingly more complicated.  We therefore leave construction of the detailed such expressions for future work.

\acknowledgments

We thank Xi Dong for many discussions on the topic of gravitational entropies.
This research was supported by NSF grants PHY-2107939 and PHY-2408110, and by funds from the University of California.
ZW was also supported by the DOE award number DE-SC0015655.
\appendix

\section{The delta-function at the splitting surface in JT gravity}
\label{app:Rdelta}

This appendix briefly reviews the computation of delta-function curvature contributions from the splitting surface following \cite{Louko:1995jw,Neiman:2013ap,Colin-Ellerin:2020mva}, though we consider the case where the splitting surface lies on a timefold (as did \cite{Colin-Ellerin:2020mva}) and we note that the choice of sign for imaginary parts in \cite{Neiman:2013ap} is opposite to ours (see below).  To be definite, we will consider this computation in $2d$ JT gravity.  However, the computation for Einstein-Hilbert gravity in any dimension is essentially the same due to the standard fact that the action for a small region near the conical singularity will receive negligible contributions from smooth terms in the metric (and that the non-smooth terms are associated with the pair of directions normal to the singular surface).

To proceed,  we  excise  a small disk ($\mathcal{U}$) containing the splitting surface $\Upsilon$ from the  manifold $\mathcal{M}$ shown in figure \ref{fig:replica}.   We wish to compute the associated action $S_\Upsilon = \int_{\mathcal{U}} \dd^2x \, \eta \sqrt{-g} \Phi R,$ where $\eta=1$ on a ket spacetime and $\eta=-1$ on a bra spacetime.
Our argument will be based on a generalization of the Gauss-Bonnet theorem to complex metrics following \cite{Louko:1995jw}.
The main subtlety is that the relevant integrals involve $\sqrt{-g}$ for the bulk metric $g$ and $\sqrt{\pm h}$ for the induced metric $h$ on the boundary
$\partial \mathcal{U}$ of our disk.  These square roots introduce branch cuts and associated choices of signs.  Some of these signs are determined by internal consistency of the formalism, but there is also  an important sign that must simply be regarded as part of the definition of the action; i.e., it must be chosen on some physical grounds.   While our discussion is closely related to that of \cite{Colin-Ellerin:2020mva}, we add further comments on several issues to clarify various subtleties.

\subsection{Definitions and conventions}
\label{subsec:DefConv}

\begin{figure}
    \centering    
    	\includegraphics[width=0.5\linewidth]{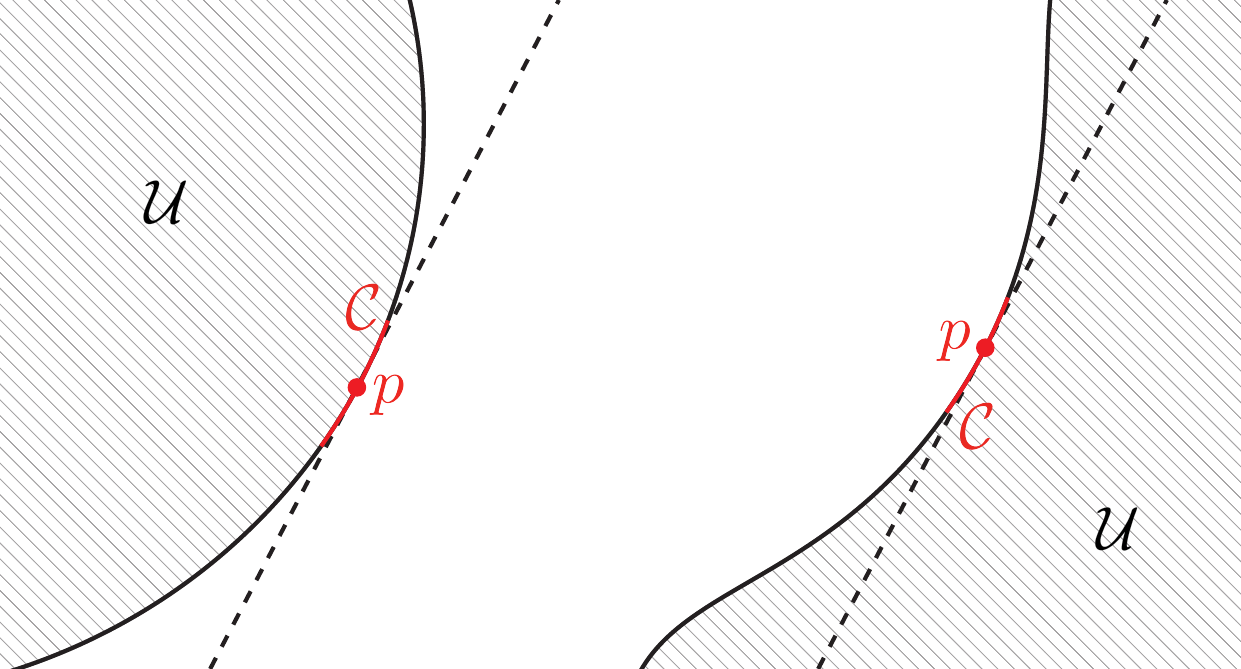}
    \caption{
{\bf Left:} On any small piece ${\cal C}$ (red) of $\partial \mathcal U$ on which there is an isolated point $p$ at which the tangent becomes null and is locally outside   $ \mathcal U$, there is a  contribution $-i\eta\frac{\pi}{2}\Phi(p)$ to $\int_{\mathcal{C}}d x\, \sqrt{|h|} \Phi K$ and thus a contribution $-i\eta\pi$ to \eqref{eq:CBG}. 
{\bf Right:} On any small piece ${\cal C}$ (red) of $\partial \mathcal U$ on which there is an isolated point $p$ at which the tangent becomes null and is locally inside   $ \mathcal U$, there is a  contribution $i\eta\frac{\pi}{2} \Phi(p)$ to $\int_{\mathcal{C}}d x\, \sqrt{|h|} \Phi K$ and thus a contribution $i\eta\pi$ to \eqref{eq:CBG}.
  }
    \label{fig:Kpoles}
\end{figure}

In \cite{Louko:1995jw}, the above signs were chosen  to be consistent with deforming the singular Lorentzian metric to a smooth-but-complex  geometry on which standard matter path integrals converge; i.e., by the inputs that lead to what is now known as the Kontsevich-Segal-Witten (KSW) criterion \cite{Kontsevich:2021dmb,Witten:2021nzp}.  With such choices, the Gauss-Bonnet theorem for general complex metrics and any region $\mathcal U$ takes the form

\begin{equation}
\label{eq:CBG}
\int_{\mathcal{U}} \dd^2x \,  \sqrt{-g}  R + 2\int_{\partial \mathcal{U}}\dd x\, \sqrt{|h|}K = -4\pi \eta i\chi({\mathcal{U}}).
\end{equation}
Here it is important that simple poles in $\sqrt{|h|}K$ (which arise, for example, in real Lorentz-signature spacetimes at places where $\partial \mathcal{U}$ becomes null) are treated with an $i\epsilon$ prescription, and that the sign of $\epsilon$ depends on $\eta$.  The upshot of this prescription is that, when $\partial \mathcal U$ and $\Phi$ are smooth, isolated points $p$ at which the tangent line to $\partial \mathcal U$ is null and $\mathcal U$ is locally convex (so that the tangent line is locally outside   $ \mathcal U$) contribute $-i\eta\frac{\pi}{2}\Phi(p)$ to $\int_{\partial \mathcal{U}}\dd x\, \sqrt{|h|}\Phi K$ and thus $-i\eta\pi$ to \eqref{eq:CBG}, while
isolated points $p$ at which a null tangent line  is locally inside   $\mathcal U$ (so that we may call $\mathcal U$ locally concave) make opposite-signed contributions; see figure \ref{fig:Kpoles}.  Indeed, since \eqref{eq:CBG} requires the real part of the left-hand-side to vanish in all spacetimes,  in real Lorentz-signature spacetimes the imaginary contributions to
$2\int_{\partial \mathcal{U}}\dd x\, \sqrt{|h|}K$ from such poles sum to $ -4\pi \eta i\chi({\mathcal{U}})$.

References \cite{Louko:1995jw,Neiman:2013ap} consider only what we call ket-spacetimes, but \eqref{eq:CBG} includes the corresponding result for bra-spacetimes.  The sign of the right-hand side is the one that must be chosen on physical grounds (or, more precisely, so as to be consistent with standard physics conventions regarding distinctions between $i$ and $-i$).  In particular, the explicit appearance of $\eta$ in \eqref{eq:CBG} arises from the use of $e^{i\eta S}$ on bra/ket-spacetimes (so that the requirement that matter path integrals converge gives an $\eta$-dependent answer)\footnote{However, \cite{Neiman:2013ap} nevertheless chooses the opposite sign.}.  One may equivalently derive the factor of $\eta$ by requiring that ket and bra versions of the same spacetime give complex-conjugate integrands for the path integral.  These arguments also explain why the contributions to $2\int_{\partial \mathcal{U}}d x\, \sqrt{|h|}K$ from poles in $\sqrt{|h|}K$ are defined using an $\eta$-dependent $i\epsilon$ prescription.

We wish to understand the analogue of  \eqref{eq:CBG} for Schwinger-Keldysh spaceimes.  Cases where $\mathcal U$ lies entirely in a bra- or ket-piece of the spacetime reduce immediately to \eqref{eq:CBG} which, for such cases, we may multiply by $\eta$ and write in the form 
\begin{equation}
\label{eq:UepsrgR2}
\int_{\mathcal{U}} \dd^2x \, \eta \sqrt{-g}  R + 2\int_{\partial \mathcal{U}}\dd x\,\eta \sqrt{|h|} K = -4\pi i \chi({\mathcal{U}}).
\end{equation}

\begin{figure}
    \centering    
    	\includegraphics[width=0.25\linewidth]{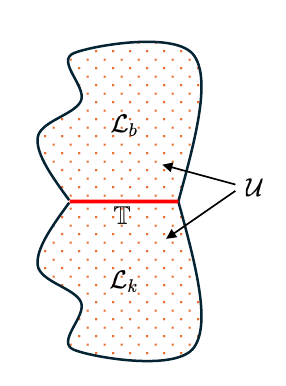}
    \caption{    A Schwinger-Keldysh retracing spacetime $\mathcal U$ can be constructed from bra and ket copies $\mathcal L_b, \mathcal L_k$ of a smooth real Lorentzian spacetime-with-boundary $\mathcal L$. To do so, one sews together $\mathcal L_b$ and $\mathcal L_k$  along corresponding pieces $\mathbb T$ (red) of their boundaries.   For such $\mathcal U$, the expression \eqref{eq:wPhiterms} is invariant under smooth deformations of $\mathcal L$.
  }
    \label{fig:retrace}
\end{figure}

We would also like to define the various terms in \eqref{eq:UepsrgR2} for cases where $\mathcal U$ intersects the timefold $\mathbb T$.   To do so, it is useful to recall that we are interested in the quantity 
$\int_{\mathcal{U}} \dd^2x \, \eta \sqrt{-g} \Phi R$ in the context of a Schwinger-Keldysh path integral and that, as a result, it should satisfy a {\it retracing identity}. To define this term, suppose that we have a smooth real Lorentz-signature spacetime $\mathcal L$, perhaps with boundary.  Let us now construct a Schwinger-Keldysh spacetime ${\mathcal U}$ with boundary      $\partial {\mathcal U}$ from two copies of $\mathcal L$.  We take one copy to be a bra-spacetime and the other to be a ket-spacetime.  We then sew the two together along some $\mathbb T \subset \partial \mathcal L$; see figure \ref{fig:retrace}.  Our retracing identity then requires the quantity

\begin{equation}
\label{eq:wPhiterms}
\int_{\mathcal{U}} \dd^2x \,  \sqrt{-g} \eta \Phi R + 2\int_{\partial \mathcal{U}}\dd x\, \sqrt{|h|}\eta \Phi K
\end{equation}
to be invariant under smooth deformations of $\mathcal L$.

As suggested above, we will continue to use the symbol $\mathbb T$ to denote the interface between the bra and ket copies of $\mathcal L$ even when this interface fails to be spacelike.  While the extrinsic curvature and dilaton gradient can be arbitrary at $\partial \mathcal L$, the condition that $\mathbb T$ is defined by sewing together corresponding parts of the bra and ket boundaries implies that they will be SK-continuous at $\mathbb T$; see again the discussion around figure \ref{fig:normals} for the definition of SK-continuity.  

For spacetimes 
with SK-continuous gradients, and where $\mathbb T$ is a codimension-1 manifold, this observation motivates us to define the $\eta \Phi R$ term in \eqref{eq:wPhiterms} to be simply the sum of the corresponding integrals in the bra and ket parts. I.e., when the gradients are SK-continuous, there is no delta-function-like contribution from $\mathbb T$ itself.  

\begin{figure}[h!]
    \centering    
\includegraphics[width=0.7\linewidth]{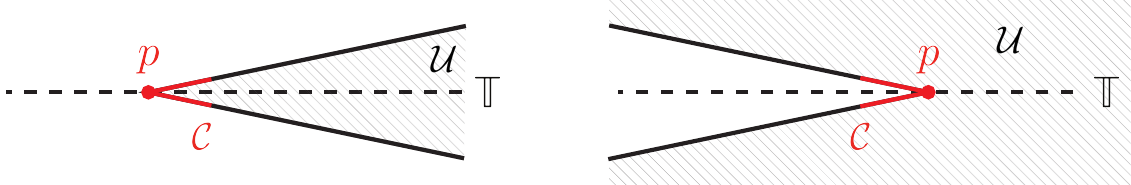}
    \caption{When an otherwise-spacelike $\mathcal C$ (red) segment of $\mathcal U$ (solid black) with SK-continuous tangent intersects a spacelike $\mathbb T$  (dashed black line) at a point $p$ where $\Phi$ is real, there are imaginary contributions to $\int_{\mathcal{C}}d x\, \sqrt{|h|}\eta \Phi K$. {\bf Left:} When $\mathcal{U}$ is locally convex at $p$, the contribution is
    $-i \eta \pi \Phi(p)$.  {\bf Right:} When $\mathcal{U}$ is locally concave at $p$, the contribution is
    $+i \eta \pi \Phi(p)$. }
    \label{fig:Tints}
\end{figure}

Similarly, when the tangent to $\partial \mathcal U$ is SK-continuous at such a $\mathbb T$, and when $\Phi$ is real at this intersection, we define the real part of the $\eta \Phi K$ term in \eqref{eq:wPhiterms} to be the sum of the corresponding bra and ket parts; i.e., there is again no delta-function-like contribution to the real part from the intersection with $\mathbb T$. We also declare there to be no imaginary 
delta-function-like contribution at points where timelike portions of an SK-continuous $\partial \mathcal U$ intersect spacelike portions of $\mathbb T$, or where 
spacelike portions of an SK-continuous $\partial \mathcal U$ intersect timelike portions of $\mathbb T$.
With these conventions, the retracing identity can immediately be seen to hold in the cases where it has been defined.  Still assuming $\Phi$ to be real at the intersection $\mathbb T \cap \partial \mathcal U$, we then define $\int_{\partial \mathcal{U}}\dd x\, \sqrt{|h|}\eta \Phi K$ to receive imaginary contributions $\mp i\eta \pi2$ from other types SK-continuous intersections with $\mathbb T$ as needed to maintain the validity of our retracing identity more generally; see figure \ref{fig:Tints}.

\begin{figure}[h!]
    \centering    
    	\includegraphics[width=\linewidth]{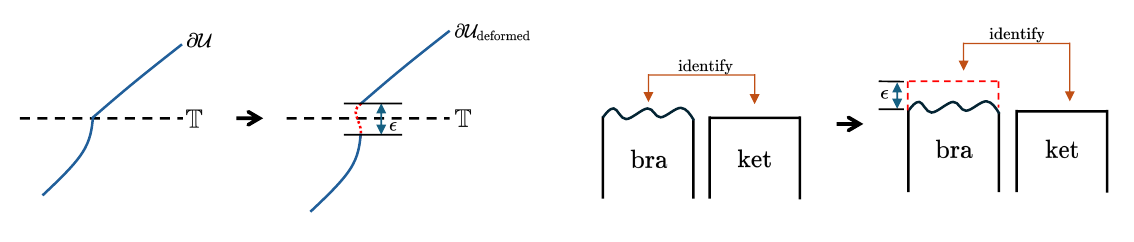}
    \caption{When one-sided limits of derivatives are well-defined at $\mathbb T$, we can define the integrals in \eqref{eq:wPhiterms} as limits of the results for SK-continuous spacetimes. {\bf Left:} The term $\int_{\partial \mathcal{U}}\dd x\, \sqrt{|h|}\eta \Phi K$ can be defined by deforming  $\partial \mathcal U$ to become SK-continuous in a region of size $\epsilon>0$ near $\mathbb T$ and then taking $\epsilon \rightarrow 0$.   
      {\bf Right:} 
      The term       $\int_{\mathcal{U}} \dd^2x \,  \sqrt{-g} \eta \Phi R$
      can be defined by deforming the fields to become SK-continuous in a region of size $\epsilon>0$ near $\mathbb T$ and then taking $\epsilon \rightarrow 0$.   
}
    \label{fig:SKreg}
\end{figure}

Thus far, we have defined the terms in \eqref{eq:wPhiterms} only when appropriate derivatives are SK-continuous at $\mathbb T$.  However, limits of spacetimes with SK-continous derivatives can lead of course to gradients with additional step-functions.  We may thus use this fact to extend the above definitions to spacetimes where one-sided limits of derivatives are well-defined at $\mathbb T$  but where SK-continuity does not hold.  To do so, we simply define each integral to be the limit of the corresponding integrals for the associated family of SK-continuous spacetimes; see figure \ref{fig:SKreg}.   One may check that  the factor of $\eta$ in both terms ensures that the same result is obtained whether the deformation is performed on the bra side of $\mathbb T$, on the ket side, or on a combination of the two.

\subsection{Complex Gauss-Bonnet}
\label{subsec:CGB}

A useful property of the definitions in section \ref{subsec:DefConv} is that they yield a complex Gauss-Bonnet theorem of the form \eqref{eq:UepsrgR2} for general $\mathcal U$, no matter how it intersects $\mathbb T$.  This result can be derived from the complex Gauss-Bonnet theorems \eqref{eq:CBG} for separate bra and ket spacetimes by studying how  both sides of \eqref{eq:UepsrgR2} change when we decompose   $\mathcal U$ into bra and ket parts $\mathcal U_b, \mathcal U_k$ and then consider \eqref{eq:UepsrgR2} (or equivalently 
\eqref{eq:CBG}) separately for each part.  Here $\mathcal U_b, \mathcal U_k$ are just the parts of $\mathcal U$ lying respectively in the bra and ket parts of the spacetime.  

\begin{figure}
    \centering    
    	\includegraphics[width=0.7\linewidth]{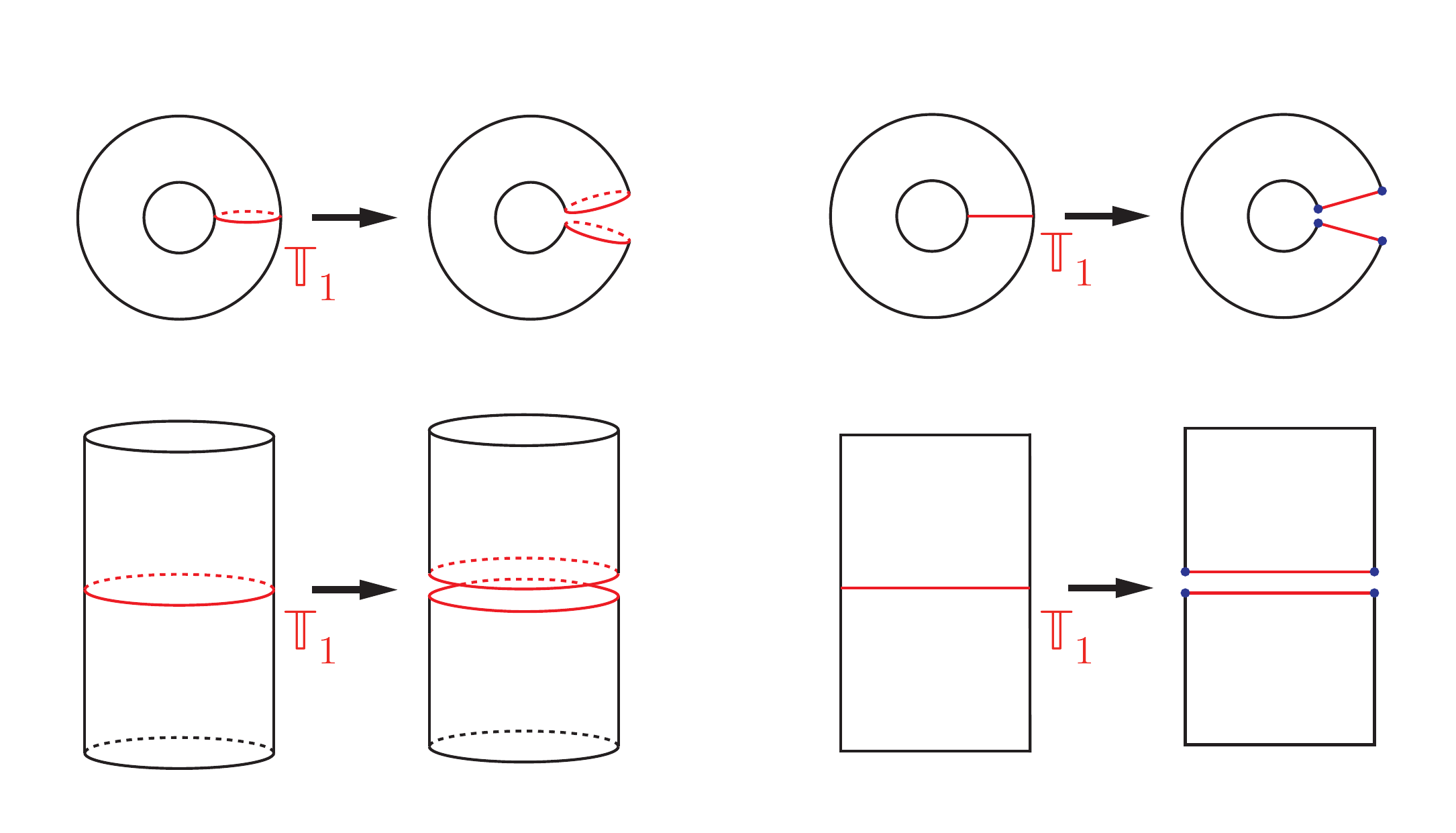}
    \caption{Cutting $\mathcal U$ along a connected piece $\mathbb T_1$ of the timefold $\mathbb T$.   
      {\bf Left:} When $\mathbb T_1$ is a circle, the cut will change the number of connected components $c$, handles $h$, and boundaries $b$ in one of two ways,  $(c,h,b) \rightarrow (c,h-1,b+2)$ (as in the top panel) or $(c,h,b) \rightarrow (c+1,h,b+1)$ (as at bottom).  But in either case the Euler character $\chi = 2c-2h -b$ remains unchanged.  Similarly, the contributions of the new boundaries (red) to $\int_{\partial \mathcal{U}}\dd x\, \sqrt{|h|}\eta  K$ must cancel by SK-continuity.  {\bf Right:} When $\mathbb T_1$ is a line segment, there are two ways in which the number of connected components $c$ and the number of boundaries $b$ may change, 
      $(c,h,b) \rightarrow {(c,h,b-1)}$   (as in the top panel) or $(c,h,b) \rightarrow (c+1,h,b+1)$ (as at bottom).  In either case $\chi = 2c-2h -b$ increases by $1$.  Contributions to $\int_{\partial \mathcal{U}}\dd x\, \sqrt{|h|}\eta  K$ from the new boundary-pieces (red) along $\mathbb T_1$ must cancel by SK-continuity, but there are additional imaginary contributions $-i \pi/2$ at each of the $4$ new corners (blue dots). As a result, in all cases the cut makes the same change to each of the two sides of \eqref{eq:UepsrgR2}.}
    \label{fig:cuts}
\end{figure}

We may thus also construct 
$\mathcal U_b, \mathcal U_k$  by slicing $\mathcal U$ apart along $\mathbb T$.  To study this process in detail, let us write $\mathcal U \cap \mathbb T$ as the union of a collection of disjoint connected codimension-1 manifolds $\mathbb T_i$. 
Starting with the original intact $\mathcal U$, we may make cuts along each $\mathbb T_i$ in turn.  For our two-dimensional spacetimes, each $\mathbb T_i$ is either a circle or a line segment.  Cutting along a circle adds two more circular boundaries to $\partial \cal U$, and either removes a handle from $\mathcal U$ or increases the number of connected components by $1$.  Since the Euler character of a surface with $c$ connected components, $h$ handles, and $b$ boundaries is $\chi = 2c-2h-b$, the increase in $b$ cancels against either the increase of $c$ or the decrease in $h$ and $\chi$ remains constant.  Similarly, using a regulating SK-continuous spacetime if necessary, we find that the left-hand-side of \eqref{eq:UepsrgR2} is also unchanged. This is manifest for the bulk term since the cut changes $\mathcal U$ only on a set of measure zero.  Furthermore, the two new boundary terms simply cancel against each other by SK-continuity; see figure \ref{fig:cuts} (left).

A similar argument holds when $\mathbb T_i$ is a line segment.  There are then 2 cases to consider as shown at right in figure \ref{fig:cuts}.  In the first case, the cut does not change the number of connected components.  Instead, it reduces by $1$ the number of boundaries.  Thus $\chi$ increases by $1$.  In the 2nd case, the cut {\it increases} the number of boundaries by $1$ but it also increases the number of connected components by $1$.  Thus $\chi$ again increases by $1$ and the right-hand-side of \eqref{eq:UepsrgR2} changes by $-4\pi i$.  In either case, the bulk term on the left-hand-side of \eqref{eq:UepsrgR2} is unchanged for the same reason as above.  Furthermore, the contributions to the boundary term from the new pieces of $\partial \mathcal U$ along $\mathbb T_i$ again cancel just as above.  

However, when cutting along a $\mathbb T_i$ that is a line segement, there are also new contributions to the boundary term from the $4$ new corners shown in figure \ref{fig:cuts}.  Rounding out the corners and comparing with figure \ref{fig:Kpoles} shows that each corner contributes $-i\pi/2$
to $\int_{\partial \mathcal{U}}d x\, \sqrt{|h|}\eta  K$.  As a result, after taking into account the explicit factor of $2$ in the boundary term, we find the left-hand side of  \eqref{eq:UepsrgR2} to change by $-4\pi i$ in perfect agreement with the right-hand side.  Since \eqref{eq:UepsrgR2} holds separately for $\mathcal U_b, \mathcal U_k$, it now follows that it also holds for general $\mathcal U$ as desired.  This agrees with the analysis of \cite{Colin-Ellerin:2020mva}, where \eqref{eq:UepsrgR2} was obtained by direct analytic continuation from a Euclidean geometry to a Schwinger-Keldysh spacetime in which both bra and ket parts were nontrivial.

\subsection{The action of the splitting surface}

We are now ready to compute the integral $\int_{\mathcal U} \dd^2x \eta \sqrt{-g} R$ for a small disk $\mathcal U$ around the splitting surface, which we denote here by $\Upsilon$.  In the limit of small $\mathcal{U}$, contributions to $\eqref{eq:UepsrgR2}$ from the smooth part of $\sqrt{-g} R$ may be ignored.  We may thus perform the computation in a Schwinger-Keldysh spacetime for which each bra or ket piece is just a region of Minkowski space.  

Since we allow an arbitrary conical singularity at the splitting surface, extrinsic curvatures in this spacetime are not SK-continuous.    We could thus use the prescriptions of section \ref{subsec:DefConv} to deform the desired spacetime to one that satisfies SK-continuity and where the splitting surface lies below the timefold.  We could then compute the result in that context.  However, we will instead simply use \eqref{eq:UepsrgR2} to compute the result directly.

\begin{figure}
	\centering
	\includegraphics[width=0.5\linewidth]{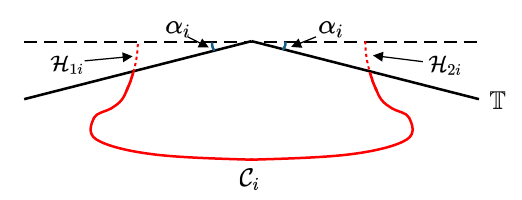}
	\caption{ $\mathcal{C}_i$ is the part of our $\partial \mathcal{U}$ in the $i$th ket/bra spacetime.   For small $\mathcal U$ we may take $\mathcal{C}_i$ to be a curve in Minkowski space.   We also choose the curve $\mathcal{C}_i$ to be hyperbolic near $\mathbb{T}$ so that it intersects $\mathbb{T}$ orthogonally.  The hyperbolic pieces of $\mathcal{C}_i$ are then connected by any smooth curve.  For $\alpha_i >0$, we  can extend the bra/ket spacetime up to the (smooth) dashed line and define an extended curve
$\widehat{\mathcal{C}}_i$ by adding in additional hyperbolic pieces ${\cal H}_{1i}, {\cal H}_{2i}$ as indicated (dotted lines). Since the endpoints of $\widehat{\mathcal C}_i$ lie on the dotted line we may associated the extended curve with rapidity $\widehat{\alpha}_i=0$. }
	\label{fig:conical}
\end{figure}

To do so, it is useful to write $\partial \mathcal{U} = \cup_{i=1}^{2n} \mathcal{C}_i$, where each $\mathcal{C}_i$ is the part of $\partial \mathcal{U}$ in a given bra or ket spacetime.   Since the path integral proposed in \cite{Marolf:2022ybi} allows arbitrary conical singularities at $\Upsilon$, we must consider the case where the region below the timefold $\mathbb{T}$ in the $i$th ket/bra spacetime is a wedge associated with a general boost-angle $\alpha_i$ as shown in figure \ref{fig:conical}. (This slightly extends the arguments of \cite{Colin-Ellerin:2020mva} which did not consider such $\alpha_i$.) Let us therefore take each $\mathcal C_i$ to be of the form shown in figure \ref{fig:conical}, having hyperbolic parts near $\mathbb {T}$ that will be orthogonal to $\mathbb {T}$, but with these hyperbolic pieces then connected by a smooth curve.  It will be useful below to first consider the case where $\alpha_i=0$ for all $i$, and then to generalize to $\alpha_i \neq 0$ by making use of the extensions shown in figure \ref{fig:conical}.

For the case $\alpha_i=0$, we can use two of the curve $\mathcal{C}_i$ to bound a region $\mathcal D$ in Minkowski space that is homeomophic to a disk. In this case we may write.

\begin{equation}
\int_{\partial \mathcal{D}}\dd x\, \sqrt{|h|} K = 2\int_{ \mathcal{C}_i}\dd x\, \sqrt{|h|} K.
\end{equation}
But since $\mathcal{D}$ is a disk, we have $\chi(\mathcal{D})=1$ and \eqref{eq:CBG} yields
\begin{equation}
\eta \int_{ \mathcal{C}_i}\dd x\, \sqrt{|h|} K = -\pi  i \ \ \ {\rm for} \ \ \ \alpha_i=0,
\end{equation}
where we have chosen to put the factor of $\eta$ on the left.  Evaluating \eqref{eq:UepsrgR2} for the case where all $\alpha_i$ vanish thus yields
\begin{equation}
\label{eq:UepsrgR3}
\int_{\mathcal{U}} \dd^2x \, \eta \sqrt{-g}  R
=4n\pi i -4\pi i \chi({\mathcal{U}}) = 4(n-1)\pi i \ \ \ {\rm for} \ \ \ \alpha_i=0,
\end{equation}
where we have used the fact that that we take $\mathcal U$ homeomorphic to a disk so that $\chi(\mathcal U)=1$.

Let us now consider the case where $\alpha_i\neq 0$ for some $i$.  As shown in figure \ref{fig:conical}, if we think of $\mathcal C_i$ as living in Minkowski space, for $\alpha_i >0$ our $\mathcal{C}_i$ can be related to another curve $\widehat{\mathcal{C}}_i$ in the same Minkowski space but for which the tangent vectors at the two endpoints have no relative boost (i.e., for which  $\widehat{\alpha}_i=0$).  The curve $\widehat{\mathcal{C}}_i$ is constructed from $\mathcal{C}_i$ by gluing appropriate hyperbolic segments $\mathcal{H}_{1i}, \mathcal{H}_{2i}$ to the endpoints of $\mathcal C_i$.  But for such segments it is easy to compute the integrals
\begin{equation}
\int_{ \mathcal{H}_{1i}}\dd x\, \sqrt{|h|} K = \int_{ \mathcal{H}_{2i}}\dd x\, \sqrt{|h|} K  = \alpha_i,
\end{equation}
so that \eqref{eq:UepsrgR3} becomes
\begin{equation}
\label{eq:UepsrgR4}
\int_{\mathcal{U}} \dd^2x \, \eta \sqrt{-g}  R
=4(n-1)\pi i -\sum_{i=1}^{2n} 4\eta_i \alpha_i,
\end{equation}
where $\eta_i= +1$ for ket spacetimes and $\eta_i=-1$ for bra spacetimes.
As a result, in the limit where $\mathcal U$ is a small disk in a continuous dilaton field we also have
\begin{equation}
\label{eq:UepsrgR5}
i\int_{\mathcal{U}} \dd^2x \, \eta \sqrt{-g} \Phi R
= -4(n-1)\pi  \Phi_\Upsilon - i \Phi_\Upsilon \sum_{i=1}^{2n} 4\eta_i \alpha_i,
\end{equation}
where $\Phi_\Upsilon$ is the value of the dilaton at the splitting surface.

It turns out that it will be convenient to describe the parameters $\alpha_i$ in terms of the contribution from a Gibbons-Hawking term evaluated along the timefold $\mathbb{T}$.  In particular, for the $i$th bra/ket spacetime $\mathcal{M}_i$ we find
\begin{equation}
2\int_{ \mathbb{T}} \dd x\, \sqrt{|h|} \Phi K_i = -4 \alpha_i\Phi_\Upsilon + 2\int_{ \mathbb{T}_{\text{smooth}}} \dd x\, \sqrt{|h|} \Phi K_i ,
\end{equation}
where $\mathbb{T}_{\text{smooth}}$ is part of the timefold away from $\Upsilon$ and symbol $K_i$ denotes the extrinsic curvature when considered as a boundary of the region $\mathcal{M}_i$. Furthermore, the final $\mathbb{T}_{\text{smooth}}$ term on the right can be neglected if we consider only contributions from inside some small $\mathcal{U}$ (so that $\mathbb T_{\text{smooth}}$ is small as well). For small $\mathcal{U}$ we thus find
\begin{equation}
\label{eq:UepsrgR6}
i\int_{\mathcal{U}} \dd^2x \, \eta \sqrt{-g} \Phi R + 2 i \sum_i \eta_i \int_{ \mathbb{T}\cap \mathcal{U}} \dd x\, \sqrt{|h|} \Phi K_i
= -4(n-1)\pi \Phi_\Upsilon  .
\end{equation}
As a result, so long as the action of a given ket/bra spacetime $\mathcal{M}_i$ is defined to include a Gibbons-Hawking term at the timefold $\mathbb{T}$, the only additional contribution from the region near $\Upsilon$ is the term $-4(n-1)\pi \Phi_\Upsilon$. It is this observation that leads to \eqref{eq:SUps}.

\section{Path integral computation of $\langle \Phi_h; \Upsilon|\Phi; \tilde T \rangle$}
\label{app:PIprop}
\label{sec:onecopy}

\begin{figure}
	\centering
\includegraphics[width=0.4\textwidth]{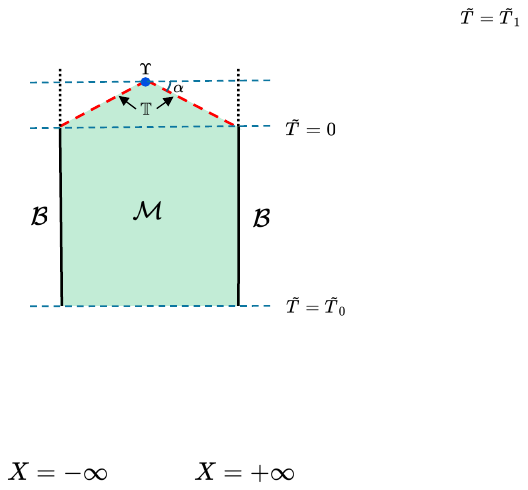}
	\caption{Spacetime configurations $\mathcal M$ are shown that contribute to the transition amplitude $\BraKet{\Phi; \tilde T}{\Phi_h;\Upsilon}$. For real spacetimes, the surfaces $\mathbb T$ and ${\tilde T}=\tilde T_0$ are Cauchy slices on which we respectively fix $\Phi_h$ and $\Phi_{\tilde{T}}$. The timefold $\mathbb T$ is chosen to pass through the extremal-dilaton point and intersects the asymptotic boundaries at time $\tilde{T}=0$, while $\tilde{T}= \tilde T_0$ is simply the slice where $\tilde{T}$ takes the value $\tilde T_0$. The parameter $\alpha$ denotes the boost angle between $\mathbb T$ and a symmetrically-oriented geodesic represented by the dashed line. On shell, the value of $\tilde T$ on the dashed line will be equal to $\delta$ and $\alpha$ will satisfy \eqref{eq:alphasol}.}
\label{Fig:action}
\end{figure}

We now present a semiclassical path integral computation of the quantity $\langle \Phi_h; \Upsilon|\Phi; \tilde T \rangle$ that we evaluated in section \ref{sec:scJT} using canonical methods.
The path integral of interest is the JT analogue of the one-ket part of the path integral of section \ref{sec:RTRenyi}.  In particular, in the leading semiclassical approximation we claim that
\begin{equation}
\label{eq:Kprod}
\langle \Phi_h; \Upsilon|\Phi; \tilde T \rangle\approx \sum_{\mathrm{saddles}} e^{i\eta S_{\rm dyn}[\mathcal{M}]},
\end{equation}
where $\mathcal{M}$ is the spacetime region shown in figure \ref{Fig:action}.  In writing \eqref{eq:Kprod} we have neglected the topological term in the JT action.  Since we include only disk contributions here, the topological term would contribute the same multiplicative factor to every term and would thus provide only an overall normalization.

As explained in section \ref{subsec:JTCCS},  the boundary conditions for our ket spacetime $\mathcal M$ involve a surface $\Sigma$ that runs along geodesics from $t_L=0$ on the left boundary to the extremal point $\Upsilon$, and then along another geodesic to $t_R=0$ on the right boundary.  We choose this surface $\Sigma$ to coincide with the timefold $\mathbb T$.

We also fix the extremal dilaton value $\Phi_h$ and the dilaton and metric profiles on a surface $\tilde T=\tilde T_0$, so in particular we fix $\Phi_{\tilde T} = \Phi$ at $\tilde T = \tilde T_0$.  In addition,  as shown in figure \ref{Fig:action}, in a real ket spacetime there is an interesting rapidity parameter $\alpha$ associated with the way in which the geodesics from the left and right boundaries intersect at $\Upsilon$.  The definition of $\alpha$ then generalizes to complex spacetimes by analytic continuation.  As described in section \ref{subsec:JTCCS}, imposing the equations of motion in the places the extremal-dilaton point $\Upsilon$ at $\tilde T=\delta$.

For simplicity, we assume below that the surface ${\tilde T}=\tilde T_0$ lies to the past of $\Upsilon$.   This implies $\tilde T_0<0$. Since $R=-2$ on shell, the Einstein-Hilbert term vanishes, and we only need to evaluate the remaining Gibbons-Hawking term. Since $K=0$ on any constant-$\tilde T$ surface, and since it also vanishes on the timefold $\mathbb{T}$ away from the dilaton-extremum $\Upsilon$, the Gibbons-Hawking term receives only contributions from the extremal point, the asymptotic timelike boundaries $\mathcal{B}$, and the corners where the spacelike and timelike boundaries meet.

As shown in figure \ref{Fig:action},
the future boundary of $\mathcal{M}$  will generally have a corner at the extremal-dilaton point $\Upsilon$.  The tangent to this boundary will thus have a discontinuity at $\Upsilon$ associated with a boost by some rapidity $2\alpha$.
Since the integral of $K$ over a small region containing the corner gives the boost of the tangent (see e.g. \cite{Neiman:2013ap}),
the contribution to $S_{dyn}[\mathcal{M}]$ from the corner at the dilaton-extremum $\Upsilon$  is
\begin{equation}
\label{eq:SUpscorner}
    S_{\Upsilon-\text{corner}}=2 \Phi_h (2\alpha).
\end{equation} 
This real contribution to $S[\mathcal{M}]$ should not be confused with the contribution  \eqref{eq:SUps} from a splitting surface to the path integral weight of a general R\'enyi geometry.

As stated above, our timefold is composed of two semi-infinite geodesics.  Furthermore,  as shown in figure \ref{Fig:action}, each geodesic intersects the surface $\Tilde T=\delta$ at a boost angle $\alpha$.  Recalling that the surface $\Tilde T=\delta$ is also the surface $t=0$ defined by the unshifted Schwarzschild time $t$, we see that each geodesic can be conveniently expressed in unshifted Schwarzschild coordinates as $t=\pm \alpha/r_s$, with the $+$ ($-$) sign referring to the left (right) piece of the timefold. Each geodesic thus reaches its respective asymptotic boundary at time $\tilde T=\delta-\alpha /r_s=\delta-\alpha \phi_b/\Phi_h$.  But since the location of the timefold is to be determined by the condition that these intersections lie at $\tilde T=0$, we find
\begin{equation}
\label{eq:alphasol}
\alpha = r_s \delta = \frac{\Phi_h \delta}{\phi_b}.
\end{equation}

To compute contributions to $S[\mathcal{M}]$ from the asymptotic boundaries, we first regulate these boundaries by deforming them to the finite surface $r=r_c$ for some constant $r_c$.  On such surfaces we have
\begin{equation}
    K=\frac{r_{c}}{\sqrt{r_{c}^{2}-r_{s}^{2}}}=1+\frac{1}{2} \frac{r_{s}^{2}}{r_{c}^{2}}+\cdots.
\end{equation}
Taking the limit $r_c \rightarrow \infty$ then yields the asymptotic boundary contribution to the Gibbons-Hawking term in the form
\begin{equation}
    S_{\text{b}}=2\int_\mathcal{B} \sqrt{h} \Phi (K-1)=2 \times 2 \int_{\tilde T}^{\tilde T_1} \dd\tilde T \, \sqrt{r_c^2-r_s^2} \left(\phi_b r_c\right)\left(\frac{1}{2} \frac{r_s^2}{r_c^2}+\ldots\right) =\frac{2 \Phi_h^2}{\phi_b}(\tilde{T}_1-\tilde T),
\end{equation}
where the second factor of $2$ in the second step comes from including both the left and right boundaries.

What remains is to compute the contribution from the four corners where spacelike and timelike boundaries meet.
The upper two corners are defined by the orthogonal intersection of the cutoff surface $r=r_c$ with a half geodesic at Schwarzschild time $t=\text{const.}$.  As a result, the real part of the contribution from those corners vanishes; see e.g \cite{Hayward:1993my}.
While there is a non-zero imaginary contribution $\int_{\text{corner}} K=-i\frac{\pi}{2}$ from each such corner (see e.g. \cite{Neiman:2013ap}, though this reference uses the opposite sign convention for the imaginary parts),
this contribution is independent of $\Phi_h$, $\Phi_{\tilde{T}}$ and $\tilde T$, and can thus be absorbed into the overall normalization of the path integral. 

The two lower corners require more careful analysis.  We will again absorb the imaginary contributions ${\rm Im} \, \int_{\text{corner}} K=-i\frac{\pi}{2}$ into the overall normalization of the path integral.  To compute the more interesting real part, we first need to construct the normal vectors to $\mathcal B$ and the Cauchy slice $\tilde T=\tilde T_0$. The normal to the Cauchy slice is
\begin{equation}
    n_a= \sqrt{1+X^2} \dd T_a .
\end{equation}
We should then study $\mathcal B$ by introudcing a cutoff version of this surface at $r=r_c$ for which
\begin{equation}
    X^2-\frac{r_c^2}{r_s^2 \cos^2 T}+1=0,
\end{equation}
so that the normal covector is
\begin{equation}
    m_a  = \frac{1}{\sqrt{N}} \left( -\frac{2r_c^2 \tan T}{r_s^2 \cos^2 T}\dd T_a+2X \dd X_a\right),
\end{equation}
for some normalization factor $N$.  Using results from \cite{Hayward:1993my} then gives the real part of the contributions from the lower corners as follows:
\begin{equation}
\label{eq:Sc}
    S_{\text{lower \ corners}}=2 \times 2\int_{\text{corner}} \sqrt{|q|} \Phi K \dd x = 4 \phi_b r_c \sinh ^{-1} (n \cdot m)= 4\phi_b r_s \sin T + \mathcal{O}(r_c^{-2}),
\end{equation}
where the integral is over a tiny region of the boundary containing a corner and
where the first factor of 2 accounts for the fact that there are two such corners, and that both corners contribute equally.

Taking the limit $r_c\rightarrow \infty$ then yields
\begin{equation}
    S_{\text{lower \ corners}}=4 \phi_b r_s \sin T= 4\Phi_h \tanh\left(\frac{\Phi_h}{\phi_b} (\tilde T-\delta)\right),
\end{equation}
so that the full dynamical action takes the form
\begin{eqnarray}\label{eq:LorentzianAction}
    S_{dyn}[\Phi,\Phi_h;\tilde{T}]: = \frac{2\Phi_h^2}{\phi_b}\left(2\delta_\pm -\tilde T\right) +4\Phi_h \tanh\left[\frac{\Phi_h}{\phi_b} (\tilde T-\delta_\pm)\right] \cr
    = \frac{2\Phi_h^2}{\phi_b} \tilde T \pm \left( 4\Phi_h \cosh^{-1} \left[\frac{\Phi_h}{\Phi} \right] - 4 \sqrt{\Phi_h^2 - \Phi^2}\right).
\end{eqnarray}
Here we have used \eqref{eq:alphasol}.  We also remind the reader that $\Phi$ is 
the value of $\Phi_{\tilde T_0}$ that sets the boundary conditions for $\mathcal M$, 
and we have used \eqref{eq:phitDef}  to define $\delta_\pm$ in terms of $\Phi_h$ and $\Phi$; i.e., we have
\begin{equation}
\label{eq:deltaPP}
    \delta_\pm= \tilde T \pm \frac{\phi_b}{\Phi_h}\cosh^{-1}\frac{\Phi_h}{\Phi}.
\end{equation}
This notation makes explicit the fact that, since $\cosh x$ is an even function on $\mathbb{R}$,  there are two possible values of $\delta$  for every choice of $\Phi,\Phi_{\tilde{T}},\tilde T$.
As a result,  for any given $\Phi_h$, $\Phi$, and $\tilde T$, there are two saddle point contributions to the right-hand-side of \eqref{eq:Kprod} that we will need to consider.  This is a result of the fact that every classical solution to JT gravity on a strip has a $\mathbb{Z}_2$ time-reversal symmetry.  This symmetry means that for given $\Phi_h$, the dilaton profile specified by $\Phi_{\tilde{T}} = \Phi$ occurs on two distinct constant $\tilde T$ slices (which, in the un-tilded global coordinates would like at $\pm T$ for some $T$; see figure \ref{fig:TwoSidedBH}). But a given solution has a well-defined value of $\delta$, so the formula to compute $\delta$ on the upper slice from $\Phi_{\tilde{T}}$ and $\tilde T$ must differ from the formula for the corresponding computation on the lower slice. This difference leads to the two signs in \eqref{eq:deltaPP}, which correspond to the choice of the upper or lower slice.  As a result,  at fixed $\Phi_t,\Phi_{\tilde{T}},\tilde T$ there are two possible values of $\delta$, depending on whether $\Sigma_{\tilde T}$ lies to the past of $\Upsilon$ or to its future.

As desired, taking into account the relations \eqref{eq:phitDef}, \eqref{eq:defQ}, and \eqref{eq:QT}, using \eqref{eq:LorentzianAction} to sum over this pair of saddles in \eqref{eq:Kprod} gives results that agree with \eqref{eq:PtPh} when the $Q$ integral is performed at leading-order in the saddle-point approximation.

\section{The Hartle-Hawking state in the $\Phi_{\tilde T}$ basis}
\label{sec:HHcalcs}
\label{sec:StatePrepartion}
\label{sec:setupHHW}

We include here a calculation of the wavefunction $\langle \Phi;\tilde T|\mathrm{HH} \rangle$ of the Hartle-Hawking state $|\mathrm{HH}\rangle$ of JT gravity with inverse temperature $\beta$.  Our starting point is the result
\begin{equation}
\BraKet{\Phi_h;\Upsilon}{\mathrm{HH}} = \Psi_{\mathrm{HH}}[\Phi_h]= \exp \left(2\pi \Phi_0+2\pi \Phi_h -\frac{\beta}2 \frac{\Phi_h^2}{\phi_b}\right),
 \label{eq:HJHH}
 \end{equation}
which was computed in \cite{Harlow:2018tqv} from a semiclassical Euclidean path integral.
Here the inverse temperature $\beta$ is the circumference of the thermal circle that defined the boundary conditions for the path integral.  As noted in section \ref{sec:scJT}, consideration of time-reversal symmetry shows that our conventions for $|\Phi_h;\Upsilon\rangle$ are consistent with the conventions of \cite{Harlow:2018tqv}.

Together with \eqref{eq:HJHH}, the results of section \ref{sec:scJT} give the following expression for the Hartle-Hawking wavefunction in terms of the basis $|\Phi; {\tilde T}\rangle$ at leading order in the semiclassical approximation \eqref{eq:sclim}:
\begin{equation}
\begin{split}
    \BraKet{\Phi;{\tilde{T}}}{\mathrm{HH}} &=\int \dd \Phi_h \dd Q \langle\Phi;{\tilde{T}}|Q;\tilde{T}\rangle\langle Q;\tilde{T}|\Phi_h\rangle\langle\Phi_h|HH\rangle\\
    &=\int \dd \Phi_h \dd Q
    \,\sqrt{\cosh\frac{Q}{4}}
    \exp \left(4i\Phi \sinh\frac{Q}{4}\right)\exp\left(-iQ\Phi_h-i\frac{2\Phi_h^2}{\phi_b}\tilde{T}\right)\Psi_{\rm HH}[\Phi_h]\\
    &=\int \dd \Phi_h \dd Q
    \,\sqrt{\cosh\frac{Q}{4}}
    \exp\left[i\left(4 \Phi \sinh{\frac{Q}{4}} -Q\Phi_h-\frac{2\Phi_h^2}{\phi_b}\tilde{T}\right)+\left(2\pi\Phi_0+2\pi\Phi_h-\frac{\beta}{2}\frac{\Phi_h^2}{\phi_b}\right)\right].
    \label{eq:HHpt4}
    \end{split}
\end{equation}

We begin by performing the $\Phi_h$ integral in equation (\ref{eq:HHpt4}). This is a Gaussian integral, so the one-loop-corrected saddle point result is exact. Taking a derivative of the logarithm of the integrand of equation (\ref{eq:HHpt4}) with respect to $\Phi_h$ and setting it equal to zero identifies the saddle point
\begin{equation}\label{eq:saddlephih}
    \Phi_{h}^{\text{SP}}=- \phi_b \frac{2\pi i + Q}{4\tilde{T}-i\beta}.
\end{equation}
The result of the integral is thus
\begin{equation}
    \BraKet{\Phi;{\tilde{T}}}{\rm HH} \approx \int \dd Q \sqrt{\cosh\frac{Q}{4}}\exp\left\{\phi_b\left[\frac{i}{2}\left(8 \frac{\Phi}{\phi_b} \sinh{\frac{Q}{4}} -\frac{\big(2\pi- i Q\big)^2}{4\tilde{T}-i \beta}\right)\right]\right\},
    \label{eq:HHpt5}
\end{equation}
where the symbol $\approx$ denotes the fact that, for simplicity, we have dropped the normalization factor (the one-loop correction) associated with the width of the Gaussian.  This factor is just an overall constant, and is also subleading in the semiclassical limit \eqref{eq:sclim}.  We have written the exponent with an explicit overall factor of $\phi_b$ in order to make features of this limit more manifest.

Note that the $Q$ integral is of the form $\int \dd Q g(Q) e^{\phi_b f(Q)}$ where $g(Q),f(Q)$ are of order $1$ as $\phi_b \rightarrow \infty$.  The saddle points of this integral are thus determined by setting $\partial_Q f$ to zero, which yields
\begin{equation}\label{eq:saddleQ}
    \frac{2\pi-iQ}{-4\tilde{T}+i\beta} +i \frac{\Phi}{\phi_b} \cosh\frac{Q}{4}=0.
\end{equation}
Interestingly, there is always a saddle at $Q=-2\pi i$.  But since $\sqrt{\cosh(-\pi i/2)}=0$, the integrand of \eqref{eq:HHpt5} in fact vanishes at this point and it cannot contribute.

However, there are infinitely many other solutions to \eqref{eq:saddleQ} when $\Phi_{\tilde{T}}\neq 0$. We will focus on the case $\tilde{T}<0$ below.  Since equation \eqref{eq:saddleQ} is transcendental, we will study these other saddles numerically.

\begin{figure}
    \centering
    \includegraphics[width=\linewidth]{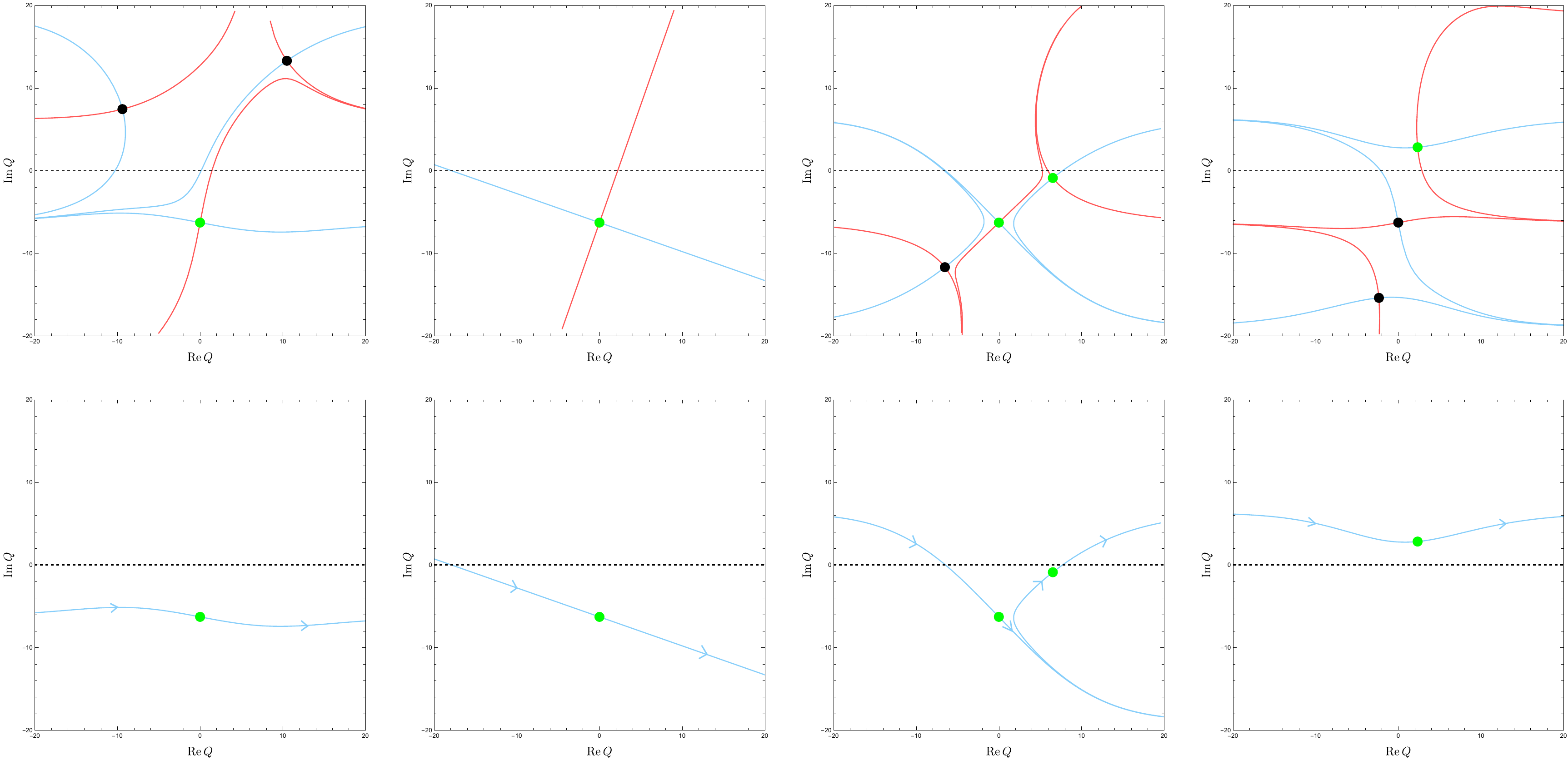}
     \caption{ {\bf Top row}: Saddle points (dots) and their steepest descent (light blue) and ascent (red) contours for the $Q$ integral with $\tilde{T}=-1,\beta=5,\phi_b=10$.   The values of $\Phi/\phi_b$ are  respectively $-0.5,0,0.5,1.5$.  The green dots are the saddles whose corresponding steepest ascent (red) contour intersects the original contour of integration (the real line, shown as a dashed black line). The first two columns show $\Phi\le 0$, in which case only the trivial saddle $Q=-2\pi i$ has $n_p\neq0$. For small positive $\Phi$ (3rd column) there is a non-trivial saddle with $n_p \neq 0$ (though the intersection number for the trivial saddle also remains non-zero).  However, above a threshold value $\Phi_{\#}(\tilde T, \beta)$ the intersection number for the trivial saddle drops to zero and only the non-trivial saddle remains. The semiclassical approximation to the integral changes continuously across $\Phi_{\#}$ since the contribution from the trivial saddle always vanishes. Our numerics gives $\Phi_{\#}\simeq 5.166(3)$ for the above parameters.      {\bf Bottom row}: Parameter values are identical to those in the top row, but only saddles with $n_p\neq 0$ are shown.  The descent contours passing through these saddles are shown as blue lines.  The original integration contour (dashed black line) can be deformed to a combination of these contours with orientations indicated by the arrows.      The third panel is subtle as the deformed contour is given by first following the blue curve from the left edge to the lower right of the figure and then following the other $U$-shaped blue curve so as to partially retrace (and cancel) the previous path before exiting through the right edge. The parameters in this panel are very close to threshold value $\Phi_{\#}(\tilde T, \beta)$ at which a Stokes ray develops and above which the lower saddle ceases to contribute.}
    \label{fig:StatePreparation}
\end{figure}

For each choice of $\beta,\tilde{T},\Phi$, we first find the various solutions to equation (\ref{eq:saddleQ}) in the complex $Q$ plane.  We then find the contours of constant phase originating from each saddle by requiring the imaginary part of $f(Q)$ to be constant. Finally we determine whether a given such contour is descending  or ascending by comparing the real part of $f(Q)$ with its value at the saddle. The results are shown in figure \ref{fig:StatePreparation}, with ascent contours colored red and descent contours colored light blue.  As described in appendix \ref{sec:SPA for GPI}, the contribution of a given saddle is multiplied by the intersection number $n_p$ of the corresponding ascent contour with the original contour of integration, which in our case is the real line. For example, the first panel in the first row of figure \ref{fig:StatePreparation} shows saddles defined by negative values $\Phi$.  For the trivial saddle at $Q=-2\pi i$, the ascent contour intersects once with the real line. Thus, (if the integrand did not vanish there) this saddle would contribute to our integral with a factor of $\pm 1$.  On the other hand, the ascent contours from the two saddles with larger ${\rm Im} \, Q$ do not intersect the real line thus do not contribute to our integral.  Thus the semiclassical approximation to our wavefunction simply vanishes at negative $\Phi$ (as one would expect from the fact that there are no allowed classical solutions with $\Phi_{\tilde T}<0$). 

In figure \ref{fig:StatePreparation}, saddles $p$ with non-zero $n_p$ are shown in green while saddles with $n_p=0$ are shown in black.
The multiplicity of solutions of \eqref{eq:saddleQ} arises from the $8\pi i$ periodicity of $\cosh \frac{Q}{4}$.  One generally expects that only the saddles closest to the real axis can have non-zero $n_p$, and this is indeed confirmed by numerical investigations.  Indeed,  the only saddles for which we find $n_p\neq 0$ have $-2\pi\le\Im Q<2\pi$. Thus we restrict ourselves to this region in presenting our numerical results. Figure \ref{fig:StatePreparation} shows only the case $\beta=5,\,\tilde{T}=-1$, though other choices of $\beta, \tilde T$ (with $\tilde T<0$) give similar results.

As we increase $\Phi$ toward zero from below, it is clear from \eqref{eq:saddleQ} that all non-trivial saddles recede to infinity (so that for $\Phi=0$ the only remaining saddle is at $Q=-2\pi i$). The interesting cases are thus the two right-most panels with $\Phi>0$.
For small positive $\Phi$ one finds a non-trivial saddle with $n_p \neq 0$, though the trivial saddle also remains relevant.  However, there is a threshold value $\Phi_{\#}(\tilde T, \beta)$ at which the ascent curve from the trivial saddle ends at the non-trivial saddle.  This situation is said to describe a Stokes ray; see e.g. \cite{Witten:2010cx}.  Above $\Phi_{\#}$ the intersection number $n_p$ vanishes for the trivial saddle. Nonetheless, the semiclassical approximation to the integral changes continuously across $\Phi{}_{\#}$ since the contribution from the trivial saddle always vanishes.

As a result, for $\Phi_{\tilde{T}}>0$ the initial state is given by
\begin{equation}
    \BraKet{\Phi;{\tilde{T}}}{HH} = \exp\left\{\phi_b\left[\frac{i}{2}\left(8 \frac{\Phi}{\phi_b} \sinh{\frac{Q^{\text{SP}}}{4}} -\frac{\big(2\pi- i Q^{\text{SP}}\big)^2}{4\tilde{T}-i \beta}\right)\right]\right\},
    \label{eq:HHpt6}
\end{equation}
where $Q^{\text{SP}}$ is the unique non-trivial solution to equation (\ref{eq:saddleQ}) with $-2\pi<\Im Q^{\text{SP}}<2\pi$. When $\Phi_{\tilde{T}}\le 0$, we have $\BraKet{\Phi_{\tilde{T}}}{\rm HH}=0$ to all orders in the semiclassical expansion.

\section{Saddle-point methods: a brief review}
\label{sec:SPA for GPI}

Consider an integral of the form
\begin{equation}\label{eq:originalIntegral}
    \int_{\Gamma} \dd z\, f(z)\exp[\lambda g(z)],
\end{equation}
where $\lambda$ is a constant, $f(z)$ and $g(z)$ are complex-valued functions, and $\Gamma$ is an appropriate contour of integration through the complex plane.   We wish to understand the asymptotic behavior of this integral at large (say, positive) real values of $\lambda$.  In this limit we may hope that the integral is controlled by stationary points (aka saddles) of the function $g(z)$.  We are especially interested in the case where the symbol $z$ denotes a large collection of complex coordinates, so that the contour $\Gamma$ is a high-dimensional manifold.

In general, the saddles of $g(z)$ do not lie on $\Gamma$. One must then ask if the contour $\Gamma$ can be deformed so as to pass through the relevant saddles in an appropriate manner (i.e., so that the integral can be approximated by an integral over the corresponding steepest descent contour).  While it can be complicated to study this question directly, the
analysis is greatly simplified by making use of tools from the
mathematics literatures of Morse theory and/or Picard-Lefshetz theory \cite{FAs,FP,AGV,BH,BH2,H}.  We review these tools briefly below.  Our presentation largely follows that  of \cite{Witten:2010cx}, to which we refer the reader for further details.

We will begin with the case of a one-dimensional integral.
Let us therefore consider a stationary point $p \in \mathbb{C}$ of the function $g(z)$.  There are two contours of interest associated with $p$: the steepest descent contour $\mathcal{J}_p$ (also called the downward flow) and the steepest ascent contour $\mathcal{K}_p$ (also called the upward flow). These contours are obtained by using the magnitude of $e^{\lambda g}$, together with the metric defined by the line element $\dd s^2 = \dd z \dd\bar z$, to generate a gradient flow and by then following that flow upward or downward from $p$.  The relevant flow equations are then
\begin{equation}
\label{eq:1dflow}
\frac{\mathrm{d} z}{\mathrm{~d} t}=\pm \frac{\partial \overline{g}}{\partial \bar{z}}, \quad \frac{\mathrm{d} \bar{z}}{\mathrm{~d} t}=\pm \frac{\partial g}{\partial z},
\end{equation}
 where the positive/negative signs correspond to $\mathcal{K}_p$, $\mathcal{J}_p$ respectively. An important property of \eqref{eq:1dflow} is that they require $\operatorname{Im} g$ (and thus the phase of $\exp[\lambda g(z)]$) to be constant along a flow line:
\begin{equation}
\frac{\mathrm{d} \operatorname{Im} g}{\mathrm{d} t}=\frac{1}{2 i} \frac{\mathrm{d}(g-\overline{g})}{\mathrm{d} t}=\frac{1}{2 i}\left(\frac{\partial g}{\partial z} \frac{\mathrm{d} z}{\mathrm{~d} t}-\frac{\partial \overline{g}}{\partial \bar{z}} \frac{\mathrm{d} \bar{z}}{\mathrm{~d} t}\right)=0.
\end{equation}

\begin{figure}
    \centering
    \includegraphics[width=0.3\linewidth]{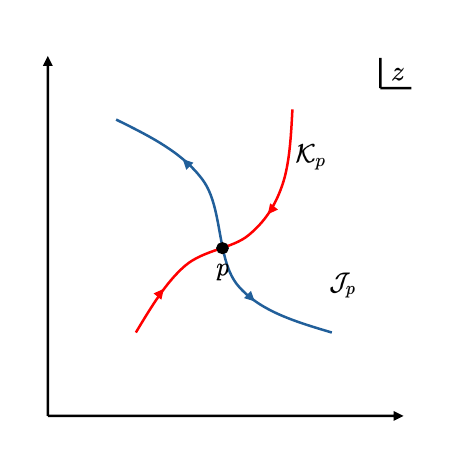}
    \caption{$\mathcal{J}_p$ denotes the steepest descent contour, while $\mathcal{K}_p$ denotes the steepest ascent contour. }
    \label{fig:flows}
\end{figure}

Since the gradient of $g$ vanishes at $p$, a nontrivial flow cannot reach $p$ at any finite value of $t$.  Instead,
we are interested in flows that approach the point $p$ as $t \rightarrow -\infty$.
  Since $g(z)$ is analytic, the Hessian of $g$ at generic saddles will have one positive eigenvalue and one negative eigenvalue.  This means that the flow near $p$ has the general form shown in figure \ref{fig:flows}, with one stable direction (along which the forward flow moves points toward $p$) and one unstable direction (along which the forward flow moves points away from $p$).  As a result, there are precisely two ascending flows and two descending flows that reach $p$ at $t=-\infty$.  The contour $\mathcal{J}_p$ is the union of the descending flows from $p$, while the contour $\mathcal{K}_p$ is the union of the ascending flows from $p$.

Rather than solving the flow equations directly,  when there is only one complex variable it is generally simpler method is just to use the fact that the phase is constant along both the ascent and descent contours and to thus merely solve $\Im \lambda g(z)=\Im \lambda g(z_0)$ where $z_0$ is the saddle point of $\lambda g(z)$.  While the solution is then the union $\mathcal{J}_p \cup \mathcal{K}_p$, it is straightforward to then simply check which curves describe ascending flows and which curves describe descending flow.

The important result from \cite{FAs,FP,AGV,BH,BH2,H} is that, without changing the value of  the integral, the original contour $\Gamma$ can be deformed to a contour $\tilde \Gamma$ consisting of $n_p$ copies of each descent curve $\mathcal{J}_p$, where $n_p$ is the (signed) intersection number of $\mathcal{K}_p$ with $\Gamma_R$.  We can thus diagnose whether or not a stationary point of the action contributes to the semiclassical approximation of the path integral by determining whether  $\mathcal{K}_p$ intersects nontrivially with the original contour $\Gamma$.  In the large $\lambda$ limit we can approximate the integral over each $\mathcal{J}_p$ by a Gaussian to find \begin{equation}
    \int_\Gamma \dd z\, f(z)\exp[\lambda g(z)]\sim \sum_p n_p \sqrt{\frac{2\pi}{\lambda |g''(z_p)|}} f(z_p) \exp[\lambda g(z_p)],
\end{equation}
where $z_p$ is the relevant saddle point of the exponent in equation (\ref{eq:originalIntegral}).
Here since the factors in front of the exponential are $\mathcal{O}(1)$ correction and are only related to normalization, we will just drop them unless $f(z_0)$ is $0$.

Similar results hold for higher-dimensional integrals, though in the higher-dimensional case there is an infinite set of ascending flows $p$ as well as an infinite set of descending flows.  The union over all ascending flows from $p$ defines $\mathcal{K}_p$, while union over all descending flows defines $\mathcal{J}_p$. For an integral over a contour with $d$ real dimensions in the $d$-dimensional complex plane $\mathbb{C}^d$, the resulting contours $\mathcal{K}_p$ and $\mathcal{J}_p$ each have the same real dimension as $\Gamma$. The key result is again that, without changing the value of the integral, $\Gamma$ can be deformed  to the union over $p$ of $n_p$ copies of $\mathcal{J}_p$, where $n_p$ is the intersection number of $\mathcal{K}_p$ with $\Gamma_R$.  Alternatively, it is always possible to write such higher-dimensional integrals as multiple one-dimensional integrals, and to perform the integrals one by one.  The parameter $d$ is of course infinite for the full gravitational path integral, but the corresponding result followed by treating the path integral as the  $d\rightarrow \infty$ limit of a finite-dimensional integral.

~~~~~~~~~~~~
\addcontentsline{toc}{section}{References}
\bibliographystyle{JHEP}
\bibliography{references}

\end{document}